\documentstyle[12pt]{article}

\catcode`\@=11

\@addtoreset{equation}{section}

\begin{document}

\baselineskip 24pt

\begin{titlepage}
\hfill
\vbox{
\hbox{SHEP 95-17}
\hbox{OUTP-95-19P}
\hbox{hep-ph/9505326} }

\begin{center}
{\Large{\bf
Resolving the Constrained
Minimal and Next-to-Minimal
Supersymmetric Standard Models }}
\medskip \\
S. F. King \footnote{Email \quad \tt king@soton.ac.uk}
\\
{\it Physics Department, University of Southampton, Southampton, SO9 5NH, U.K.}
\\ \vspace{0.05in}
\bigskip
P. L. White \footnote{Email \quad \tt plw@thphys.ox.ac.uk}
\\
{\it Theoretical Physics, University of Oxford,
1 Keble Road, Oxford, OX1 3NP, U.K.} \\
\vspace{0.1in}

{\bf Abstract} \end{center} \setcounter{page}{0}
{We perform a detailed analysis of the  next-to-minimal supersymmetric
standard model (NMSSM), imposing the constraints of two-loop gauge
coupling unification, universal soft supersymmetry breaking and the
correct pattern of electroweak symmetry breaking. We compare our
results with those for the minimal supersymmetric standard model (MSSM)
using closely related techniques and, as far as possible, a common set
of input and output variables.
In general, in the constrained NMSSM,
there are much stronger correlations
between parameters than in the constrained MSSM,
and we map out the allowed parameter space.
We also give a detailed discussion of
how to resolve the two models experimentally, concentrating primarily
on the prospects at LEPII.
We find that, for top mass
$\stackrel{>}{\sim}$150GeV, the constrained NMSSM is only viable in
regions where its spectrum is in general
very similar to that of the MSSM, although there are exceptions
which we explore. For example, in some corners of parameter space,
the lightest CP-even Higgs boson in the constrained NMSSM may be
detected at LEPII with singlet-diluted couplings which may
allow it to be distinguished from that of the MSSM.
However, if small universal gaugino mass $M_{1/2}$
is required, then we expect a standard model-like Higgs boson
which may be detected at LEPII, together with
a very characteristic Higgs and SUSY spectrum.
We also study fine-tuning in the
constrained NMSSM,
which is typically a more severe constraint than in the MSSM, and
give a simple analytical discussion of the potential and spectrum.}

\end{titlepage}

\section{Introduction}

Supersymmetry (SUSY) \cite{SUSY} is a well studied extension to the
standard model which reproduces the successes of the standard model,
including its radiative corrections, to the current very high
level of accuracy. The existence of TeV-scale SUSY would greatly ease the
naturalness problems associated with the construction of a  grand
unified theory (GUT) of the strong and electroweak interactions. The
basic idea of GUTs is that the gauge couplings, which govern the
strength of the strong and electroweak interactions at low energy, are
actually equal to some unified coupling $g_X$ at some very high scale
$M_X$ due to their renormalisation group (RG) running \cite{GQW}.  The
original motivation for SUSY broken at a TeV was to help to stabilise
the Higgs mass against GUT scale quadratic radiative corrections. The
potent combination of SUSY and GUTs has recently found some indirect
experimental support due to the accurate measurement of the strong and
electroweak couplings  on the Z pole by LEP.  These measurements are
inconsistent with gauge coupling unification if a standard model desert
is assumed, but are consistent with unification if a SUSY desert above
the TeV scale is assumed \cite{Amaldi}. This does not of course
constitute real evidence for either SUSY or GUTs, but it does provide
sufficient motivation for the detailed studies which have been made
since this observation.

In the first post-LEP analyses \cite{Amaldi} the whole SUSY spectrum
was either assumed to be degenerate at the scale $M_{SUSY}$, or smeared
around this scale. In reality the SUSY partners may have a complicated
spectrum, parametrised by a large number of soft SUSY-breaking
parameters, which may spread over one or two orders of magnitude of
masses. In order to reduce the number of independent soft SUSY-breaking
parameters one may appeal to supergravity or superstring scenarios
which involve the notion of SUSY breaking in a hidden sector coupled
only gravitationally to our observable sector. By this means, in the
minimal supersymmetric standard model (MSSM), one ends up with just
four independent soft SUSY breaking parameters: $m_0$, $M_{1/2}$, $A_0$
and $B_0$ corresponding to the universal soft scalar, gaugino,
trilinear and bilinear couplings, respectively. Using these four
parameters together with the bilinear Higgs term $\mu_0$ and the top
quark Yukawa coupling, which plays an important role in driving
electroweak symmetry breaking \cite{ir}, several groups have performed
an RG analysis whose goal is to predict the SUSY and Higgs spectrum
and then use this spectrum as the basis of a more reliable estimate of
gauge coupling unification by running the gauge couplings through the
various thresholds from the Z mass $M_Z$ up to a unification scale
$M_X$ \cite{RR,all}. Apart from the constraints of gauge coupling
unification and correct electroweak symmetry breaking, there are
various other phenomenological and cosmological constraints  which may
be applied and recent studies have  concluded that it is possible to
satisfy all these constraints simultaneously \cite{all}.

In this paper we shall consider both the constrained MSSM
discussed above and
a slightly different low
energy SUSY model, but one which is equally consistent with
gauge coupling unification, namely the so-called next-to-minimal
supersymmetric standard model (NMSSM) \cite{NMSSM1,NMSSM2,NMSSM3}.
The basic idea of the NMSSM is to add just one extra gauge singlet
superfield $N$ to the spectrum of the MSSM, and to
replace the $\mu$-term in the
MSSM superpotential with a purely cubic superpotential,
\begin{equation}
\mu H_1H_2 \longrightarrow \lambda NH_1H_2 - {k\over 3} N^3.
\end{equation}
The motivation for this ``minimal non-minimal'' model is
that it solves the so-called
$\mu$-problem of the MSSM \cite{mu}
in the most direct way possible by
eliminating the $\mu$-term altogether,
replacing its effect by the vacuum expectation value (VEV)
$<N>=x$, which may be naturally related to the usual Higgs VEVs
$<H_i>=\nu_i$.
There are other solutions to the $\mu$-problem \cite{mu}. Also, the
inclusion of singlets may cause the destabilisation of the hierarchy if
there are strong couplings to super-heavy particles such as Higgs
colour triplets \cite{stabone}.  Recently similar effects have been
shown to result from non-renormalisable operators suppressed by powers
of the Planck mass  \cite{stabtwo}. The dangerous non-renormalisable
operators would require that gravity violate the $Z_3$ symmetry which
is respected by the renormalisable operators of the theory. Our view is
that since these effects are model-dependent the NMSSM is as well
motivated as the MSSM and should be studied to the same level of
approximation. Only by so doing may the two models be
phenomenologically compared when (or if) Higgs bosons and SUSY
particles are discovered.

The present paper is not the first to discuss the effect of unification
and electroweak symmetry breaking constraints in the NMSSM. The
original analysis of ref.\cite{NMSSM3}, has recently been up-dated
\cite{ell,ell2,ell3,ekw0}. However only one of these analyses
\cite{ekw0} considered the effect of low-energy Higgs and SUSY particle
threshold effects on gauge coupling unification. As in the MSSM, such
effects play an important role in determining the unification
parameters, and in constraining the parameter space of the model. In
this analysis \cite{ekw0}, and the present analysis, we input the gauge
couplings $g_1(M_Z)$ and $g_2(M_Z)$ and run them up through the 24 SUSY
and Higgs thresholds to find $M_X$ and $g_X$, which must of course be
consistent with our original input values. By iterating this procedure
we obtain solutions which satisfy both the requirements of correct
electroweak symmetry breaking and coupling constant unification {\it
simultaneously}.

The result of these recent analyses \cite{ell,ell2,ell3,ekw0} is that
the constrained NMSSM is always quite close to the MSSM limit. In
effect the extra singlet decouples both from the Higgs sector and the
neutralino sector, making the problem of resolving the NMSSM from the
MSSM extremely difficult. Although there are already a large number of
analyses of the constrained MSSM in the literature \cite{all},
based on different techniques, and using different input and output
parameters, none of these analyses may be easily be extended to include
the constrained NMSSM. In the present paper we shall introduce a set of
input and output variables for the constrained MSSM which most closely
resembles those which are appropriate for our analysis of the
constrained NMSSM, thereby enabling both models to be dealt with on the
same footing, and enabling the results of both models to be
meaningfully compared.  The purpose of the present paper is therefore
to analyse the constrained MSSM and NMSSM models in a unified way,
which enables a comparison of the results of the two models, and hence
to address the important phenomenological question of how the two
models may be resolved experimentally.

The outline of the remainder of the paper is as follows.  In section 2
we introduce the NMSSM and the MSSM.  In section 3 we discuss our
unifying approach and methods for the two models. In section 4 we apply
these methods to the  MSSM and discuss the basic features of this
model. In section 5 we describe the results of our analysis of the
NMSSM. Section 6 addresses the  phenomenological question of how the
two models might be resolved experimentally at forthcoming colliders,
and section 7 concludes the paper. Appendices A and B supply
renormalisation group equations and radiatively corrected mass
matrices.

\section{The MSSM and NMSSM}

The MSSM has, in addition to the usual matter and gauge particle
content, a Higgs sector containing two Higgs doublets $H_1$ and $H_2$
\cite{SUSY}. The superpotential for the MSSM is of the
form
\begin{equation}
W_{MSSM} = h_u Q H_2 u^c + h_d Q H_1 d^c + h_e L H_1 e^c
              + \mu H_1 H_2,
\end{equation}
and the most general soft--breaking potential with our conventions is
\begin{eqnarray}
V_{MSSM}^{SOFT} & = & m_Q^2 |\tilde{Q}|^2 + m_u^2 |\tilde{u^c}|^2
              + m_d^2 |\tilde{d^c}|^2 + m_L^2 |\tilde{L}|^2
              + m_e^2 |\tilde{e^c}|^2 + m_{H_1}^2 |H_1|^2 \nonumber \\
        & + & m_{H_2}^2 |H_2|^2
            +  \frac{1}{2} (  M_1 \bar{\lambda_1} \lambda_1
                            + M_2 \bar{\lambda_2} \lambda_2
                            + M_3 \bar{\lambda_3} \lambda_3 )
                  \nonumber \\
        & - & (  h_u A_u \tilde{Q} H_2 \tilde{u^c}
               + h_d A_d \tilde{Q} H_1 \tilde{d^c}
               + h_e A_e \tilde{L} H_1 \tilde{e^c}
               - B \mu H_1 H_2 + h.c. ),
\end{eqnarray}
where gauge and generation indices are understood, $H_1 H_2 = H_1^0
H_2^0 - H_1^- H_2^+$, with $H_1^T = (H_1^0,H_1^-)$ and $H_2^T = (H_2^+,
H_2^0)$. The chiral superfields $Q$ contain the left--handed quark
doublets; $L$ the left--handed lepton doublets; $u^c$, $d^c$ and $e^c$
the charge conjugates of the right--handed up--type quarks,
right--handed down--type quarks and right--handed electron--type
leptons respectively. In the potential (rather than the superpotential)
we employ the usual convention that scalar components of Higgs
superfields are denoted by the same symbol as the corresponding
superfield, but that the scalar components of matter superfields are
tilded. $\lambda_1$, $\lambda_2$, $\lambda_3$ are the gauginos
corresponding to the $U(1)$, $SU(2)$, $SU(3)$ gauge groups
respectively, and are here Majorana fermions.  The low--energy spectrum
of the MSSM contains two CP--even Higgs scalars, one CP--odd Higgs
scalar, and two charged Higgs scalars.

The $\mu H_1 H_2$ term and its associated $\mu B H_1 H_2$ term are
necessary in order to ensure that the correct pattern of electroweak
symmetry breaking occurs. Thus the VEVs of $H_1$ and $H_2$ may be
taken to be of the form
\begin{equation}
\label{vevs}
<H_1> = \left( \begin{array}{c} \nu_1 \\ 0 \end{array} \right),
\: \: \:
<H_2> = \left( \begin{array}{c} 0 \\ \nu_2 \end{array} \right),
\end{equation}
where $\nu_1$ and $\nu_2$ are positive reals, $\sqrt{\nu_1^2 + \nu_2^2}
= \nu = 174$ GeV, and we define $\tan \beta = \nu_2 / \nu_1$. It might
be expected that the natural scale for $\mu$ should be the Planck mass,
or at least many orders of magnitude greater than the soft masses, but
in fact this does not allow correct electroweak symmetry breaking. This
is the $\mu$ problem. There has been extensive discussion of this
problem in the literature, including attempts to solve it
\cite{mu}.

The introduction of a gauge singlet superfield $N$ eliminates the
necessity for a dimensionful coupling in the superpotential since its
VEV plays the role of this coupling. We then obtain the NMSSM
\cite{NMSSM1,NMSSM2,NMSSM3}. The superpotential for the NMSSM is
\begin{equation}
\label{nmssmsuperpotential}
W_{NMSSM} = h_u Q H_2 u^c + h_d Q H_1 d^c + h_e L H_1 e^c + \lambda N
              H_1 H_2 - \frac{1}{3} k N^3.
\end{equation}
The general soft--breaking potential in our conventions is given by
\begin{eqnarray}
V_{NMSSM}^{SOFT} & = & m_Q^2 |\tilde{Q}|^2 + m_u^2 |\tilde{u^c}|^2
              + m_d^2 |\tilde{d^c}|^2 + m_L^2 |\tilde{L}|^2
              + m_e^2 |\tilde{e^c}|^2 \nonumber \\
         & + &  m_{H_1}^2 |H_1|^2 + m_{H_2}^2 |H_2|^2
              + m_N^2 |N|^2 + \frac{1}{2}
                          (  M_1 \bar{\lambda_1} \lambda_1
                           + M_2 \bar{\lambda_2} \lambda_2
                           + M_3 \bar{\lambda_3} \lambda_3 )
                  \nonumber \\
        & - & (  h_u A_u \tilde{Q} H_2 \tilde{u^c}
               + h_d A_d \tilde{Q} H_1 \tilde{d^c}
               + h_e A_e \tilde{L} H_1 \tilde{e^c} \nonumber \\
        &   &  + \lambda A_{\lambda} N H_1 H_2
               + \frac{1}{3} k A_k N^3 + h.c. ).
\label{softbreak}
\end{eqnarray}
The low--energy spectrum of the NMSSM contains three CP--even Higgs
scalars, two CP--odd Higgs scalars, and two charged Higgs scalars.

The cubic term in $N$ in the NMSSM superpotential is necessary in
order to avoid a $U(1)$ Peccei--Quinn symmetry which, when the fields
acquire their VEVs, would result in a phenomenologically unacceptable
axion in the particle spectrum. However, there still remains a
${\hbox{Z$\!\!$Z}}_3$
symmetry, under which all the matter and Higgs fields $\Phi$ transform
as $\Phi \rightarrow \alpha \Phi$, where $\alpha^3=1$.  This
${\hbox{Z$\!\!$Z}}_3$
symmetry may be invoked to banish such unwanted terms in the
superpotential as $H_1 H_2$, $N^2$ and $N$, all of which would have
massive parameters associated with them.

It has recently been indicated that gauge singlet fields may induce a
destabilisation of the gauge hierarchy in the presence of
non--renormalisable operators suppressed by powers of the Planck mass
\cite{stabtwo}.  These operators can only exist, however, if gravity
violates the ${\hbox{Z$\!\!$Z}}_3$ symmetry which is respected by the
renormalisable operators of the theory. The coefficients of such
operators, if they exist, have not yet been calculated and their size
and importance is unclear.  Furthermore, if the singlet couples
strongly to superheavy particles, such as Higgs colour triplets,
destabilisation of the gauge hierarchy may result \cite{stabone}.
However, this problem is strongly dependent on the  structure of the
GUT model, and we shall therefore not discuss it here. However we note
that, after spontaneous breaking, the ${\hbox{Z$\!\!$Z}}_3$ symmetry
must result in stable domain walls at the electroweak scale
\cite{walls}, a cosmological catastrophe which can however be avoided
by allowing explicit ${\hbox{Z$\!\!$Z}}_3$ breaking by terms suppressed
by powers of the Planck mass which will ultimately dominate the wall
evolution \cite{NMSSMwalls} without affecting the phenomenology of the
model.  The important question of whether such a cosmic catastrophe can
be avoided without destabilising the hierarchy is one which we shall
not address here.

The NMSSM thus has its own particular problems, so we cannot claim that
it is superior to the MSSM. However, it seems to us that it deserves as
close a scrutiny as the MSSM in order that the models may be compared
with experiment.

Unlike the case in the MSSM, where it is possible to derive simple
constraints which test whether correct electroweak symmetry breaking
will occur (at least at tree level), the possible vacuum structure of
the NMSSM is very complicated. We must always check that a particular
selection of parameters in the low energy Higgs potential will not
result in the VEVs breaking electromagnetism.  The condition that
electromagnetism is not broken simply reduces to requiring that the
physical charged Higgs mass squared be non--negative \cite{NMSSM3}. It
can be shown, at tree level, that spontaneous ${CP}$
violation does not occur in a wide range of SUSY models including the
NMSSM \cite{nocpviol}. Given that these conditions are satisfied, we
are left with a choice of VEVs for $H_1$ and $H_2$ as in Eq.
(\ref{vevs}) and with $<N>=x$, where $\nu_2$ is positive real
and $\nu_1$ and $x$ are real \cite{NMSSM3}. We shall define
$\tan\beta$ as before, and $r=x/\nu$.

In the squark and slepton sectors, there exists the possibility of
VEVs breaking electromagnetism or colour (or both).
Derendinger and Savoy \cite{NMSSM2} have
formulated simple conditions which determine in which regions of
parameter space such VEVs do not occur. The
condition that we have no slepton VEVs is
\begin{equation}
A_e^2 < 3 (m_e^2 + m_L^2 + m_{H_1}^2),
\label{slepvev}
\end{equation}
where the parameters are defined by the soft SUSY breaking potential in
Eq. (\ref{softbreak}).  This constraint is derived from the tree--level
potential under certain approximations, and should be tested at a scale
of order $A_e/h_e$, a typical VEV (for the slepton case).  A similar
condition on squark parameters will ensure the absence of
colour--breaking squark VEVs:
\begin{equation}
A_t^2 < 3 (m_T^2 + m_Q^2 + m_{H_2}^2).
\end{equation}
The reliability of the
results has been discussed in the literature \cite{NMSSM3}. We take
them as providing a coarse indication of when sparticle VEVs
are likely to occur.

In the analysis which follows, we shall assume that gauge coupling
unification {\it does} occur, that is, that $SU(3)_c \otimes SU(2)_L
\otimes U(1)_Y$ does embed in a simple group defined by one gauge
coupling constant. There are of course other models in the literature
not based on a simple gauge group, in which exact gauge coupling
unification does not occur.

Finally we note that there is a well defined limit of the NMSSM in
which the components of the  singlet decouple from the rest of the
spectrum which therefore resembles that of the MSSM (assuming no
degeneracies of the singlet with the other particles of similar spin
and CP quantum numbers which may lead to mixing effects which will
enable the NMSSM to be distinguished from the MSSM even in this limit.)
This limit is simply \cite{NMSSM3}: $k\rightarrow 0, \lambda
\rightarrow 0, x\rightarrow \infty$ with $kx$ and $\lambda x$ fixed.

\section{Methodology}

\subsection{Scaled Parameters}

The NMSSM is defined at the GUT scale, $M_X$, by the Yukawa couplings
$h_{t0}$, $\lambda_0$ and $k_0$ (we neglect all other Yukawa couplings
since they are small compared to the third generation couplings) and by
the soft parameters $m_0^2$, the universal scalar mass squared, $A_0$,
the universal trilinear coupling, and $M_{1/2}$, the universal gaugino
mass. The MSSM is similarly defined, except that $\lambda_0$ and $k_0$
are replaced by $\mu_0$, the $H_1 H_2$ superpotential coupling, and its
associated soft--breaking parameter $B_0$. Thus, the NMSSM is defined
by the set of parameters ${\cal P}_{NM} = \{ h_{t0}, \lambda_0, k_0,
A_0, m_0^2, M_{1/2} \} $ and the MSSM is defined by the set ${\cal
P}_M= \{ h_{t0}, \mu, B_0, A_0, m_0^2, M_{1/2} \}$. In the NMSSM, we
shall often find it convenient to use the variable $A_0/m_0$ in our
discussion, since this seems to classify the interesting regions of
parameter space more easily.

Below the GUT scale, the soft parameters run away from their values at
$M_X$. In the NMSSM the three soft parameters at $M_X$ evolve into 32
separate couplings below $M_X$, and in the MSSM the four soft
couplings at $M_X$ evolve into 30 separate couplings below $M_X$
(assuming no inter--generational mixing). However, since we retain
only the third generation's Yukawa couplings, then the first and
second generations' couplings will run identically. Below $M_X$ we
have soft scalar couplings $m_i^2$ (e.g. $m_{Q_3}^2$, $m_T^2$),
gaugino masses $M_i$ and soft trilinear couplings $A_i$ (e.g. $A_t$);
in the MSSM we also have $B$. The unification constraints then are
\begin{eqnarray}
g_i^2(M_X) & = & g_X^2 \\
M_i(M_X)   & = & M_{1/2} \\
m_i^2(M_X) & = & m_0^2 \\
A_i(M_X)   & = & A_0
\end{eqnarray}
In the MSSM, $B_0$, being a bilinear soft mass, need not be related to
$M_{1/2}$, $m_0^2$ or $A_0$, and we note that with our choice of
conventions $A_k(M_X)=-A_0$.

Our various phase and sign conventions in the NMSSM and MSSM are
adapted from reference \cite{NMSSM3}. The conventions in the MSSM are
fairly standard, but it is worth discussing the various conventions we
use for the NMSSM. The phases of the dimensionless couplings in the
superpotential are selected by appropriate field redefinitions and are
chosen so that all the dimensionless couplings are real. We may then
also impose that the gaugino masses are positive. We assume that $A_0$
is real, and we know that $m_0$ is real and positive. This then leads
us to VEVs $\nu_1$, $\nu_2$, $x$ which we may take to be real
\cite{nocpviol}. The problem has thus reduced to one of sign
conventions for all these quantities. In principle $\lambda_0$, $k_0$
and $A_0$ are allowed to be positive or negative, and the resulting
VEVs $\nu_1$, $\nu_2$, $x$ may each be positive or negative. However
there are 3 symmetries of the one-loop effective potential which will
lead to simplifications, namely:
\begin{eqnarray}
\lambda_0 & \rightarrow & - \lambda_0, \ \
k_0 \rightarrow - k_0, \ \ x \rightarrow -x  \\
\lambda_0 & \rightarrow & - \lambda_0, \ \
\nu_1 \rightarrow - \nu_1 \
\hbox{ (or } \nu_2\rightarrow - \nu_2 ) \\
\nu_1 & \rightarrow & -\nu_1, \ \
\nu_2 \rightarrow  -\nu_2
\end{eqnarray}
These equations show that negative VEVs can be re-interpreted as
negative $\lambda_0,k_0$. In our analysis we shall take positive
$\lambda_0$, $k_0$, and search for global minima of the effective
potential with positive $\nu_2$ and both possible signs of $\nu_1$ and
$x$. We could have required all VEVs to be positive by using the
invariance of the action under the above symmetries and then allowing
$\lambda_0$ and $k_0$ to take negative values,  but as explained in our
discussion of the electroweak potential, this is rather inconvenient
since we cannot impose such constraints on the VEVs when minimising the
potential. It has been shown that we need not consider complex VEVs
\cite{nocpviol}. To summarise, all input parameters and VEVs are taken
real and in the NMSSM only $x$, $\nu_1$, $A_0$ may be negative, while
in the MSSM only $\mu_0$, $B_0$, $A_0$ may be negative.

Suppose that we have a set of parameters (${\cal P}_{NM}$ or
${\cal P}_M$) defined at $M_X$, together with $g_X$. In general, an
arbitrary selection of parameters at $M_X$ will not result in an
acceptable pattern of electroweak symmetry breaking, both qualitatively
and quantitatively (\hbox{\it i.e.} giving the correct $Z$ mass). In
order to ensure that a given set of parameters results in correct when
electroweak symmetry breaking, we make the following observations.

Firstly, looking at the RG equations in appendix A, we see that to  one
loop \footnote{To two loops this is no longer strictly true, but the
error introduced in this is no worse than the error which we implicitly
accept in calculating the soft masses to one loop.} we have the relation
\begin{equation}
\frac{M_i(Q)}{M_{1/2}} = \frac{g_i^2(Q)}{g_X^2}.
\end{equation}
where $Q$ is the $\overline{MS}$ renormalisation scale.
This suggests the idea of recasting the RG equations for dimensionful
quantities in terms of those quantities divided by suitable powers of
$M_{1/2}$ \cite{NMSSM3}. The RG equations for $A_i$ and $m_i^2$
(and $B$ and $\mu$) become equations for $A_i/M_{1/2}$ and
$m_i^2/M_{1/2}^2$ (and $B/M_{1/2}$ and
$\mu/M_{1/2}$), while knowledge of the gauge couplings allows
us to avoid explicitly running the gaugino masses. Our GUT scale
parameters may then be replaced by the same parameters made
dimensionless by scaling with appropriate powers of $M_{1/2}$.
We denote these scaled quantities by a tilde. The parameters defining
the NMSSM are thus ${\tilde{\cal P}}_{NM} = \{h_{t0},
\lambda_0, k_0, \tilde{A}_0, \tilde{m}_0^2 \}$ and those defining the MSSM
are ${\tilde{\cal P}}_M = \{h_{t0}, \tilde{\mu},
\tilde{B}_0, \tilde{A}_0, \tilde{m}_0^2 \}$. These scaled  parameters may
then be fed into the RG equations in order to extract the scaled
low energy parameters.

By defining tilded scalar fields, that is, fields divided by
$M_{1/2}$, the tree--level Higgs potential $V_0$ becomes
$\tilde{V_0}$, where the latter potential involves only scaled
parameters. If ${\nu}_i$ (and $x$ in the NMSSM) minimise the potential
$V_0$, then ${\nu}_i / M_{1/2}$ (and $x/M_{1/2}$)
minimise the potential $\tilde{V_0}$. Knowing the scaled VEVs
$\tilde{\nu_i}$, we may then determine which value of $M_{1/2}$
is necessary to give us the correct $Z$ mass from
\begin{equation}
\frac{M_{Z}^2}{M_{1/2}^2} = \frac{1}{2} (g_1^2+g_2^2)
({\tilde{\nu_1}}^2 + {\tilde{\nu_2}}^2)
\end{equation}

\subsection{The Effective Potential}

It is well known that VEVs may not be reliably calculated
using the tree--level potential, so that it is necessary to employ the
one loop effective potential \cite{effpot1,effpot2}, given by
\begin{equation}
V_1 = V_0(Q) + \frac{1}{64\pi^2} Str {\cal M}^4 \left(
\log \frac{{\cal M}^2}{Q^2} - \frac{3}{2} \right),
\label{onelooppotential}
\end{equation}
where $Q$ is the $\overline{MS}$ renormalisation scale,
which should be selected so that the logarithms in the
effective potential are relatively small, and so we select a value of
$Q=150$ GeV. This choice of $Q$ is reasonable for $M_{1/2}$ and
$m_0$ less than about 1 TeV.
${\cal M}^2$
is the field--dependent, tree--level mass--squared matrix of those
fields whose radiative corrections we wish to include. By
field--dependent mass matrices is meant the mass matrices calculated
from the potential {\it prior} to setting the fields to their VEVs. We
include radiative corrections from loops of top quarks and
squarks. We do not include radiative corrections arising from the
rest of the spectrum since in this model such an analysis
would be computationally prohibitive. Of course it is well known
\cite{effpot2} that in order to achieve $Q$ independence it is
necessary to include the entire SUSY spectrum and to work to
all orders. There will therefore be a $Q$ dependence in our results
which we shall discuss later.
If $m_t$, $m_{\tilde{t_1}}$ and $m_{\tilde{t_2}}$ are the
field--dependent eigenvalues of the top quark and squark mass matrices
respectively,
then the contribution to the effective potential
from radiative corrections due to these states is given by
\begin{equation}
\frac{3}{32\pi^2} \left[
\sum_{i \in \{ \tilde{t_1}, \tilde{t_2} \}}
m_i^4 \left( \log \frac{m_i^2}{Q^2} - \frac{3}{2} \right)
- 2 m_t^4 \left( \log \frac{m_t^2}{Q^2} - \frac{3}{2} \right) \right].
\end{equation}
Notice that if particles and sparticles are degenerate, then their
contribution to radiative corrections vanish --- this is an instance
of the well known non--renormalisation theorem in SUSY.

Unfortunately, the presence of the logarithm in $V_1$ rather
complicates the scaling argument just presented. We may easily obtain
$\tilde{V_0}$, and therefore ${\tilde{\cal M}}^2$, but in order to
be able to evaluate the logarithm, we need to know
$M_{{1}/{2}}$. We may overcome this difficulty by noting that the
scaled one loop potential may be written as
\begin{equation}
\label{scaledpot}
\tilde{V_1} = \tilde{V_0} + \frac{1}{64\pi^2} Str {\tilde{\cal M}}^4
\left[ \log \left( \frac{{\tilde{\cal M}}^2}{Q^2} M_{1/2}^2
\right) -\frac{3}{2} \right].
\end{equation}
The problem then reduces to finding the value of $M_{1/2}$
which, when inserted in the logarithms of the scaled potential, give
the same output value of $M_{1/2}$. This is calculable numerically,
although there are typically two consistent solutions for $M_{1/2}$.

There are a number of problems with the calculation of the minimum of
the electroweak potential in the NMSSM which do not occur in the MSSM,
and it is worth highlighting them here. Firstly, there is the trivial
point that the space where we must look for solutions is simply bigger;
instead of minimising a function of two variables which can be
constrained to be positive, we are minimising a function of three, and
only one of these has its sign constrained.

More significantly, the minimum in the MSSM is unique at tree-level
(apart from the effects of the $Z_2$ symmetry taking $H_i\to -H_i$),
and this is often true even for the one loop case. For the NMSSM there
are virtually always several non-degenerate minima (up to five in some
cases), and so we must reliably calculate all of them and compare their
respective values of the potential to decide which one is preferred.
Sometimes, but not always, some or all of these minima are unphysical
in the sense that one or more of $\nu_1$, $\nu_2$, $x$ is zero.
\footnote{Note that this means that if we had used different phase
conventions such that all of the VEVs were positive, we would still
have had to allow different signs of VEVs in the potential calculation
to ensure that the minimum with positive VEVs was deeper than any of
the minima with one or more VEVs negative, since there is no phase
choice which will make {\it every} minimum occur at positive VEVs
simultaneously.}

This problem of multiple minima in turn leads to some serious technical
problems with reliably calculating the true deepest minimum of the
potential, since there may be minima with very different VEVs but
similar values of the potential. In this case we would not expect minor
changes in our input parameters or approximations to greatly change the
VEVs at the different minima, but we might well find that such changes
could affect which of the minima were deepest, and so lead to very
substantial changes in output. Fortunately, we do not find that this
structure is common through most of parameter space, but we do find
that there are certain regions where altering the input parameters
leads to the failure of electroweak breaking, and clearly the exact
place where this occurs is likely to be sensitive to minor changes. We
shall discuss the implications of this when we come to study the
parameter space of the NMSSM electroweak potential in detail later.

\subsection{Calculation of the Spectrum}

With knowledge of the VEVs and $M_{1/2}$ we may calculate the spectrum
of states induced by our parameters. In appendix B we give all relevant
scalar and fermion mass matrices. In the scalar sector, we calculate
one loop radiative corrections to all tree--level mass matrices which
arise from loops of top and bottom quarks and squarks, and thus are
controlled by $h_t$ and $h_b$. In particular, we calculate radiative
corrections to all Higgs boson masses and to top and bottom squark
masses.

In the fermionic sector, only the top and bottom quarks' masses are
controlled by $h_t$ and $h_b$.  Since we take $Q=150$ GeV, and as $m_t
\sim 150-200$ GeV, the error induced by evaluating $m_t(Q)$ rather than
$m_t(m_t)$ is small, and so this is what we do.\footnote{Note that the
mass which we quote for the top quark is the pole mass.} Similarly we
evaluate all SUSY masses at the scale Q, ignoring the effects of
decoupling which have been calculated in the MSSM and are known to be a
few percent \cite{pierce} except for very heavy states.

\subsection{Overall Procedure}

Let us now discuss our explicit numerical techniques. In either the
MSSM or the NMSSM, we start with a set of 5 input parameters $\tilde P$
as discussed above. From this we must derive $M_X$, $g_X$, $M_{1/2}$,
$\alpha_3(M_Z)$ and the entire low energy spectrum. We can find the
other parameters for some given value of $M_{1/2}$ as follows. We first
guess $M_X$, $g_X$, allowing us to calculate the full low energy
spectrum in terms of scaled variables (and hence of unscaled variables,
making use of $M_{1/2}$). Imposing correct unification of the low
energy couplings after running up through all the thresholds allows us
to iteratively refine our guess of $M_X$ and $g_X$ and to calculate
$\alpha_3(M_Z)$. Even when this has converged, however, our initial
choice of $M_{1/2}$ will not in general be consistent with electroweak
symmetry breaking and the correct value of $M_Z$. We must therefore
repeat the entire process with new values of $M_{1/2}$ until we have
found those values which are consistent with those derived from the
Higgs potential as discussed above.

Although we do not include the effects of $h_{b0}$ and $h_{\tau 0}$,
their inclusion is in principle straightforward, and they are included
in the analytical results in the Appendices. As might be expected,
their effect is negligible when $\tan\beta$ is small. We have explored
the effect of including such corrections for larger $\tan\beta$, but
found that they tend to make our numerical calculations prohibitively
slow, and so we do not include them.

This procedure is inevitably highly computationally intensive in the
NMSSM, primarily because of the difficulties implicit in numerically
minimising the Higgs potential, but is relatively simple for the MSSM.
However, it is still less efficient for the latter than the
``ambidextrous'' approach used by many authors \cite{all} where the
input parameters are $\{ m_0, M_{1/2}, A_0, m_t, \tan\beta \}$. Here
the fact that $B$ and $\mu$ do not appear in the RGEs for other
parameters means that they can be regarded effectively as low energy
parameters to be exchanged for $m_t$ and $\tan\beta$, and there is thus
no need for such repeated minimisation of the potential to achieve
convergence. In the NMSSM $B$ and $\mu$ are replaced by $\lambda$ and
$k$, and these appear in the RG equations for many of the other
parameters, preventing us using this technique.

\subsection{Physical Constraints}

We now turn to a discussion of the physical constraints. Apart from the
constraints of the correct pattern of electroweak symmetry breaking and
gauge coupling unification mentioned in section 3, the particle
spectrum resulting from an otherwise valid point in parameter space
must be subjected to certain phenomenological and cosmological
constraints. We have already mentioned that slepton VEVs and squark
VEVs must not be permitted.

Searches for pair production of sparticles coupling to the $Z$ have
been undertaken at LEP \cite{gencon}, with the following results:
\begin{itemize}
\item{Charged sleptons and sneutrinos have masses exceeding 43 GeV}
\item{Top squarks have masses exceeding 43 GeV}
\item{Charginos have masses exceeding 47 GeV}
\item{Charged Higgses have masses exceeding 45 GeV}
\end{itemize}
Squarks other than top squarks, together with the gluino, must have
masses in excess of 100 GeV. Constraints on the squark spectrum from
flavour changing neutral currents are not considered \cite{hagelin}.

In the MSSM the lightest neutralino must have a mass in excess of 18
GeV \cite{inobnd}. In the NMSSM this constraint may be greatly relaxed,
with even neutralino masses of zero acceptable for some regions of
parameter space since the lightest neutralino may contain a large
singlet component \cite{NMSSM3,NMSSMchicons}.

The only cosmological constraint which we impose is that the lightest
SUSY particle (LSP) should be a neutralino. We do not impose the
constraint that $\Omega\le 1$, since the calculation of the relic
density is complicated in this model, and has been considered elsewhere
in the MSSM \cite{ppbp} and the NMSSM \cite{NMSSMcosmology}.

In the neutral Higgs sector of the MSSM the constraints on the lightest
CP--even Higgs scalar, $h$, are complicated and depend on the
suppression of its couplings to the $Z$ relative to those in the standard
model, with a bound of up to 60GeV \cite{LEP1}. In the NMSSM, since
this state may contain a significant proportion of the singlet field,
its coupling to the $Z$ will be diluted. This complicates the
constraint and will be discussed in detail in section 7, while it has
been considered without the imposition of unification constraints in
Ref. \cite{NMSSMhiggscon}.  We do not consider further constraints on
the Higgs sector from the decay $b \rightarrow s \gamma$, since the
magnitude of these is uncertain; for a review of this topic, see for
example, \cite{bsg}. In any case it will turn out that in all the
interesting regions of parameter space that the charged Higgs scalars
are quite heavy.

In order that the spectra be heavy enough to evade all the constraints,
it is generally only necessary that $M_{1/2}$ be large enough; if it
falls below around 50--100GeV then many of them start being violated
simultaneously. Similarly, the experimental value of $\alpha_3(M_Z)$ is
so uncertain that we virtually always find that our results are within
experimental errors. More significant bounds arise from the value of
the top mass, which has recently been measured to be about 180 GeV
\cite{CDF} with an error of order 10 GeV. We shall conservatively only
require $m_t>140$ GeV in presenting our numerical results. However we
shall emphasise those regions of parameter space which can yield heavy
top quark masses in the preferred range 160-200 GeV.

The only constraint on the dimensionless couplings is that of avoiding
triviality, so that $h_{t0}$, $\lambda_0$, $k_0$ are all $<3$,
although, as we shall see, there are tight constraints on them for
given values of the other parameters from correct electroweak symmetry
breaking.

These constraints, and in particular the values of $m_t$ and
$\alpha_3(M_Z)$ will in general be discussed in the text when we come
to consider the various different regions of parameter space in which
they may be violated.

\subsection{Naturalness Constraints}

Finally, we note that there is a further constraint which is inspired
by naturalness rather than phenomenology \cite{RR,ft,CC}. This is the
so-called fine-tuning constraint, which is usually expressed as the
requirement that the value of $M_Z$ derived from the electroweak
potential should not be too sensitive to small changes in $h_{t0}$. In
practice, this amounts in the MSSM to ensuring that the supersymmetric
spectrum is not too heavy relative to the electroweak breaking scale.
We follow \cite{RR} in defining
\begin{equation}
c_{h_t}=\vert \frac{\delta M_Z / M_Z}{\delta h_{t0} / h_{t0}}  \vert
\end{equation}
which allows us to quantify the fine-tuning inherent in any given set
of parameters.

In the NMSSM we also define $c_{\lambda}$ and $c_k$ analogously to
$c_{h_t}$ as follows.
\begin{equation}
c_{\lambda}=\vert \frac{\delta M_Z / M_Z}{\delta \lambda_0 / \lambda_0}  \vert
\end{equation}
\begin{equation}
c_{k}=\vert \frac{\delta M_Z / M_Z}{\delta k_0 / k_0}  \vert
\end{equation}

We shall not impose any explicit restriction on these
parameters because the amount of fine-tuning to be tolerated is
essentially a matter of taste, particularly in the NMSSM where it is
not clear to which dimensionless  parameter any such constraint should
be applied, and instead shall discuss the  extent of fine-tuning
further when we come to our results for each of the two models.

\section{MSSM Results}

It will be helpful to discuss the MSSM in some detail. Although a
number of studies of this model have been done \cite{all}, none has
recently considered it with our choice of variables, chosen entirely at
the GUT scale, which are more enlightening for purposes of comparison
with our analysis of the NMSSM.

\subsection{The Spectrum}

We begin by noting that there are a number of common themes to analyses
of this sort which apply for both models. Firstly, the value of
$\alpha_3(m_Z)$ which allows unification is around 0.126 (0.118) for
$M_{1/2}$ of order 100GeV (1TeV) at two loops with $m_0=M_{1/2}$; it is
lower at one loop, by typically around 10\%. Of course, this is rather
an oversimplification, and for values of $m_0$ far from $M_{1/2}$ there
will be strong dependence on $m_0$, together with dependence on $\mu$
and $A$, but the primary determinant of $\alpha_3(m_Z)$ is that of the
scale of the gaugino thresholds. A detailed discussion of the
dependence of the couplings on the various thresholds is given in
reference \cite{langpol}.

To illustrate this, we display a plot of contours of constant
$\alpha_3$ and $m_t$ in the $m_0-M_{1/2}$ plane in Figure 1. Here the
value of $\alpha_3(M_Z)$ decreases as we increase the effective scale
of the SUSY breaking masses, as discussed before. It is noticable here
that the $m_0-M_{1/2}$ plane is not completely covered with contours;
this is because correct electroweak symmetry breaking does not occur in
some regions, while only in part of the plane do we find acceptable top
masses. Graphs generated by other authors often have smaller excluded
regions, because they are for fixed $\tan\beta$ and $m_t$ and variable
$B_0$ and $\mu_0$, which makes correct electroweak breaking easier to
arrange. This dependence of $\alpha_3(M_Z)$ on $m_0$ and $M_{1/2}$ is
exactly the same for the NMSSM.

The typical spectrum is also fairly similar in both models. In general,
all the scalar superpartners have masses of around $m_0$ (although for
small $m_0$, $M_{1/2}$ corrections can dominate). Masses are rather
larger for squarks because of QCD effects, but the stop squark is in
general rather lighter thanks to the corrections from large $h_t$. The
gauginos, as already noted, have masses given by
$M_i=(g_i/g_X)^2M_{1/2}$, and so the gluino tends to have a mass of
around three times $M_{1/2}$, while the lightest chargino and lightest
neutralino have masses less than or order of $M_{1/2}$ and
$\frac{1}{2} M_{1/2}$ respectively. The case for the neutralino and
chargino is greatly complicated by the fact that these have mass
eigenstates which are mixtures of gaugino and Higgsino. The Higgs
masses are typically of the same order as the other scalars, except for
the lightest CP-even Higgs state which is constrained to be lighter
than an upper bound in both the MSSM \cite{MSSMbound} and the NMSSM
\cite{tlNbound,NMSSMbound}.

\subsection{Electroweak Symmetry Breaking}

Turning now specifically to the case of the MSSM, the main difference
between this model and the NMSSM is its relatively simple Higgs
structure. At tree level, ignoring the one-loop effective potential, it
is possible to analytically find the minimum of the potential, and we
shall simply quote some of the results here. These can be found in, for
example, \cite{SUSY}. We define  $m_1^2=m_{H_1}^2+\mu^2$,
$m_2^2=m_{H_2}^2+\mu^2$ and $m_3^2=B\mu$. Firstly, in order that the
minimum of the potential not be at the origin, we have
$m_1^2m_2^2<m_3^4$, while to prevent the potential being unbounded from
below $m_1^2+m_2^2+2m_3^2>0$. We must avoid breaking electromagnetism
which implies $m_3^2<0$.  These results tend to favour the case where
only one of $m_1^2$ and $m_2^2$ is negative or where both are positive
but one is  rather less than the other. Of course, this is the reason
why  radiative electroweak symmetry breaking via $h_t$ works \cite{ir},
as it  drives $m_2^2$ to be smaller (possibly negative), while $m_3^2$
can  easily be set negative.

We now note that, from the minimisation conditions of the potential,
\begin{equation}
\sin 2\beta=\frac{-2m_3^2}{m_1^2+m_2^2},
\label{tb}
\end{equation}
and so we see that large $\tan\beta$ will occur in the region where
$m_3^2$ is small, at least relative to $m_1^2+m_2^2$. The second
minimisation condition, expressed in terms of $M_Z^2$, is
\begin{equation}
M_Z^2=\frac{2m_1^2-2m_2^2\tan^2\beta}{\tan^2\beta-1}
=\frac{2m_{H_1}^2-2m_{H_2}^2\tan^2\beta}{\tan^2\beta-1}-\mu^2
\label{zeqn}
\end{equation}
which suggests that in general $M_Z^2$ will be of order $-m_2^2$
for reasonably large values of $\tan\beta$.

In numerical studies including the one-loop effective potential, we
find that electroweak symmetry breaking works perfectly well for
reasonably large $h_t$ (to drive $m_2^2$ small enough) so long as $B_0$
and $\mu_0$ are of opposite sign (in fact $B_0\sim 0$ also works
because of radiative corrections which drive $B$ negative by the
electroweak scale). Furthermore, selecting smaller $B$ allows us to
make $\tan\beta$ as large as we like, so long as $h_b$ is neglected.
When $h_b$ is included \cite{copw}, it becomes large for large
$\tan\beta$, driving $\tan\beta$ down and thus making it difficult to
obtain very large $\tan\beta$ without fine-tuning the parameters.

The dependence on $\mu_0$ is rather more complicated; although it might
appear from Eq. (\ref{tb}) that small $\mu_0$ gives larger $\tan\beta$,
in fact the dependence of $m_1^2$ and $m_2^2$ on $\mu$ means that the
effect of varying it is much more complicated and  depends on which
region of parameter space we study. Since large $m_0^2$ means large
$m_1^2$, we see that pushing up $m_0$ beyond a few times $M_{1/2}$
tends to increase $\tan\beta$.

In drastic contrast to the case in the NMSSM, $A_0$ seems to be
relatively unimportant to electroweak symmetry breaking. It can be
picked as small as we wish, or as large as is consistent with avoiding
the constraints on slepton vevs discussed earlier.

\subsection{Fine-Tuning}

Finally we turn to the dependence on $h_t$. This is crucial as $h_t$
has the dominant effect on $m_2^2$, forcing it to become negative and
so driving electroweak symmetry breaking \cite{ir}. The behaviour of
$M_{1/2}$ as $h_{t0}$ is varied is complicated, since changes in
$M_{1/2}$ affect the potential in several ways: by directly changing
the values of the soft masses (for fixed scaled values of the other
dimensionful quantities); by affecting the logarithms in the one loop
potential; and by changing the range over which we run and the
couplings (since altering $M_{1/2}$ will significantly change $M_X$ and
$g_X$). We show a typical graph of the dependence of $M_{1/2}$ on
$h_{t0}$ in Fig.2. This kind of structure often occurs in the NMSSM
too, although there the behaviour can be more complicated.

A simpler dependence on $h_{t0}$, and one which has been studied by a
number of authors interested in the fine-tuning behaviour, is that of
$M_Z$ on $h_{t0}$ for fixed values of the soft masses (and hence of
$M_{1/2}$ in our notation) \cite{ft,CC}. From the RG equations in
Appendix A and the tree-level analysis of the last sub-section, it can
be seen that if $h_t$ is too large $-m_2^2$ will fall too rapidly
ultimately becoming too negative, leading the tree-level
potential to have very large VEVs. For small $h_t$, $m_2^2$ will not be
driven negative, and electroweak symmetry breaking will not occur. We
are thus reduced to an intermediate zone of $h_t$, and the behaviour of
$M_Z$ as $h_t$ is increased is to rise rapidly from zero to a very
large value (often infinity at tree level) driven by the large value of
$-m_2^2$.

A crude understanding of the fine-tuning problem may be given as
follows. The RG equation for $m_2^2$ shows that, assuming $h_t^2$ terms
dominate, the value of $m_2^2$ approaches a fixed point of order minus
a few times a typical squark mass squared, resulting in $M_Z$ of order a
typical squark mass (unless this large negative $m_2^2$ makes the
potential unbounded from below). For large squark masses, the value of
$M_Z$ can only be reduced by carefully reducing $h_t$ so that this
fixed point is not aproached too closely. This is the fine-tuning
problem.

It should be noted that, although this discussion has been at
tree-level, the behaviour at the one loop level is qualitatively
similar, although the one loop contributions reduce the fine-tuning
\cite{CC}. This is simply because they introduce into Eq.(\ref{zeqn})
extra terms of opposite sign to those which are present at tree level,
and which also grow with increasing values of the squark masses.

In order to illustrate the level of fine-tuning in the MSSM, in Fig.3
we show the effect on the $Z$ mass of changing $h_{t0}$ with $\tilde
m_0=\tilde\mu_0=-\tilde B_0=1$ and $A_0=0$, with $M_{1/2}=$50, 100,
300, 500, 1000GeV. For the curve shown in Fig.3, the value of $c_{h_t}$
is 2.3,4.5,25,49,83 for $M_{{1}/{2}}=$ 50, 100, 300, 500, 1000GeV at
$M_Z$=91.2GeV. These very large values of $c_{h_t}$ typically lead
authors to restrict their values of the soft masses to less than 1TeV
or so. The fine-tuning becomes severe only as the squark masses becomes
significantly larger than $M_Z$; for example, if $M_Z$ were 200GeV then
$c_{h_t}=50$ would not be reached until $M_{1/2}$ were 1300GeV, instead
of 500GeV. Lastly we remark that for very large $M_{1/2}$, we find a
cutting out of the data in the curves, caused by the transition between
two minima, one at the origin and one at non-zero VEVs.

\section{NMSSM Results}

In order to study the NMSSM parameter space, we first selected a grid
of points with $h_{t0}=0.5,1,2,5$, $\lambda_0=0.01-2.0$,
$k_0=0.01-2.0$,  $\tilde m_0=0.2-5.0$, and $A_0/m_0=-4$ to 4
(as we shall see later it is convenient to scale $A_0$ by $m_0$).
Within this parameter range we can classify the successful parts of
parameter space into two distinct regions:

(i) $k_0>\lambda_0>h_{t0}$

(ii) $h_{t0}>\lambda_0>k_0$

Although the range of parameters above covers most of the parameter
space it turns out that there are other phenomenologically interesting
parts of parameter space outside this region. In particular there is a
third region of parameter space with values of the couplings $k_0$,
$\lambda_0$, $\tilde m_0$ which are significantly smaller than those
considered above, and which is asociated with {\em very} large values
of $r$. Small $\lambda$, $k$ and large $r$ implies that the NMSSM is
close to the MSSM limit in which the singlet completely decouples from
the rest of the spectrum. Although not logically distinct from region
(ii) above which also approaches the MSSM limit, this third region is
so close to it that we shall refer to it as the ``deep MSSM limit'' of
the model and discuss it separately.

The results in the case of the NMSSM are naturally more complicated
than for the case of the MSSM. The primary constraint is that of
correct electroweak symmetry breaking, which we shall find restricts us
to only a relatively small part of parameter space. The fundamental
problem with the electroweak potential in the NMSSM is that the minimum
of the potential is often ruled out because, for example, it has
$\nu_1=0$ or $\nu_2=0$. This does not usually occur in the MSSM,
because (at tree-level) the existence of a $\mu$ term of the correct
sign guarantees that if one of $\nu_1$ and $\nu_2$ is non-zero, so is
the other.

In order to find correct electroweak breaking in the NMSSM, we must
thus ensure that $N$ is driven to have a VEV. We find in practice that
this can be done in two ways. Firstly, it is possible to force $m_N^2$
negative by having large Yukawa couplings in the Higgs sector.
Secondly, it is possible to have electroweak breaking with $m_N^2>0$ by
having very large trilinear terms, and hence large $A_k$. We shall
discuss each of these two disconnected regions of correct electroweak
symmetry breaking separately below, before turning to a fuller
discussion of our uncertainties and approximations, and of the
phenomenological implications of our findings.

\subsection{$k_0>\lambda_0>h_{t0}$ Region}

The first region which we shall discuss is that which has electroweak
symmetry breaking driven by Higgs sector Yukawa couplings. Here we have
$\lambda_0$, $k_0$ large, typically with $\lambda_0+k_0>2$, thus
driving the Higgs masses squared negative, and we find regions of
correct electroweak symmetry breaking for values of $|A_0/m_0|$ down to
zero. This is in contradiction to the results of \cite{ell,ell2},
although we note that this may be partially explained by the cut on the
top mass used in \cite{ell3}.

Typically, we find that varying $\lambda_0$ and $k_0$ does not have a
great impact in this region. Reducing either of these leads to a
gradual increase in $\tan\beta$ which can thus be made to take any
value from around 3 to as large as we like (until our approximations
begin to break down), while they can be increased arbitrarily so long
as $k_0$ remains rather larger than $\lambda_0$. Varying $h_{t0}$,
however, leads to more dramatic changes, since by tuning this we can
select any value of $M_{1/2}$ we choose. Unfortunately, we generally
find that in order to obtain values of $M_{1/2}$ less than
many TeV it is necessary to reduce $h_{t0}$ until the top mass
is very small, and so only very little of this data survives.

The dependence of the output on $A_0/m_0$ is straightforward;
$M_{1/2}$ has a minimum for some value typically between 0 and -1.
Electroweak breaking fails for $|A_0/m_0|$ greater than some value
typically $\stackrel{<}{\sim}2$. Dependence on $\tilde m_0$ is less
crucial to correct electroweak breaking, but it seems that $\tilde
m_0>1$ is preferred. As appears to occur everywhere, $|r|$ is roughly
directly proportional to $M_{1/2}$ for fixed values of the other
parameters, and here can be less than 1 for very small $M_{1/2}$. We
note that changing the sign of $A_0$ has as its main effect changing
the sign of $r$.

In Fig.4 we show a typical plot of the mass spectrum in this region.
The lightest CP-even Higgs is within the reach of LEPII, and over some
of the $M_{1/2}$ range the lightest chargino is also observable at
LEPII. Note that the top quark mass is fairly constant at 147 GeV over
the whole range, a value which is typical of the largest achievable in
this region, and so future more stringent bounds on the top quark mass
could exclude this region completely. For
$M_{1/2}\stackrel{<}{~}100$GeV the chargino mass violates its
experimental bound.

\subsection{$h_{t0}>\lambda_0>k_0$ Region}

The second region of parameter space which we shall discuss is that
where the  VEV of $N$ is driven to be non-zero by the effect of $A_k$.
This region is immediately more promising than before since it allows a
large top quark Yukawa coupling, and hence can have larger top quark
masses. This region was also discussed in ref.\cite{ell,ell3,ekw0} and
is discussed here again for completeness. We find that here
$h_{t0}>\lambda_0>k_0$, and that we are typically in a region where
$|r|>>1$ and $\lambda_0,k_0<<1$. We find that only a restricted range
of $|A_0/m_0|$ around 3 can give consistent electroweak breaking
without generating slepton VEVs \cite{ell}; the reasons for this can be
understood analytically and will be discussed in Section 6.

We now turn to a discussion of the behaviour of the solutions as we
adjust the various input parameters. As in the  $k_0>\lambda_0>h_{t0}$
region, we find that increasing $h_t$ causes $M_{1/2}$ to increase
arbitrarily. However, unlike in the former region, here we can avoid
the constraint of small top mass by tuning $\lambda_0$ and $k_0$, since
in this region $M_{1/2}$ is a very sensitive function of these
variables. In fact we find we find that for any given values of the
other parameters, the requirements of electroweak breaking restrict us
to a very small range of $k_0$ (or $\lambda_0$), leading to a
correlation between these parameters in this region.

To illustrate the correlation between $\lambda_0$ and $k_0$, in Fig.5
we show contours of $h_{t0}$ in the $\lambda_0-k_0$ plane with
$\tilde{m}_0=2$, $M_{1/2}=500$ GeV and $A_0/m_0=-3$. Although we could
have chosen other values of $M_{1/2}$, the qualitative features of
Fig.5 would be unaltered, and the value of $k_0$ would only change by a
few per cent, since in this region the value of $M_{1/2}$ is a very
sensitive function of $\lambda_0$ and $k_0$. The fact that for each
line $\lambda_0/k_0$ is virtually constant can be understood from the
electroweak potential near the MSSM limit, as discussed in Section 6.
The $h_{t0}=0.5,1,2,3$ contours correspond to $m_t\sim 145, 175, 185,
190$ GeV, respectively, with $|\tan\beta|$ varying from 2-7 and $|r|$
from 100-30, from the lowest to the highest value of $k_0$
respectively. Values of $k_0$ beyond the ends of the plot are forbidden
by the requirements of electroweak symmetry breaking, although the
exact ranges of acceptable $k_0$ could be altered by varying our choice
of $A_0$ and $\tilde m_0$ slightly. Similarly, $m_t$ would be altered
if we use a different $M_{1/2}$ primarily due to the resulting change
in $\alpha_s(M_Z)$. Changing the sign of $A_0$ leads to qualitatively
rather similar output data with a change in the sign of $r$.

As mentioned above, in this region we need to have $|A_0/m_0|\sim 3$.
In fact we find that negative $A_0$ works best (because thanks to the
form of the RG equations this leads to a less restrictive slepton VEVs
constraint), while we can increase the maximum permitted value of
$|A_0/m_0|$ somewhat by reducing $\tilde m_0$. There is also a
correlation between $h_{t0}$ and $\tilde m_0$; for $h_{t0}=1(3)$ and
$A_0/m_0=-3$ we find acceptable solutions for $\tilde m_0\approx 1-2(2-5)$.

Typically we find quite large values of $|\tan\beta|$ in this region as
well, but do not in general find any obvious correlation with the other
input parameters. $|r|$ is generally directly proportional to $M_{1/2}$
for fixed values of the other parameters, and as we approach the origin
in the $\lambda_0-k_0$ plane it becomes very large. This corresponds to
a very close approach to the MSSM limit.

In Fig.6 we plot the spectrum as a function of $M_{1/2}$ for the case
$h_{t0}=2$, $A_0/m_0=-3$, $\tilde{m}_0=5$, $\lambda_0=0.4$, which is a
typical point in this region of parameter space. Note that the
experimental constraint that the charginos are heavier than 47 GeV
implies that $M_{1/2}\stackrel{>}{\sim}70$ GeV in this case. For
$M_{1/2}\approx 100 (1000)$ GeV, which is controlled by choosing
$k_0=0.275(0.300)$, we find $|\tan\beta | \approx 6(8)$, $|r| \approx
13(120)$, $\alpha_s(M_Z)\approx 0.121(112)$, $M_X=2.1(0.99)\times
10^{16}$ GeV, $g_X=0.71(0.69)$. The lightest CP-even Higgs boson has
standard model-like couplings ($R_{ZZh}>0.99$ everywhere) and for
$M_{1/2}\stackrel{<}{\sim}100$ GeV is in the LEP2 range, as are the
lighter chargino and neutralinos which have a much stronger $M_{1/2}$
mass dependence. The top quark mass ranges from $m_t=193-184$ GeV,
being smaller for larger $M_{1/2}$ due to $\alpha_s(M_Z)$ being
consequently smaller. For $M_{1/2}\stackrel{<}{\sim}100$ GeV, the
lightest stop and gluino are not too much heavier than the top quark,
although the remaining sparticles and Higgs bosons are significantly
heavier than the top quark.

One of our most interesting results is that for choices of parameters
outside a ``safe'' range $h_{t0}\approx 1.5-3$, $|A_0/m_0|\approx 3$,
$\tilde{m}_0\stackrel{>}{~}3$, we find that as $M_{1/2}$ is reduced the
data suddenly cuts out below some critical value of $M_{1/2}$. This
corresponds to some new (often unphysical) set of VEVs becoming
preferred below some $M_{1/2}$ value. The effect is somewhat $Q$
dependent and so deserves a detailed discussion.

\subsection{Uncertainties in the Electroweak Potential}

We now study the effects of varying the $\overline{\hbox{MS}}$ scale
$Q$. Although in principle none of our results should depend on this,
in practice there are a number of inaccuracies caused by the fact that
the logarithms in our potential may not be small. It is not necessarily
possible to guarantee that all the logarithms are small simultaneously,
since we need to investigate the relative depths of multiple minima at
significantly different values of the VEVs and hence of the terms
appearing in the logarithm. A $Q$ dependence is also introduced by our
neglect of all but the effects of the $h_t$ Yukawa coupling in the
effective potential, since inclusion of all the spectrum would be
prohibitively computationally intensive, and by our evaluation of
masses at tree-level.

We find that in practice even relatively large changes of $Q$ (up to an
order of magnitude) do not have a very large impact on the value of the
VEVs, except in cases where the scaled VEVs are very small which
implies very large $M_{1/2}$ and so is outside the region of greatest
physical interest. However, the impact on which of the VEVs is deepest
is not always negligible. We find that in regions where consistent
electroweak breaking cannot be achieved below some critical value of
$M_{1/2}$ this value may be reduced by a reduction of $Q$. However,
even this sensitivity to $Q$ is not too severe; for example, a
reduction of $Q$ to around 10GeV is necessary in order to obtain
arbitrarily small values of $M_{1/2}$ for $\tilde m_0=1$. Thus we
confirm that there is only a restricted safe range where arbitrarily
small values of $M_{1/2}$ are possible, even allowing for changes in
$Q$.

We illustrate the existence of the cut-out of data in Fig.7, where we
plot the spectrum as a function of $M_{1/2}$ for the case $h_{t0}=0.5$,
$A_0/m_0=-4$, $\tilde{m}_0=0.5$, $\lambda_0=0.1$. These parameters are
outside the safe range $h_{t0}\approx 1.5-3$, $|A_0/m_0|\approx 3$,
$\tilde{m}_0\approx 3-5$, in which small $M_{1/2}$ can be achieved
independently of the choice of the renormalisation scale $Q$, and so we
plot the spectrum for two choices of $Q$. The main effect of changing
$Q$ is to change the value of $M_{1/2}$ at which the data cuts out. For
$Q=150(25)$ GeV, we find the minimum values $M_{1/2}=300(125)$ GeV
corresponding to $m_0=150(62.5)$ GeV. As in Fig.6, the lightest CP-even
Higgs boson has standard model-like couplings, and for a given
$M_{1/2}$ is even lighter than in Fig.6. For the smaller $M_{1/2}$
values which we can obtain with $Q=25$ GeV the lighter chargino and
neutralinos may be in the LEP2 range. Compared to Fig.6, the
left-handed sleptons are now much lighter (due to the smaller value
$\tilde{m}_0=0.5$) while the lighter stop is much heavier (due to the
smaller value $h_{t0}=0.5$). The top quark mass has a maximum value of
$m_t=160$ GeV for $M_{1/2}=300$ GeV and here $|\tan\beta |=3.8$. The
gluino in Fig.7 is now the heaviest sparticle, whereas in Fig.6 it was
one of the lighter ones. In general the Higgs and sparticle masses in
Fig.7 are focussed into a narrower band of    masses than in Fig.6,
which is a simple result of having $\tilde{m}_0=0.5$ rather than
$\tilde{m}_0=5$.

\subsection{The Deep MSSM Limit: $h_{t0} > \lambda_0 \gg k_0$ }

This region of parameter space is characterised by very small values of
the parameters $\tilde m_0\ll 0.2$, $k_0\ll \lambda_0 < 0.01$, which we
refer to as the deep MSSM limit of the model, since in this region we
find that $|r|\gg 100$. This region is nothing more than an extreme
limit of the region just discussed, but deserves a separate discussion
since there is some interesting physics in this region.

We have seen that in the constrained NMSSM one of the CP-even Higgs
bosons is almost pure gauge singlet. Such a decoupled Higgs boson might
be  expected since the constrained NMSSM is close to the MSSM limit. In
the analysis of ref.\cite{ell3} the decoupled Higgs boson may have a
mass either less than or greater than the lighter of the other two
CP-even Higgs bosons and when the would-be decoupled Higgs is close in
mass to the lighter physical Higgs, strong mixing can occur, leading to
two weakly coupled Higgs bosons. However, given the range of parameters
considered so far, we find that the decoupled CP-even Higgs boson is
always substantially heavier than the lighter of the two physical
CP-even Higgs bosons. Similarly the CP-odd Higgs bosons are much
heavier than the lightest CP-even Higgs boson. The reason for this
difference is simply that in ref. \cite{ell3} the range of parameters
considered exceeds the range discussed so far here. In order to bring down
the mass of the CP-even singlet sufficiently one requires
$\tilde{m}_0\ll 0.2$, $\lambda_0< 0.01$ and $k_0\ll \lambda_0$,
corresponding to extremely large values of $r\gg 100$.  In this region
of parameter space, the singlet CP-even Higgs boson does indeed become
much lighter, leading to the strong mixing effect mentioned above. It
is ironic that this effect only seems to take place in the deep MSSM
limit of the constrained NMSSM parameter space, making the NMSSM more
easily resolvable from the MSSM in the region of parameter space where
the two models are formally most similar.  \footnote{In the
unconstrained NMSSM the effect also  occurs away from the MSSM limit as
discussed in ref.\cite{ekw}.}

In order to illustrate this effect, in Fig.8 we show the spectrum as a
function of $M_{1/2}$, for a point in this region. Note that the
lightest CP-even Higgs bosons (dashed lines) interfere for
$M_{1/2}\approx 2000$ GeV. In Fig.9 we show the amplitude of singlet
$N$ contained in the lightest two CP-even massive states, corresponding
to Fig.8 (the heaviest CP-even has singlet component zero throughout).
For $M_{1/2}\ll 2000$ GeV, the lightest CP-even Higgs boson is almost
pure singlet, whilst for $M_{1/2}\gg 2000$ GeV the second lightest
CP-even Higgs boson is almost pure singlet. Thus a simple
interpretation of this effect is that as $M_{1/2}$ is steadily
increased, the singlet CP-even Higgs mass rapidly rises,``crosses'' the
physical CP-even Higgs boson line, and  continues to rapidly increase
with $M_{1/2}$. In the crossing region the two CP-even states will
strongly mix leading to two states with diluted couplings. We shall
discuss the phenomenological implications of this in some detail in
Section 7.

\subsection{Fine-Tuning in the NMSSM}

In Table 1, we give the fine-tuning parameter $c_{h_t}$ for a typical
point in each of the three most interesting regions of parameter space
and a range of values of $M_{1/2}$, together with the corresponding
data for the MSSM point already discussed in Section 4.3. These points
are defined below. In Tables 2,3 we also show the alternative fine-tuning
parameters $c_{\lambda}$, $c_{k}$ for the NMSSM for each of the three
cases.

\begin{itemize}
\item {
NMSSM1: $h_{t0}=2$, $\lambda_0=0.4$, $\tilde m_0=5$, $A_0=-3m_0$, a point
in the safe range where arbitrarily small $M_{1/2}$ can be achieved,
and with the same parameters as Fig.6}
\item {NMSSM2: $h_{t0}=0.5$, $\lambda_0=0.1$, $\tilde m_0=0.5$, $A_0=-4m_0$
a point outside the safe range, with the same parameters as Fig.7}
\item {NMSSM3: $h_{t0}=0.5$, $\lambda_0=0.005$, $\tilde m_0=0.02$, $A_0=-5m_0$,
a point in the deep MSSM limit, with the same parameters as Fig.8}
\item {MSSM : the MSSM with $\mu_0=-B_0=m_0=M_{1/2}$ for comparison, with
the same parameters as Fig.3}
\end{itemize}

\vspace{0.25in}

\noindent{ Table 1. Fine tuning parameter $c_{h_t}$
as a function of $M_{1/2}$ for the MSSM and the
three examples of NMSSM parameter space defined in the text.}

\begin{tabular}{|c|c|c|c|c|} \hline
$M_{1/2}$(GeV) & MSSM & NMSSM1 & NMSSM2 & NMSSM3 \\ \hline
50   & 2.3 &    4.2 &  -- &  -- \\ \hline
100  & 4.5 &   13   &  -- &  -- \\ \hline
300  & 25  &   60   &  18 &  -- \\ \hline
500  & 49  &  140   &  37 &  -- \\ \hline
1000 & 83  &  390   &  75 &  59 \\ \hline
2000 & 130 & 1200   & 130 & 110 \\ \hline
\end{tabular}

\vspace{0.25in}

\noindent{Table 2. Fine tuning parameter $c_{\lambda}$
as a function of $M_{1/2}$ for the three examples of
the NMSSM defined in the text.}

\begin{tabular}{|c|c|c|c|} \hline
$M_{1/2}$(GeV) & NMSSM1 & NMSSM2 & NMSSM3 \\ \hline
50   &    9.9 &  -- &  -- \\ \hline
100  &   30   &  -- &  -- \\ \hline
300  &  160   &  23 &  -- \\ \hline
500  &  360   &  42 &  -- \\ \hline
1000 & 1000   &  70 &  79 \\ \hline
2000 & 3300   & 160 & 130 \\ \hline
\end{tabular}

\vspace{0.25in}

\noindent{Table 3. Fine tuning parameter $c_{k}$ as a function
of $M_{1/2}$ for the three examples of the NMSSM defined in the text.}

\begin{tabular}{|c|c|c|c|} \hline
$M_{1/2}$(GeV) & NMSSM1 & NMSSM2 & NMSSM3 \\ \hline
50   &   10   &  -- &  -- \\ \hline
100  &   30   &  -- &  -- \\ \hline
300  &  160   &  22 &  -- \\ \hline
500  &  370   &  42 &  -- \\ \hline
1000 & 1100   &  80 &  80 \\ \hline
2000 & 3500   & 130 & 130 \\ \hline
\end{tabular}

The main points to note from these tables are that, as usual,
fine-tuning increases with increasing $M_{1/2}$ and $m_0$. Many authors
choose to restrict $c_{h_t}$ to be less than a value in the range
10-100, and this would clearly be very restrictive, although would
still not eliminate all of any of the different regions, despite the
lack of data with small $M_{1/2}$ in some of them.

We also note that the fine-tuning in $\lambda_0$ and $k_0$ are the
same. This can be seen to be a result of the fact that the VEVs
$\nu_1$, $\nu_2$ depend only on $\lambda$ and $k$ only through the
combination $\lambda/k$ near the MSSM limit, as discussed below.
Increasing $\tilde m_0$ increases $c_k$ more than $c_{h_t}$, which
again can be understood through the analytical discussion in Section 6.

We illustrate this behaviour with graphs of $M_Z$ against $k_0$ in two
of these cases. Fig.10 is for the parameters NMSSM1 above and shows the
simplest behaviour. Here $M_Z$ rises very rapidly from zero to of order
$m_0$ and then levels off. It is clearly reasonable to suggest that the
natural value of $M_Z$ is the value in this plateau region. Fig.11 is
for parameters NMSSM2 above, and shows a richer behaviour. As before,
there is a plateau region. However instead of the value of $M_Z$ simply
falling rapidly to 0 as $k_0$ is decreased, it changes value
discontinuously. This discontinuity can occur because a change in $k_0$
can lead to a change in which of the various minima is deepest. This is
simply the behaviour where the data cuts out in Fig.7, where it is not
possible to find solutions for smaller values of $M_{1/2}$ consistent
with $M_Z=91.2$GeV.

\subsection{Summary of NMSSM Results}

Since this has been a lengthy and important section, we shall briefly
summarise the results before continuing. Crudely speaking, the various
constraints reduce the parameter space to two allowed regions (i)
$k_0>\lambda_0>h_{t0}$, and (ii) $h_{t0}> \lambda_0 >k_0$. We found
that region (i) generally involves a rather small Yukawa coupling,
leading to a top quark mass smaller than about 150 GeV, which may be
considered rather too small in the light of current measurements of the
top quark mass \cite{CDF}, and so we shall not discuss it further.

Region (ii) involves an arbitrarily large top quark Yukawa coupling,
leading to values of $m_t$ up to about 200 GeV. With
$h_t\stackrel{>}{\sim}1$, other parameters in this region are
constrained to be: $\tilde{m}_0 \stackrel{>}{\sim}  1$, $|A_0/m_0| \sim
3$, leading to large values of $r\gg 1$, with $r$ proportional to $m_0$
for given $k_0$. Thus this region is always close to the MSSM limit.
Many of its features can be understood analytically as discussed in the
next section, and it is noticable that $\lambda_0$ and $k_0$, which
must always be quite small here have no great impact on the spectrum
except to fix $M_{1/2}$.

One restricted subset of region (ii) can be regarded as a ``safe''
region,  characterised by fairly restrictive parameters $h_{t0}\approx
1.5-3$, $|A_0/m_0|\approx 3$, $\tilde{m}_0\stackrel{>}{~}3$, where it
is always possible to achieve small $M_{1/2}$. Outside this ``safe''
region the data suddenly cuts out below a critical $M_{1/2}$ value.
The spectrum for light $M_{1/2}$ is thus always quite similar to that
shown in Fig.6, and since the fine-tuning parameters for this region
(NMSSM1 in Tables 1-3) tend to favour small values of
$M_{1/2}\stackrel{<}{~}100$ GeV, this is perhaps the most interesting
region of the model/ For example, if we consider $M_{1/2}\simeq
100$GeV, at LEPII we would expect to see a standard-model like Higgs
boson with a mass of about 100 GeV. In addition there may be three SUSY
particles accessible to LEPII: a chargino and two neutralinos. The
heavier neutralino and the chargino become closely degenerate as
$M_{1/2}$ is increased. Since the chargino has a strong $M_{1/2}$
dependence its accessibility to LEPII is not guaranteed, but the more
seriously one takes the fine-tuning constraints the more likely its
discovery will seem. We have varied all the parameters over the
``safe'' region, and the characteristic spectrum is never very
different from that shown in Fig.6. The most significant differences
occurs if we take larger $\tilde m_0$, which increases all the scalar
masses, but has no significant impact on the lightest states except to
increase the lightest Higgs mass through radiative corrections.

If we wish to reduce $\tilde m_0$, we must reduce $h_{t0}$ and hence
the top mass; and we must also accept that there is then a minimum
value of $M_{1/2}$ below which correct electroweak symmetry breaking
does not occur. The corresponding spectrum is thus less interesting
phenomenologically, with a typical example shown in Fig. 7.

Another subset of region (ii) was defined as the deep MSSM limit,
namely: $h_{t0} > \lambda_0 \gg k_0$.  Although simply an extreme limit
of region (ii), this region of parameter space involves {\em huge}
values of $|r|\gg 100$, corresponding to quite large values  of
$M_{1/2}$. This deep MSSM region typically involves $\tilde m_0\ll 1$,
$k_0\ll \lambda_0 < 0.01$, while the associated fine-tuning  parameters
corresponding to Fig.8 (NMSSM3 in Tables 1-3) all exceed 50 for the
minimum $M_{1/2}$ value in this example. The reason why this rather
peculiar choice of parameters is nevertheless interesting is that here
we may decrease $m_0$ sufficiently far that the dominantly singlet
state CP-even Higgs boson becomes light. Hence we may find the
phenomenon of interference between the lightest two CP-even Higgs
bosons (one of which is a would-be decoupled state) leading to
interesting phenomenological effects which will be discussed more fully
in Section 7.

\section{The MSSM Limit of the NMSSM}
\subsection{The Electroweak Potential}
Since so much of our most interesting data is in regions which have
$\lambda_0,k_0<<1$ and $|x|>>\nu$, the limit where the NMSSM approaches
the MSSM, it is worthwhile to discuss in some detail the potential and
spectrum in this limit. We begin by looking at the full tree-level
potential
\begin{eqnarray}
V &=& \frac{g_1^2+g_2^2}{8}(\nu_1^2-\nu_2^2)^2
    + \lambda^2 x^2 (\nu_1^2+\nu_2^2)
    + \lambda^2 \nu_1^2 \nu_2^2
    + k^2 x^4 \\ \nonumber
  && - 2 \lambda k \nu_1 \nu_2 x^2
     - 2 \lambda A_{\lambda} \nu_1 \nu_2 x
     - \frac{2}{3} k A_k x^3 \\ \nonumber
  && + m_{H_1}^2\nu_1^2  + m_{H_2}^2\nu_2^2 + m_{N}^2x^2
\end{eqnarray}
In the limit, the minimisation condition for $N$ takes the form \cite{ell}
\begin{equation}
x(2k^2x^2-kA_kx+m_N^2)=0
\end{equation}
which implies
\begin{equation}
4kx=A_k\pm\sqrt(A_k^2-8m_N^2)
\label{xeqn}
\end{equation}
where we select the positive sign for the square root if $A_k>0$ to
ensure that we have selected a minimum and not a maximum. For
simplicity we shall assume this throughout the following discussion; if
$A_k<0$ nothing is altered except this minus sign throughout. Eq.
(\ref{xeqn}) implies that $A_k^2>8m_N^2$.
(Note that $A_k^2>9m_N^2$ ensures that the minimum with non-zero
$x$ is deeper than the minimum at the origin.)
The constraint of
Eq.(\ref{slepvev}) that the sleptons should not acquire VEVs implies
$A_0^2\stackrel{<}{\sim}3m_0^2$, with larger values of $|A_0/m_0|$
being permitted for negative $A_0$ and smaller $\tilde m_0$ as can be
seen from the RG equations. Thus using the fact that in this limit
$A_0\sim A_k$ and $m_N\sim m_0$ we find ourselves restricted to a range
of values of $|A_0/m_0|$ around 3 or slightly larger \cite{ell}. Note
that Eq.(\ref{xeqn}) implies that for fixed $\tilde A_0$ and $\tilde
m_0$, $k\tilde x$ is also fixed, and so $x$ is proportional to
$M_{1/2}$. Furthermore, it is clear that reducing $k$ will lead to an
increase in $x$ and so a very close approach to the MSSM limit.

In this region, the model mimics the MSSM, with
\begin{eqnarray}
\mu\equiv\lambda x \nonumber \\
B\equiv-A_{\lambda}-kx
\label{muBdef}
\end{eqnarray}
as can be simply read off from the lagrangian. Here the value of $kx$
and hence effectively $B$ is set by Eq. (\ref{xeqn}) for given values
of the soft parameters; while  $\lambda x$ ($\mu$) is set by the Z mass
through Eq. (\ref{zeqn}). We note that empirically we find that in
general $\lambda x$ is at least a few hundred GeV in the deep MSSM
limit, which leaves us close to a simple limit of the MSSM where
$\mu>>\nu$.

An interesting feature of the MSSM limit is that we may neglect all
occurences of $\lambda$ and $k$ except those of form $\lambda x$ and
$kx$, and $kx$ can be removed by use of Eq. (\ref{xeqn}), leaving the
only dependence on $\lambda$ and $k$ that through $\lambda/k$ as
mentioned in Section 5.5. We may also come to understand the
fine-tuning behaviour better by considering equation \ref{zeqn}. Here
we may consider
\begin{equation}
\mu^2\equiv\lambda^2 x^2=\left ( \frac{\lambda}{k} \right )^2 (kx)^2
\end{equation}
and since $kx$ is fixed by Eq.(\ref{xeqn}), and ignoring the
dependence of $\tan\beta$ on $\lambda$ and $k$, we find that
$c_{\lambda}$ and $c_k$ derived by differentiating with respect to the
low energy value of $\lambda$ or $k$ is given by
$c_{\lambda}=c_k=(\lambda x)^2/M_Z^2$. The use of $\lambda_0$ rather
than $\lambda$ has the effect of adding a correction factor of  $
\frac{\delta \lambda / \delta \lambda_0}{\lambda_0 / \lambda} $ which
is of order 1 (and is in fact equal to 1 if we neglect all the Yukawa
terms in the running of $\lambda$, which is a reasonable approximation
only for very small $h_t$). At low energy the three  fine-tuning
scenarios NMSSM1,2,3 discussed above have $\lambda x/M_{1/2}$ of around
4,1,1 respectively, which explains the relative sizes of the different
parameters in NMSSM2 and 3 very well. This result is more sensitive to
the closeness to the MSSM limit than others presented here, since we
are neglecting the effects of changing $\lambda$, $k$ on $m_{H_1}^2$,
$m_{H_1}^2$ which is only valid for extremely small $\lambda$, $k$, and
so we do not expect perfect agreement when we consider points such as
NMSSM1 which have relatively large $\lambda$ and $k$.

\subsection{MSSM Equivalent Parameter Sets}
In order to illustrate the close resemblance between the MSSM and the
NMSSM near the limit, we show in Fig. 12 a plot of the spectrum versus
$M_{1/2}$ in the MSSM, with all parameters the same as that in Fig. 6,
and $\tilde\mu_0=-7.5$, a value chosen purely in order to allow us to
mimic the spectrum of Fig. 6. It is clear that the spectrum is
virtually identical, with the exception that the (decoupled) singlet
states are no longer present. Generating such equivalent points in the
MSSM is {\it always} possible near the MSSM limit, since we may simply
read off the low energy values of $\mu$ and $B$ from Eq.
(\ref{muBdef}), run them up to the GUT scale, and thus have a set of
parameters which is guaranteed to give a similar spectrum. Of course,
this relies on the fact that $\lambda$ and $k$ are so small that they
have no impact on the other terms in the RG equations, which is
clearly a good approximation even here, where $\lambda$ and $k$ are
relatively large for this region.

Similarly we present Fig. 13, which is an MSSM version of Fig. 8. Here
the data cuts out for some minimum value of $M_{1/2}$ just above 500GeV
because $\tan\beta$ blows up, and this behaviour occurs in both the
MSSM and NMSSM. Here the spectra are again extremely similar except in
the region where the singlet and non-singlet Higgs are virtually
degenerate. In this region, the two Higgses would have suppressed
couplings to the standard model particles, and so it would be possible
in principle to distinguish the two models, despite the fact that we
are in a region which is extremely close to the MSSM limit. This
behaviour will be discussed in detail in Section 7. Elsewhere, the
extra singlet states in the NMSSM would be virtually undetectable.

By contrast, Fig. 14 shows the MSSM equivalent to Fig. 7, in which the
data does not stop at $M_{1/2}<300$GeV. This is because the data in the
NMSSM version stops because of an alternate minimum which does not
exist for the MSSM. This is an example of the fact that, although the
MSSM can always closely mimic the NMSSM (except where singlet and
non-singlet states are nearly degenerate), the converse is not true;
there are regions where the deepest minimum in the NMSSM is one of the
non-physical minima which does not exist in the MSSM.

We thus conclude that ultimately the most interesting region of the
NMSSM is that where there is a corresponding point in MSSM parameter
space with very similar spectrum and electroweak potential. The
exception is of course the spectrum of singlet states, which can
sometimes mix with physical states to a detectable degree. However,
because of the extremely restrictive nature of the NMSSM eletroweak
potential, there are some sets of MSSM which cannot be imitated
in the NMSSM.

\subsection{Singlet Mass Spectrum}
Now we shall very briefly discuss the mass spectrum in this limit.
Since this consists essentially of the usual MSSM spectrum supplemented
by some almost completely decoupled singlet states we shall just
present some simple results for the singlet states.

The mass of the singlet CP-even Higgs scalar is simply given by the 33
component of the Higgs mass matrix in appendix B as
\begin{equation}
m^2_{h_N}=\frac{1}{4}
       \bigl(A_k\sqrt(A_k^2-8m_N^2) - A_k^2 + 8m_N^2 \bigr )
          + {\cal O}(\frac{\nu^2}{x^2})
\end{equation}
where we have substituted for $x$ using Eq.(\ref{xeqn}). Given that
$A_k$ is a few times larger than $m_0$, we can immediately see that
the mass of the CP-even singlet Higgs will be proportional to $m_0$.
Thus to get the singlet mass down to around the mass of the lighter
physical CP-even Higgs or below, which is necessary if we wish to have
any chance of detecting it, we must have small $m_0$. Since
$M_{1/2}\stackrel{>}{\sim}70$GeV (to keep the chargino clear of the LEP
limit), the ``safe'' region with $m_0\stackrel{>}{\sim}3M_{1/2}$ is
uninteresting, and we must consider a region with smaller $\tilde m_0$
and hence relatively large $M_{1/2}$. This is why we shall find
detectable singlet states (of course, the ``singlet'' state has some
mixing with physical states, which make it couple to the Z) only for
regions with $m_0<1$. We note that empirically we find that we are
fairly close to the limit in which the MSSM CP-even Higgs boson
approaches its mass bound, with fairly large $\tan\beta$ and $\mu$.

Similarly, the CP-odd singlet has mass
\begin{equation}
m^2_{A_N}=\frac{3}{4}A_k \bigl ( A_k - \sqrt(A_k^2-8m_N^2) \bigr )
          + {\cal O}(\frac{\nu^2}{x^2})
\end{equation}
which implies that
\begin{equation}
\frac{m_{h_N}^2}{m_{A_N}^2}=\frac{1}{3}\sqrt(1-8\frac{m_N^2}{A_k^2})
\end{equation}
 From this expression we see that $m_{h_N}/m_{A_N}$ lies in the range
0.4-0.55 for $|A_0/m_0|$ between 3.5 and 6, although it can fall off to
smaller values as $|A_0/m_0|$ approaches its minimum value consistent
with electroweak symmetry breaking. This explains the relation between
the lightest CP-even and lightest CP-odd states shown in reference
\cite{ell3}. The fact that the lightest CP-odd which is {it not}
primarily singlet is always heavier than around 150GeV is explained by
the relatively large value of $\lambda x$, since the mass of the MSSM
CP-odd Higgs is given by $\mu B(\tan\beta+\cot\beta)$ at tree level.

The relationship between CP-odd and CP-even Higgs masses is clearly
seen in Fig.15, where we show a simple scatter plot of the mass of the
lighter CP-odd Higgs boson against the lightest and second lightest
CP-even Higgses. As we would expect, there is always a physical CP-even
Higgs scalar in the region of the NMSSM Higgs bound together with
another CP-even which is primarily singlet. The CP-odd scalar, which is
predominantly singlet everywhere in this figure, has a mass of around
twice that of the singlet CP-even, leading to the diagonal line of data
running through the figure, while the physical Higgs scalar has a mass
in the range $\sim110-140$GeV regardless of the singlet masses, giving
the vertical band of data on the figure.

Finally, we mention the singlet neutralino. This has a mass of $2kx$,
which is around 2-5$m_0$; given that the bino has a mass of typically
around 0.5$M_{1/2}$, we expect to find that the singlet is the lightest
neutralino when $\tilde m_0\stackrel{<}{\sim}0.15$, which is consistent
with our numerical results.

\section{Resolving the Constrained MSSM and NMSSM}

Having discussed the results for both the MSSM and the NMSSM in general
terms, in this section we focus on specific phenomenological aspects of
the constrained  NMSSM which may enable it to be distinguished from the
(constrained) MSSM at future colliders. This question is far from
trivial, since, as we have discovered, the constrained NMSSM is always
close to the MSSM limit, and so it will present a considerable
experimental challenge to resolve the two models. In particular, as we
have shown above, when we are reasonably close to the MSSM limit (as we
always are, except in the low top mass $k_0>\lambda_0>h_{t0}$ region)
it is always possible to construct a set of MSSM parameters which will
mimic ${\it any}$ NMSSM spectrum except for the singlet states. Note
that Higgs bosons do not decay into singlet Higgs bosons in regions of
parameter space close to the MSSM limit.

We shall thus discuss the two possible ways in which it may be possible
to resolve the constrained NMSSM from the MSSM once SUSY or Higgs
bosons have been discovered. The first and most obvious of these is
singlet dilution. The extra singlet degrees of freedom may mix with the
physical degrees of freedom, leading to extra states with diluted
production and decay couplings.  This will always happen for regions
of parameter space of the NMSSM away from the MSSM limit, but the
effect is usually smaller and smaller as this limit is approached. An
important exception is when there are degeneracies of the singlet
states with physical states, leading to strong mixing effects which
persist even in the deep MSSM limit. In general this effect relies on
the  accidental degeneracy of two states, where away from the
degeneracy one of the states is essentially singlet, and the other
state is physical and has the same spin and CP quantum numbers as the
singlet. This effect may be observed when the CP-even singlet becomes
degenerate with a physical CP-even Higgs boson, when the  CP-odd Higgs
singlet becomes degenerate with the physical CP-odd Higgs bosons, or
when the singlet Higgsino becomes degenerate with the  physical
Higgsinos (or physical neutralino states which contain a significant
Higgsino component).

The second method for distinguishing the two models is by finding
regions of parameter space in the MSSM where the spectrum cannot be
mimicked in the NMSSM, and we shall also discuss this possibility
below.

\subsection{CP-even Higgs Bosons at LEP}

We have already seen that the successful regions of parameter space of
the constrained NMSSM mean that the model is always close to the formal
MSSM limit in which the singlet components decouple.  Furthermore, over
much of the parameter space\cite{ell3,ekw0} this leads to the lightest
CP-even Higgs boson being in the range 70-140 GeV and having couplings
which are so close to those of the standard model Higgs boson as to
make it practically indistinguishable. A similar expectation arises
from the MSSM in the large CP-odd mass limit where all the Higgs boson
states become heavy and decoupled, apart from the lightest CP-even
Higgs boson which mimics the standard model Higgs boson \cite{Haber}.
Thus the discovery of a standard model-like Higgs boson at  LEPII is
consistent with both the MSSM and the NMSSM, and would not tell us very
much about supersymmetry.

However there exist regions of parameter space of the constrained NMSSM
in which it may be possible to experimentally distinguish the lightest
CP-even  Higgs boson in the NMSSM from that of the MSSM or the standard
model.  For example, in the constrained NMSSM, a (would-be) decoupled
singlet CP-even Higgs boson may become degenerate with a physical
CP-even Higgs boson, leading to strong mixing and hence two weakly
coupled states. The purpose of this sub-section is to discuss the
properties and prospects of discovery of such mixed states at LEP. Some
work along these lines has already been done \cite{ell3}. Specifically
scatter plots of the Higgs masses against the Z couplings for the
lightest CP-even Higgs boson have been  made for a representative data
sample. Most of the data sample was seen to lie outside of the
discovery range of LEPI or LEPII. Other scatter plots were also
considered including one for the second lightest CP-even Higgs
\cite{ell3}, but not including a  discussion of its Z couplings.
The decay couplings of the two Higgs bosons were also not
considered in ref.\cite{ell3}.

The CP-odd Higgs bosons play no role at LEPII since for masses less
than around 150GeV the lighter one has a mass roughly equal to twice
that of the lightest CP-even Higgs boson, as discussed earlier, and is
in any case dominantly gauge singlet. The heavier (physical) CP-odd
Higgs boson turns out to be heavier than 150 GeV \cite{ell3}.

The purpose of this section is try to build up a better picture of the
implications which might be drawn from the discovery of non-standard
Higgs bosons at LEP. After discussing the production and decay
couplings of the two lightest CP-even Higgs bosons relevant for LEP, we
shall discuss the effect of Higgs  crossing seen in Figs.8,9  in a
little more detail by plotting both Higgs masses against their
respective physical couplings. We shall then discuss three different
non-standard model Higgs scenarios at LEPI and LEPII, and show examples
of constrained NMSSM parameter space which can yield Higgs events
within experimental reach for each scenario.

We begin the discussion by summarising the
production and decay couplings relevant for LEP
of the two lightest CP-even
Higgs bosons in the NMSSM,
following closely the notation of ref.\cite{NMSSM3}.
We first write down the real orthogonal matrix $U_{ij}$ which
relates the CP-even mass eigenstates $(h_1,h_2,h_3)$
(where by definition the masses are ordered as $m_{h_1}<m_{h_2}<m_{h_3}$)
to the original CP-even states
\begin{equation}
(H_1,H_2,N)\equiv
\sqrt{2}(\Re (H_1^0) - \nu_1,
\Re (H_2^0) - \nu_2,
\Re (N) - x)
\end{equation}

\begin{equation}
\left( \begin{array}{c}
h_1 \\
h_2 \\
h_3  \end{array} \right)
 =
     \left( \begin{array}{ccc}
     U_{11} & U_{12} & U_{13} \\
     U_{21} & U_{22} & U_{23} \\
     U_{31} & U_{32} & U_{33}
     \end{array} \right)
\left( \begin{array}{c}
H_1 \\
H_2 \\
N   \end{array} \right)
\end{equation}
Following ref.\cite{NMSSM3}, we now define the relative couplings $R$
compared to the standard model couplings as: coupling $= R \times $
standard model coupling, where the relative $ZZ$ production couplings
of the lightest and second lightest CP-even Higgs bosons, and the
$\bar{u}u$ and $\bar{d}d$ decay couplings of these bosons (all three
generations will have their couplings suppressed by the same amount),
are given below.

\begin{eqnarray}
R_{ZZh_1} & = & \cos \beta U_{11} + \sin \beta U_{12} \nonumber \\
R_{ZZh_2} & = & \cos \beta U_{21} + \sin \beta U_{22} \nonumber \\
R_{\bar{u}uh_1} & = & \frac{U_{12}}{\sin \beta} \nonumber \\
R_{\bar{d}dh_1} & = & \frac{U_{11}}{\cos \beta} \nonumber \\
R_{\bar{u}uh_2} & = & \frac{U_{22}}{\sin \beta} \nonumber \\
R_{\bar{d}dh_2} & = & \frac{U_{21}}{\cos \beta}
\end{eqnarray}

These couplings should be compared to those in the MSSM:
\begin{equation}
\left( \begin{array}{c}
h_1 \\
h_2 \end{array} \right)
 =
     \left( \begin{array}{cc}
     \cos \alpha & - \sin \alpha \\
     \sin \alpha & \cos \alpha
     \end{array} \right)
\left( \begin{array}{c}
H_1 \\
H_2 \end{array} \right)
\end{equation}

\begin{eqnarray}
R_{ZZh_1} & = & \cos (\beta +\alpha) \nonumber \\
R_{ZZh_2} & = & \sin (\beta + \alpha) \nonumber \\
R_{\bar{u}uh_1} & = & \frac{-\sin \alpha}{\sin \beta} \nonumber \\
R_{\bar{d}dh_1} & = & \frac{\cos \alpha}{\cos \beta} \nonumber \\
R_{\bar{u}uh_2} & = & \frac{\cos \alpha}{\sin \beta} \nonumber \\
R_{\bar{d}dh_2} & = & \frac{\sin \alpha}{\cos \beta}
\end{eqnarray}

We have chosen to focus on the production and decay couplings
of the lightest and second lightest CP-even Higgs bosons only,
since the third CP-even Higgs boson expected in the NMSSM will
be out of range of LEPII. We have already remarked that a
likely scenario of the MSSM and NMSSM is of
a single standard model-like Higgs boson in the LEPII range.
Now we turn our attention to the more optimistic
possibility that the lightest two CP-even Higgs have non-standard
couplings. In practice this means weaker couplings than in the
standard model. As seen earlier, this effect only
seems to occur in the deep MSSM limit of the model,
corresponding to a limited volume of parameter space
of the full NMSSM, as
discussed earlier.

In order to illustrate the effect, in Fig.16a we plot the lightest and
second lightest CP-even Higgs boson masses $m_{h_1}$ and $m_{h_2}$,
against their respective $R_{ZZh_1}$ (small crosses) and $R_{ZZh_2}$
(large crosses). Fig.16a uses a range of parameters corresponding to
those used in Figs.8,9 ({\it i.e.} in the deep MSSM limit of the model
where the Higgs singlet may become light and mix).  The correlation
between the two Higgs boson masses is the obvious one: namely the
lightest small crosses are associated with the lightest large crosses.
The values of  $k_0$ and $M_{1/2}$ increase from left to right, while
$m_t$ and $\tan \beta$ decrease in this direction. From left to right
the data thus corresponds to the effect observed in Figs.8,9 of the
singlet approaching the physical Higgs boson, mixing with it, then
crossing it -- although the behaviour is better described as a
displacement rather than a crossing since the two states are never
exactly degenerate. Also shown in Fig.16a is the region excluded by
LEPI (short dashed line) \cite{LEP1}, the discovery reach of LEPII at
an energy of 175 GeV and an integrated luminosity of 500 pb$^{-1}$
(dashed line) and the discovery reach of LEPII at an energy of 205 GeV
and an integrated luminosity of 300 pb$^{-1}$ (full line). The latter
two lines are simply calculated from the cross-section \cite{hhg} so as
to give 50 events, without imposing experimental cuts. Clearly LEPII
cannot reach any of these particular data points. However there do
exist regions of parameter space in the singlet mixing region which are
accessible to LEP, as we shall discuss.

In Fig.16b we plot the couplings
$R_{\bar{d}dh_1}$ (small crosses),
$R_{\bar{u}uh_1}$ (circles),
$R_{\bar{d}dh_2}$ (large crosses),
$R_{\bar{u}uh_2}$ (squares),
against the corresponding Higgs boson mass. Note that, for a given
Higgs boson mass, the relative $u$ and $d$ couplings are approximately
equal, although often with substantially reduced couplings relative to
the standard model. This effect is due to the fact that in this case
the Higgs mixing effect can be regarded as taking place between a
standard-model-like Higgs boson (with relative R couplings close to
unity) and a singlet. The effect of mixing with the singlet is simply
to dilute all the couplings equally, both here and for other parameters
where we do not show the quark couplings explicitly.

We distinguish three representative scenarios for LEP:

(i) A light very weakly coupled CP-even boson with $0<m_{h_1}<60$ GeV

Such a boson may be hiding in high statistics data from LEPI. If only
one Higgs boson is discovered then it may be possible  to exclude
either the minimal standard model, or the MSSM, if the production and
decay couplings of the Higgs boson are seen to be  sufficiently
non-standard. For example, a light very weakly coupled CP-even Higgs
boson in the MSSM would correspond to $R_{ZZh_1}  =  \cos (\beta
+\alpha)\ll 1$, but if its mass were less than about 40 GeV such a
boson would necessarily be accompanied by a light CP-odd Higgs boson
with $R_{ZAh} = \sin (\beta +\alpha) \approx 1$, and such a scenario is
already excluded in the MSSM. Thus the discovery of such a boson at
LEPI or LEPII would rule out the MSSM, and suggest the NMSSM.

An example of a light very weakly coupled Higgs boson accessible to
LEPI (rather than LEPII) is shown in Fig.17. In this case the light
weakly coupled Higgs boson  has a mass 34-60 GeV, and is just outside
the existing LEPI excluded region. Such a Higgs boson with mass around
40 GeV or less would therefore most likely be found when more Z pole
data is analysed, and would be a clear signal of the NMSSM. The second
lightest Higgs boson has standard model-like couplings to the Z, but
has a mass of 127-130 GeV, outside the LEPII range in this example. The
corresponding  $R_{\bar{u}uh_1}$, $R_{\bar{d}dh_1}$, $R_{\bar{u}uh_2}$,
$R_{\bar{d}dh_2}$ factors are not shown, since they are virtually
identical to the $R_{ZZh}$, $R_{ZZH}$ couplings in Fig.17

The remainder of the spectrum is unaccessible to any likely collider in
the near future, with a singlet CP-odd Higgs in the range 70-120GeV, a
singlet neutralino of 60-100GeV, and the remainder of the spectrum very
heavy because of the typically large values of $M_{1/2}$ of around
1TeV. Such general comments generally apply whenever the lightest
CP-even Higgs is light enough to be detectable at LEP.

(ii) A weakly coupled CP-even boson with $60<m_{h_1}<110$ GeV
and $R_{ZZh}<0.95$.

Such a boson may be discovered at LEPII, but since the statistics will
be lower, the couplings probed must not be so weak as in the previous
case. A heavier weakly coupled CP-even Higgs boson with mass
$m_{h_1}>60$ GeV is now allowed in the MSSM since for small $\tan
\beta$ the CP-odd boson may be heavier than 100 GeV. In this case in
order to exclude the MSSM it would be necessary to study the decay
couplings $\bar{c}ch_1$, $\bar{b}bh_1$, for example.  It may be
difficult at LEPII to study these decays of the Higgs boson to
sufficient accuracy to enable the MSSM to be excluded. Thus the
discovery  of such a boson at LEPII may exclude the standard model, and
suggest supersymmetry, but it will be more difficult to distinguish
between the MSSM and the NMSSM.

Fig.18a shows some examples of heavier weakly coupled Higgs bosons
which would be visible at LEPII. The input parameters, which are
selected so that such a Higgs boson emerges in the mass range 60-110
GeV, are over quite a broad range. The corresponding
$R_{\bar{u}uh_1}$, $R_{\bar{d}dh_1}$, $R_{\bar{u}uh_2}$,
$R_{\bar{d}dh_2}$ factors are shown in Fig.18(b). Some of this data
(for $m_{h_1}>100$ GeV) shows  relative differences between $d$ and $u$
type decays. The second lightest Higgs bosons are outside the range of
LEPII.

(iii) Two CP-even bosons with $m_{h_{1,2}}<110$ GeV

The situation improves dramatically if two CP-even Higgs bosons are
discovered at LEP.  For example, the lighter one could be discovered in
high statistics data at LEPI or  in the first phase of LEPII, and the
heavier one could be found in the final phase of LEPII. Such twin
discoveries would immediately rule out the standard model, and also the
MSSM since in this model the second CP-even Higgs boson cannot be so
light. In the unconstrained NMSSM this scenario is perfectly possible
but this has  already been discussed elsewhere\cite{ekw0}. In the
constrained NMSSM, this possibility appears  to be more unlikely than
either (i) or (ii) above. The reason is clear from Figs.17, 18, where
it is seen that the second lightest Higgs boson is usually outside the
mass reach of LEPII.

Fig. 19 shows an example of how tantalisingly close a twin Higgs
discovery might be at LEPII. The lightest Higgs bosons within the LEPII
reach are associated with heavier Higgs bosons just outside the reach
of  the final phase of LEPII. Note that the top quark mass is below 150
GeV in this case. One reason for this is that the second Higgs boson
(predominantly physical) has a mass which increases with increasing top
mass due to radiative corrections, and we have required its mass to be
as small as possible. We can arrange the parameters to reduce the mass
of the second lightest Higgs boson to the LEPII range, and increase the
top quark mass, but then  we find that the lightest Higgs boson's
couplings become too weak to allow it to be produced at LEPII. Thus we
infer that this particular scenario, although spectacular, is very
unlikely in the constrained NMSSM. We do not show quark couplings
because they are nearly equal to the Higgs couplings. Note that
the lighter (heavier)
values of the lightest CP-even state are correlated with the
lighter (heavier) values of the second lightest CP-even state.

\subsection{Neutralinos}

The neutralino structure is essentially that of the MSSM supplemented
by singlet, with a mass in the range around 2-5 $m_0$ as noted above.
While it is possible for the two lightest neutralinos to be degenerate,
which we find to occur for $\tilde m_0\sim 0.1$, we find negligible
neutralino mixing even in this region. The reason for this is that, as
a look at the mass matrix (see Appendix B3) will show, there is
negligible mixing between the singlet and the bino, and we always find
that the lightest MSSM-like neutralino is pure gaugino (photino for
small $M_{1/2}$, or Bino for large $M_{1/2}$) because we never find
very small $\mu$. Hence it is unlikely that the singlet neutralinos
could ever be detected except in the case where there is mixing with a
heavier higgsino-like neutralino, which would only become of
phenomenological interest after the lighter states of the neutralino
spectrum has been discovered.

Although over some regions of parameter space, it is kinematically
possible for the lightest CP-even Higgs boson to decay to two lightest
neutralinos, in fact this will not happen because the lightest
neutralino is always pure gaugino except in regions where it is pure
singlet and the couplings are very small. Thus invisible Higgs decays
do not occur in this model.

\subsection{Regions Where the NMSSM Cannot Exist}

In principle, once supersymmetric particles have been discovered it
would be possible to derive the values of the supersymmetric parameters
$M_{1/2}$, $A_0$, $B_0$, $m_0$, and $\tan\beta$, although to do so
would require the discovery and study of several of the supersymmetric
states. The question would then arise as to whether the spectrum was
the result of the MSSM, or of the NMSSM in a region where the two are
indistinguishable. Here we shall review a few of the features of the
constrained NMSSM which would allow it to be ruled out if the
supersymmetric spectrum did not display them.

The first, and most obvious, of these features is that of large
$|A_0/m_0|$. Unlike the case in the NMSSM, in the MSSM this can take
any value from zero up to a maximum of around 3-5 caused by the
constraint of avoiding slepton VEVs, and so if $|A_0/m_0|$ were found
to take a smaller value than around 3 we could immediately rule out the
constrained NMSSM. Of course, this would require a detailed enough
study of the spectrum to calculate all the SUSY parameters.

The second question is that of the relation between $B$ and $A_0$ given
by
\begin{equation}
B\equiv-A_{\lambda}-kx
=-A_{\lambda}-\frac{1}{4}\left(A_k-\sqrt(A_k^2-8m_N^2)\right)
\label{Beqn}
\end{equation}
Given that it is trivial to relate $A_{\lambda}$, $A_k$, and $m_N^2$ to
$A_0$ and $m_0$ in the small $\lambda_0$, $k_0$ limit, the MSSM limit
of the NMSSM implies a relation between the effective MSSM parameters
which, if violated, would allow us to rule out the constrained NMSSM.

Finally, even if the MSSM were in some region where it obeys the above
relations, it would be possible to use our knowledge of $\mu_0$, $B_0$,
and $A_0$ together with the relation $\mu\equiv\lambda x$ and
Eqs.(\ref{Beqn}),(\ref{xeqn}) to derive $\lambda$, $k$, $x$. One would
then have full knowledge of all the parameters in the effective
potential, and could simply test it to ensure that the minimum
mimicking the MSSM were indeed the deepest. An example of such a case
is that shown in Figs.7,14, where for small $M_{1/2}$ the NMSSM
minimum with the correct Z mass is not the deepest, unlike the case for
the MSSM.

If the MSSM were to pass all of these various tests, then we would
perhaps begin to suspect that it were really the NMSSM in a limit which
would be very difficult to test. However, if the NMSSM were to fail
these tests, then we would either have to rule it out altogether or
consider such possibilities as non-universal soft masses to avoid these
constraints, which would reduce the aesthetic appeal of the model.

\section{Conclusions}

In this paper we have given a comprehensive
discussion of the NMSSM constrained
by unification, correct electroweak symmetry breaking, and universal
soft parameters, and compared the results in this model to those of the
similarly constrained MSSM, where we have chosen a set of input
variables most appropriate for this comparison. In particular the input
parameters $h_{t0}$, $\tilde{A}_0$ and $\tilde{m}_0$ are common to both
models, where the tilde denotes scaling by $M_{1/2}$ which is
determined as an output parameter by the requirement of obtaining the
correct Z mass. We have also given a short discussion of how the
constraint of fine-tuning affects the model.

After briefly discussing the MSSM, which serves as a check of our
methods, we have explored the parameter space of the NMSSM consistent
with the above constraints. The constraint of correct electroweak
symmetry breaking is much more severe in the NMSSM than the MSSM. The
reason is simply that in the NMSSM there are three VEVs $\nu_1$,
$\nu_2$, $x$ to consider, and it is non-trivial to arrange for them all
to be consistently non-zero. Non-physical vacua with one or more of
these VEVs equal to zero litter the parameter space of the NMSSM, and
often the physical vacuum is competing with several non-physical vacua.
Having found a consistent physical vacuum for one choice of parameters,
moving around parameter space is rather like walking through a
minefield since suddenly one of the non-physical vacua can become the
global minimum of the effective potential. We have mapped out the
successful regions of parameter space in Section 5, and summarised the
results of our survey in 5.6. Some of these results can be understood
analytically as discussed in Section 6.

One important question is whether and how the constrained  MSSM
and NMSSM may be resolved experimentally, and we studied this in
Section 7. The main difficulty is that the allowed region of parameter
space of the NMSSM which allows a large top quark mass (referred to as
region (ii), which has $h_{t0}>\lambda_0>k_0$) is always quite close to
the MSSM limit of the model, while another region with
$k_0>\lambda_0>h_{t0}$ is less interesting because it gives very small
top masses. The problem was illustrated in Section 6.2 where Figs.12-14
show how the constrained MSSM can mimic the constrained NMSSM spectrum
of Figs.6-8 less the (in practice almost undetectable) singlet states.
The most marked difference is that between Fig.8 and Fig.13, due to the
approximate degeneracy and subsequent mixing of the lightest and second
lightest CP-even Higgs bosons in the constrained NMSSM. Away from the
region of approximate degeneracy, one of the states is physical and one
of the states corresponds to the decoupled singlet. The effect of their
mixing is thus to increase the number of physical states by one, but to
dilute the physical couplings of each of the two states. We refer to
this phenomenon as ``singlet dilution''. Paradoxically singlet dilution
only appears to occur in the deep MSSM limit of the constrained NMSSM
(although this is not true of the unconstrained NMSSM).

Despite the fact that the deep MSSM limit involves rather odd choices
of parameters with a preference for smaller top mass, very large
singlet VEV, very small $m_0/M_{1/2}$ and $k_0$, and relatively
large $M_{1/2}$, and also involves quite severe fine-tuning, the
phenomenological consequences of singlet dilution are so important that
we discussed its implications for LEP in detail in section 7. It is
possible that there is a very light very weakly coupled Higgs boson
which LEPI, with increased statistics could discover but which LEPII
would fail to find (Fig.17). Such a discovery would be an unmistakable
signal of the NMSSM, since the MSSM can never emulate this. Another
possibility is that of a weakly coupled CP-even Higgs boson outside the
range of LEPI but visible to LEPII (Fig.18a). Such a Higgs boson cannot
exist in the standard model, but can in the MSSM. In order to
differentiate between the NMSSM and MSSM, one must study the
decay couplings of the Higgs boson. As shown in Fig.18b the decay
couplings in the NMSSM are diluted by a similar amount to the
production couplings, which is a simple consequence of singlet
dilution, and provides a clear signature of the NMSSM. The
simplest and cleanest test of singlet dilution, however,
would be to discover
{\em both} the lightest two CP-even Higgs bosons at LEP, although as
illustrated in Fig.19, this possibility is outside the
regions of parameter space which we have explored.

We have seen that
the allowed regions of parameter space of the constrained NMSSM is
quite restricted. A large top mass restricts us to region (ii)
(summarised in section 5.6) where
$|A_0/m_0|\stackrel{>}{\sim}3$ is necessary.
The further requirement of small $M_{1/2}$ further
restricts region (ii) to a very small ``safe'' region of
parameter space, leading to the prediction of the
almost unique Higgs and SUSY spectrum,
illustrated in Fig.6. However, as remarked, such a spectrum
can be accurately mimicked by the MSSM, as shown in Fig.12.
It would therefore not be possible to prove the existence
of the NMSSM from a spectrum such as this,
but it would provide circumstantial evidence for the model.

Finally it is worth emphasising that  the constrained NMSSM is
relatively easy to exclude.  For example, once the SUSY spectrum has
been studied, one may infer the value of the ratio $|A_0/m_0|$ and if
it is not sufficiently large then the constrained NMSSM would be
excluded. Other similar consistency checks may also be applied on the
constrained NMSSM as discussed in Section 7.3, and these could also
serve to rule out the model. Of course if some of the constraints are
removed (for example those concerning universal soft parameters) then
the NMSSM might not be excluded. Removing the constraints of
unification and universal soft parameters returns us to the
unconstrained NMSSM, which was recently studied in ref.\cite{ekw}.

\noindent{\Large {\bf Acknowledgements}}

\noindent
We would like to thank Ulrich Ellwanger for a number of helpful
discussions, and Terry Elliott for his collaboration at an early stage
of this project.

\pagebreak

\appendix
\noindent{\Large {\bf Appendices}}

\section{Renormalisation Group Equations}

Here we reproduce the set of one--loop Renormalisation Group Equations
in the Next--to--Minimal Supersymmetric Standard Model, and in the
Minimal Supersymmetric Standard Model. These may be found in
Derendinger and Savoy \cite{NMSSM2} to one loop, and the derivation of
the two loop extension is straightforward \cite{mikeandco}

Retaining only $h_t$, $h_b$ and $h_\tau$ (the top quark, bottom quark
and tau lepton Yukawa couplings, respectively), we have the RG
equations (where $t=\log \mu$, and $\mu$ is the $\overline{DR}$
renormalisation scale) for the NMSSM are given below. The RGEs for
the dimensionless couplings are given to two loops, those for the
soft masses to one.
\begin{eqnarray}
16\pi^2\frac{d}{dt} g_1     & = & 11g_1^3 + \frac{g_1^3}{16\pi^2}\left(
                 \frac{199}{9}g_1^2 + 9g_2^2
                       + \frac{88}{3}g_3^2 \right . \nonumber \\
            &  & \left. \qquad\qquad - \frac{26}{3}h_t^2
                -\frac{14}{3}h_b^2 - 6h_{\tau}^2
                -2\lambda^2  \right ) \nonumber \\
16\pi^2\frac{d}{dt} g_2     & = & g_2^3 + \frac{g_2^3}{16\pi^2}\left(
                 3g_1^2 + 25g_2^2 + 24g_3^2 \right .\nonumber \\
            &  & \left . \qquad\qquad - 6h_t^2
                - 6h_b^2 - 2h_{\tau}^2
                -2\lambda^2  \right ) \nonumber\\
16\pi^2\frac{d}{dt} g_3     & = & (-3)g_3^3 + \frac{g_3^3}{16\pi^2}\left(
                 \frac{11}{3}g_1^2 + 9g_2^2 + 14g_3^2
                    - 4h_t^2 - 4h_b^2 \right ) \nonumber\\
16\pi^2\frac{d}{dt} h_t     & = & (6 h_t^2 + h_b^2 + \lambda^2
                   - \frac{13}{9} g_1^2
                   - 3 g_2^2 - \frac{16}{3} g_3^2) h_t \nonumber\\
            & & + \frac{h_t}{16\pi^2}\left(
                 \frac{2743}{162}g_1^4 + \frac{15}{2}g_2^4
                - \frac{16}{9}g_3^4
                + \frac{5}{3}g_1^2g_2^2
                + \frac{136}{27}g_1^2g_3^2
                + 8g_2^2g_3^2 \right.\nonumber\\
            & & \quad + (2g_1^2 + 6g_2^2 + 16g_3^2) h_t^2
                + \frac{2}{3}g_1^2h_b^2 \nonumber\\
            & &\quad -22h_t^4 - 5h_t^2h_b^2 - 5h_b^4 - h_b^2h{\tau}^2
                  \nonumber\\
            & &\left. \quad
             - \lambda^2(3h_t^2+4h_b^2+h_{\tau}^2+2k^2+3\lambda^2)
               \right ) \nonumber\\
16\pi^2\frac{d}{dt} h_b     & = & (6 h_b^2 + h_t^2 + h_\tau^2 + \lambda^2
                   - \frac{7}{9} g_1^2 - 3 g_2^2
                   - \frac{16}{3} g_3^2) h_b \nonumber\\
            & & + \frac{h_b}{16\pi^2}\left(
                \frac{287}{90}g_1^4 + \frac{15}{2}g_2^4 - \frac{16}{9}g_3^4
                + \frac{5}{3}g_1^2g_2^2
                + \frac{40}{27}g_1^2g_3^2
                + 8g_2^2g_3^2 \right . \nonumber\\
            & & \quad + (\frac{2}{3}g_1^2 + 6g_2^2 + 16g_3^2) h_b^2
                + \frac{4}{3}g_1^2h_t^2 + 2g_1^2h_{\tau}^2 \nonumber\\
            & &\quad -22h_b^4 - 5h_t^2h_b^2 - 5h_t^4
                - 3h_b^2h{\tau}^2 - 3h_{\tau}^4 \nonumber\\
            & &\left . \quad - \lambda^2(3h_b^2+4h_t^2+2k^2+3\lambda^2)
               \right ) \nonumber\\
16\pi^2\frac{d}{dt} h_\tau  & = & (4 h_\tau^2 + 3 h_b^2 + \lambda^2
                   - 3 g_1^2 - 3 g_2^2) h_\tau \nonumber\\
            & & + \frac{h_\tau}{16\pi^2}\left(
                \frac{27}{2}g_1^4 + \frac{15}{2}g_2^4
                + 3g_1^2g_2^2 \right . \nonumber\\
            & & \quad + (2g_1^2 + 6g_2^2) h_{\tau}^2
                \quad + (-\frac{2}{3}g_1^2 + 16g_3^2) h_b^2 \nonumber\\
            & &\quad -10h_{\tau}^4 - 9h_{\tau}^2h_b^2 - 9h_b^4
                - 3h_b^2h_t^2 \nonumber\\
            & &\left. \quad - \lambda^2(3h_{\tau}^2+3h_t^2+2k^2+3\lambda^2)
               \right ) \nonumber\\
16\pi^2\frac{d}{dt} \lambda & = & (4 \lambda^2 + 2 k^2 + 3 h_t^2
                   + 3 h_b^2 + h_\tau^2
                   - g_1^2 - 3 g_2^2) \lambda \nonumber\\
            & & + \frac{\lambda}{16\pi^2}\left(
                \frac{23}{2}g_1^4 + \frac{15}{2}g_2^4
                + 3g_1^2g_2^2 \right . \nonumber\\
            & & \quad + 2g_1^2\lambda^2 + 2g_1^2h_{\tau}^2
                 + \frac{4}{3}g_1^2h_t^2 - \frac{2}{3}g_1h_b^2
                 + 6g_2^2\lambda^2 \nonumber\\
            & &  + 16g_3^2 h_t^2 + 16g_3^2 h_b^2 \nonumber\\
            & &\quad -9h_t^4 - 6h_t^2h_b^2 - 9h_b^4 - 3h_{\tau}^4
                  -10\lambda^4-8k^4 \nonumber\\
            & &\quad \left . - \lambda^2(9h_t^2+9h_b^2+3h_{\tau}^2+12k^2)
               \right ) \nonumber\\
16\pi^2\frac{d}{dt} k       & = & 6 (\lambda^2 + k^2) k
               + \frac{k}{16\pi^2}\left(
               6g_1^2\lambda^2 + 18g_2^2\lambda^2 \right.\nonumber \\
            & & \left. -\lambda^2(12\lambda^2-24k^2-18h_t^2
              -18h_b^2-6h{\tau}^2) -24 k^4
               \right ) \nonumber\\
16\pi^2\frac{d}{dt} A_{u_a} & = & 6 h_t^2 (1+\delta_{a 3}) A_t
                  + 2 h_b^2 \delta_{a 3} A_b
                  + 2 \lambda^2 A_\lambda \nonumber \nonumber\\
            & - & 4 (\frac{13}{18} g_1^2 M_1 + \frac{3}{2} g_2^2 M_2
                       + \frac{8}{3} g_3^2 M_3) \nonumber\\
16\pi^2\frac{d}{dt} A_{d_a} & = & 6 h_b^2 (1+\delta_{a 3}) A_b
                  + 2 h_t^2\delta_{a 3} A_t
                  + 2 h_\tau^2 \delta_{a 3} A_\tau
                  + 2 \lambda^2 A_\lambda \nonumber \\
            & - & 4 (\frac{7}{18} g_1^2 M_1 + \frac{3}{2} g_2^2 M_2
                       + \frac{8}{3} g_3^2 M_3) \nonumber\\
16\pi^2\frac{d}{dt} A_{e_a} & = & 2 h_\tau^2 (1 + 3 \delta_{a 3}) A_\tau
                +  6 h_b^2 A_b + 2 \lambda^2 A_\lambda \nonumber \\
            & - & 6 (g_1^2 M_1 + g_2^2 M_2) \nonumber\\
16\pi^2\frac{d}{dt} A_\lambda & = & 8 \lambda^2 A_\lambda - 4 k^2 A_k
                    + 6 h_t^2 A_t
                    + 6 h_b^2 A_b + 2 h_\tau^2 A_\tau \nonumber \\
            & - & 2 (g_1^2 M_1 + 3 g_2^2 M_2) \nonumber\\
16\pi^2\frac{d}{dt} A_k     & = & 12 (k^2 A_k - \lambda^2 A_\lambda)
\nonumber\\
16\pi^2\frac{d}{dt} m_{Q_a}^2 & = & 2 \delta_{a 3} h_t^2
                    (m_{Q_3}^2 + m_{H_2}^2 + m_T^2 + A_t^2)
                  + 2 \delta_{a 3} h_b^2
                    (m_{Q_3}^2 + m_{H_1}^2 + m_B^2 + A_b^2)
                    \nonumber \\
              & - & 8 (\frac{1}{36} g_1^2 M_1^2 + \frac{3}{4} g_2^2 M_2^2
                      +\frac{4}{3} g_3^2 M_3^2)
                    + \frac{1}{3} g_1^2 \xi \\
16\pi^2\frac{d}{dt} m_{u_a}^2 & = & 4 \delta_{a 3} h_t^2
                    (m_{Q_3}^2 + m_{H_2}^2 + m_T^2 + A_t^2)
                    \nonumber \\
              & - & 8 (\frac{4}{9} g_1^2 M_1^2 + \frac{4}{3} g_3^2 M_3^2)
                    - \frac{4}{3} g_1^2 \xi \\
16\pi^2\frac{d}{dt} m_{d_a}^2 & = & 4 \delta_{a 3} h_b^2
                    (m_{Q_3}^2 + m_{H_1}^2 + m_B^2 + A_b^2)
                    \nonumber \\
              & - & 8 (\frac{1}{9} g_1^2 M_1^2 + \frac{4}{3} g_3^2 M_3^2)
                    + \frac{2}{3} g_1^2 \xi \\
16\pi^2\frac{d}{dt} m_{L_a}^2 & = & 2 \delta_{a 3} h_\tau^2
                    (m_{L_3}^2 + m_{H_1}^2 + m_\tau^2 + A_\tau^2)
                    \nonumber \\
              & - & 8 (\frac{1}{4} g_1^2 M_1^2 + \frac{3}{4} g_2^2 M_2^2)
                    - g_1^2 \xi \\
16\pi^2\frac{d}{dt} m_{e_a}^2 & = & 4 \delta_{a 3} h_\tau^2
                    (m_{L_3}^2 + m_{H_1}^2 + m_\tau^2 + A_\tau^2)
                    \nonumber \\
              & - & 8 g_1^2 M_1^2 + 2 g_1^2 \xi \\
16\pi^2\frac{d}{dt} m_{H_1}^2 & = & 6 h_b^2
                    (m_{Q_3}^2 + m_{H_1}^2 + m_B^2 + A_b^2)
                  + 2 h_\tau^2
                    (m_{L_3}^2 + m_{H_1}^2 + m_\tau^2 + A_\tau^2)
                    \nonumber \\
              & + & 2 \lambda^2
                    (m_{H_1}^2 + m_{H_2}^2 + m_N^2 + A_\lambda^2)
                   - 8(\frac{1}{4} g_1^2 M_1^2 + \frac{3}{4} g_2^2 M_2^2)
                   - g_1^2 \xi \\
16\pi^2\frac{d}{dt} m_{H_2}^2 & = & 6 h_t^2
                   (m_{Q_3}^2 + m_{H_2}^2 + m_T^2 + A_t^2)
                  + 2 \lambda^2
                   (m_{H_1}^2 + m_{H_2}^2 + m_N^2 + A_\lambda^2)
                   \nonumber \\
              & - & 8 (\frac{1}{4} g_1^2 M_1^2 + \frac{3}{4} g_2^2 M_2^2)
                  + g_1^2 \xi \\
16\pi^2\frac{d}{dt} m_{N}^2   & = & 4 \lambda^2
                    (m_{H_1}^2 + m_{H_2}^2 m_N^2 + A_\lambda^2)
                   + 4 k^2 (3 m_N^2 + A_k^2)
\end{eqnarray}
Here the subscript $a\in\{1,2,3\}$ is a generation index. We have
assumed no inter--generational mixing. $\xi$ is the
hypercharge--weighted sum of all soft masses--squared
\begin{equation}
\xi = \sum_i Y_i m_i^2,
\end{equation}
where $i$ runs over all scalar particles. Notice that if we impose the
constraint $m_i^2(Q)=m_0^2$ for some value of $Q$, typically the
unification scale $M_X$, then $\xi(Q) = 0$. If $\xi$ is 0 at one
scale, then it is 0 at all scales.

It is well known that the gaugino masses $M_i$ evolve identically to
$\alpha_i$ at one loop, so that we have the result
\begin{equation}
\frac{M_i(Q)}{M_{\frac{1}{2}}} = \frac{g_i^2(Q)}{g_X^2}
\end{equation}

The RG equations for the MSSM are identical to those for the NMSSM
with $\lambda$ and $k$ set to 0, supplemented with the following two
equations for $\mu$ and $B$:
\begin{eqnarray}
16\pi^2\frac{d}{dt} \mu & = & \mu (3 h_t^2 + 3 h_b^2
     + h_\tau^2 - g_1^2 - 3 g_2^2) \\
16\pi^2\frac{d}{dt} B   & = & -2( 3 h_t^2 A_t + 3 h_b^2 A_b + h_\tau^2 A_\tau
                             - g_1^2 M_1 - 3 g_2^2 M_2)
\end{eqnarray}

\section{Scalar and Fermion Mass Matrices}

In this appendix we give all relevant scalar and fermion mass
matrices, including one--loop radiative corrections to the Higgs and
squarks mass matrices. We neglect radiative corrections due to loops
of tau leptons and sleptons since, at low energy, the tau lepton
Yukawa coupling is approximately 3 times smaller than the bottom quark
Yukawa coupling, and, furthermore, loops of leptons and sleptons have
no colour factor associated with them. The corrections to the Higgs
masses have been calculated elsewhere
\cite{MSSMbound,notation,NMSSMbound,ekw}, and are given in the
notation of \cite{notation}.

First we briefly review the use of the one--loop effective potential
for the calculation of radiative corrections to scalar masses. The
one--loop effective potential is given by
Eq. (\ref{onelooppotential}).  The term $V_0$ is the tree--level
scalar potential, and the remaining contribution to the
right--hand--side of Eq. (\ref{onelooppotential}) comes from radiative
corrections to the scalar potential, and we denote it by $\Delta V_1$:
\begin{equation}
\Delta V_1 = \frac{1}{64\pi^2} Str {\cal M}^4 \left(
\log \frac{{\cal M}^2}{Q^2} - \frac{3}{2} \right).
\end{equation}
The supertrace is a trace over the eigenvalues of the field--dependent
mass--squared matrix weighted by spin factors $(2j+1)(-1)^{2j}$ for
particles of spin $j$ going around the loops. Thus, a scalar loop
contributes $+1$, whereas a (2-spinor) fermion loop contributes $-2$.

The tree--level mass--squared matrix $M^2$ of general scalar particles
$\phi_i$ is given by
\begin{equation}
M^2_{ij} = \frac{\partial^2 V_0}{\partial \phi_i \partial \phi_j} |_{VEVs},
\end{equation}
and the one--loop corrections $\delta M^2$ are given by
\begin{equation}
\delta M^2_{ij} = \frac{\partial^2 \Delta V_1}
       {\partial \phi_i \partial \phi_j} |_{VEVs}.
\label{notquiteright}
\end{equation}
Naturally, we must insert the correct VEVs: for one--loop calculations
we must minimise the full one--loop effective potential in order to
extract the VEVs to one--loop. Actually, Eq. (\ref{notquiteright})
neglects scalar self--energy contributions, but these are expected to
be small for the lighter scalar states.

\subsection{Higgs Mass Matrices}

The CP--even mass--squared matrix in the NMSSM, in the basis
$ \{ H_1, H_2, N \} $ is given by $M^2+\delta M^2$, where $M^2$ and
$\delta M^2$ are given by
\begin{eqnarray}
M^2 & = & \left( \begin{array}{ccc}
          M_{Z}^2 \cos^2 \beta
        & (\lambda^2 \nu^2 - \frac{1}{2}M_{Z}^2) \sin 2\beta
        & 2 \lambda^2 x \nu \cos \beta \\
          (\lambda^2 \nu^2 -\frac{1}{2} M_{Z}^2) \sin 2\beta
        & M_{Z}^2 \sin^2 \beta
        & 2 \lambda^2 x \nu \sin \beta \\
          2 \lambda^2 x \nu \cos \beta
        & 2 \lambda^2 x \nu \sin \beta
        & k x (4 k x - A_k)
\end{array} \right) \nonumber \\
 & + & \lambda x \left( \begin{array}{ccc}
       \tan \beta [A_\lambda+kx]
     & - [A_\lambda+kx]
     & - \frac{\nu_2}{x} [A_\lambda+2kx] \\
       - [A_\lambda+kx]
     & \cot \beta [A_\lambda+kx]
     & - \frac{\nu_1}{x} [A_\lambda+2kx] \\
       - \frac{\nu_2}{x} [A_\lambda+2kx]
     & - \frac{\nu_1}{x} [A_\lambda+2kx]
     & \frac{\nu_1 \nu_2}{x^2} [A_\lambda]
\end{array} \right)
\end{eqnarray}
and
\begin{equation}
\delta M^2 =
\left( \begin{array}{ccc}
  \Delta_{11}^2 & \Delta_{12}^2 & \Delta_{13}^2 \\
  \Delta_{12}^2 & \Delta_{22}^2 & \Delta_{23}^2 \\
  \Delta_{13}^2 & \Delta_{23}^2 & \Delta_{33}^2
\end{array} \right)
+
\left( \begin{array}{ccc}
   \tan \beta    &       -1         &   -\frac{\nu_2}{x} \\
      -1         &   \cot \beta     &   -\frac{\nu_1}{x} \\
-\frac{\nu_2}{x} & -\frac{\nu_1}{x} & \frac{\nu_1 \nu_2}{x^2}
\end{array} \right) \Delta^2,
\end{equation}
where $\Delta^2$ and the $\Delta_{ij}^2$ are given by
\begin{eqnarray}
\Delta^2      & = &
\frac{3}{16 \pi^2}  (\lambda x)
    \left(  h_t^2 A_t f(m_{\tilde{t_1}}^2,m_{\tilde{t_2}}^2)
          + h_b^2 A_b f(m_{\tilde{b_1}}^2,m_{\tilde{b_2}}^2) \right)    \\
\Delta_{11}^2 & = &
\frac{3}{8 \pi^2} h_t^4 \nu_2^2 (\lambda x)^2
      \left(
      \frac{-A_t+\lambda x \cot \beta}
           {m_{\tilde{t_2}}^2 - m_{\tilde{t_1}}^2}
      \right)^2
      g(m_{\tilde{t_1}}^2,m_{\tilde{t_2}}^2) +
\frac{3}{8 \pi^2} h_b^4 \nu_1^2
      \log \frac{m_{\tilde{b_1}}^2 m_{\tilde{b_2}}^2}{m_b^4}  \nonumber \\
              & - &
\frac{3}{8 \pi^2} h_b^4 \nu_1^2
      \frac{A_b(-A_b+\lambda x\tan\beta)}
           {m_{\tilde{b_2}}^2 - m_{\tilde{b_1}}^2}
      \left(
      2 \log \frac{m_{\tilde{b_2}}^2}{m_{\tilde{b_1}}^2 } -
      \frac{A_b(-A_b+\lambda x\tan\beta)}
           {m_{\tilde{b_2}}^2 - m_{\tilde{b_1}}^2}
      g(m_{\tilde{b_1}}^2,m_{\tilde{b_2}}^2) \right)          \nonumber \\
              &   &                                                     \\
\Delta_{12}^2 & = &
\frac{3}{8 \pi^2} h_t^4 \nu_2^2 (\lambda x)
      \left(
      \frac{-A_t+\lambda x \cot \beta}
      {m_{\tilde{t_2}}^2 - m_{\tilde{t_1}}^2}
      \right)
      \left(
      \log \frac{m_{\tilde{t_2}}^2}{m_{\tilde{t_1}}^2 } -
      \frac{A_t(-A_t+\lambda x\cot\beta)}
           {m_{\tilde{t_2}}^2 - m_{\tilde{t_1}}^2}
      g(m_{\tilde{t_1}}^2,m_{\tilde{t_2}}^2)
      \right)                                                 \nonumber \\
              & + &
\frac{3}{8 \pi^2} h_b^4 \nu_1^2 (\lambda x)
      \left(
      \frac{-A_b+\lambda x\tan\beta}
           {m_{\tilde{b_2}}^2 - m_{\tilde{b_1}}^2}
      \right)
      \left(
      \log \frac{m_{\tilde{b_2}}^2}{m_{\tilde{b_1}}^2 } -
      \frac{A_b(-A_b+\lambda x\tan\beta)}
           {m_{\tilde{b_2}}^2 - m_{\tilde{b_1}}^2}
      g(m_{\tilde{b_1}}^2,m_{\tilde{b_2}}^2)
      \right)                                                 \nonumber \\
              &   &                                                     \\
\Delta_{13}^2 & = &
\frac{3}{8 \pi^2} h_t^4 \nu_2^2 (\lambda x)(\lambda \nu_1)
      \left(
      \frac{-A_t+\lambda x \cot \beta}
           {m_{\tilde{t_2}}^2 - m_{\tilde{t_1}}^2}
      \right)^2
      g(m_{\tilde{t_1}}^2,m_{\tilde{t_2}}^2) +
\frac{3}{8 \pi^2} h_t^2 (\lambda x)(\lambda \nu_1)
      f(m_{\tilde{t_1}}^2,m_{\tilde{t_2}}^2)                 \nonumber \\
              & + &
\frac{3}{8 \pi^2} h_b^4 \nu_1^2 (\lambda \nu_2)
      \left(
      \frac{-A_b+\lambda x \tan \beta}
           {m_{\tilde{b_2}}^2 - m_{\tilde{b_1}}^2}
      \right)
      \left(
      \log \frac{m_{\tilde{b_2}}^2}{m_{\tilde{b_1}}^2} -
      \frac{A_b(-A_b+\lambda x \tan \beta)}
           {m_{\tilde{b_2}}^2 - m_{\tilde{b_1}}^2}
      g(m_{\tilde{b_1}}^2,m_{\tilde{b_2}}^2)
      \right)                                                \nonumber \\
              &   &                                                    \\
\Delta_{22}^2 & = &
\frac{3}{8 \pi^2} h_b^4 \nu_1^2 (\lambda x)^2
      \left(
      \frac{-A_b+\lambda x \tan \beta}
           {m_{\tilde{b_2}}^2 - m_{\tilde{b_1}}^2}
      \right)^2
      g(m_{\tilde{b_1}}^2,m_{\tilde{b_2}}^2) +
\frac{3}{8 \pi^2} h_t^4 \nu_2^2
      \log \frac{m_{\tilde{t_1}}^2 m_{\tilde{t_2}}^2}{m_t^4} \nonumber \\
              & - &
\frac{3}{8 \pi^2} h_t^4 \nu_2^2
      \frac{A_t(-A_t+\lambda x \cot \beta)}
           {m_{\tilde{t_2}}^2 - m_{\tilde{t_1}}^2}
      \left(
      2 \log \frac{m_{\tilde{t_2}}^2}{m_{\tilde{t_1}}^2 } -
      \frac{A_t(-A_t+\lambda x \cot \beta)}
           {m_{\tilde{t_2}}^2 - m_{\tilde{t_1}}^2}
      g(m_{\tilde{t_1}}^2,m_{\tilde{t_2}}^2)
      \right)                                                \nonumber \\
              &   &                                                    \\
\Delta_{23}^2 & = &
\frac{3}{8 \pi^2} h_b^4 \nu_1^2 (\lambda x)(\lambda \nu_2)
      \left(
      \frac{-A_b+\lambda x \tan \beta}
           {m_{\tilde{b_2}}^2 - m_{\tilde{b_1}}^2}
      \right)^2
      g(m_{\tilde{b_1}}^2,m_{\tilde{b_2}}^2) +
\frac{3}{8 \pi^2} h_b^2 (\lambda x)(\lambda \nu_2)
      f(m_{\tilde{b_1}}^2,m_{\tilde{b_2}}^2)                 \nonumber \\
              & + &
\frac{3}{8 \pi^2} h_t^4 \nu_2^2 (\lambda \nu_1)
      \left(
      \frac{-A_t+\lambda x \cot \beta}
           {m_{\tilde{t_2}}^2 - m_{\tilde{t_1}}^2}
      \right)
      \left(
      \log \frac{m_{\tilde{t_2}}^2}{m_{\tilde{t_1}}^2} -
      \frac{A_t(-A_t+\lambda x \cot \beta)}
           {m_{\tilde{t_2}}^2 - m_{\tilde{t_1}}^2}
      g(m_{\tilde{t_1}}^2,m_{\tilde{t_2}}^2)
      \right)                                                \nonumber \\
              &   &                                                    \\
\Delta_{33}^2 & = &
\frac{3}{8 \pi^2} h_t^4 \nu_2^2 (\lambda\nu_1)^2
      \left(
      \frac{-A_t+\lambda x \cot \beta}
           {m_{\tilde{t_2}}^2 - m_{\tilde{t_1}}^2}
      \right)^2
      g(m_{\tilde{t_1}}^2,m_{\tilde{t_2}}^2)                 \nonumber \\
              & + &
\frac{3}{8 \pi^2} h_b^4 \nu_1^2 (\lambda\nu_2)^2
      \left(
      \frac{-A_b+\lambda x\tan\beta}
           {m_{\tilde{b_2}}^2 - m_{\tilde{b_1}}^2}
      \right)^2
      g(m_{\tilde{b_1}}^2,m_{\tilde{b_2}}^2)
\end{eqnarray}
and the functions $f$ and $g$ are defined by
\begin{eqnarray}
f(m_1^2,m_2^2) & = &
      \frac{1}{m_1^2-m_2^2}
      \left(  m_1^2 \log \frac{m_1^2}{Q^2}
            - m_2^2 \log \frac{m_2^2}{Q^2}
            - m_1^2 + m_2^2 \right) \\
g(m_1^2,m_2^2) & = &
      \frac{1}{m_1^2-m_2^2}
      \left( (m_1^2+m_2^2) \log \frac{m_2^2}{m_1^2}
            + 2 (m_1^2 - m_2^2) \right)
\end{eqnarray}
In these expressions, $m_{\tilde{t_1}}$, $m_{\tilde{t_2}}$,
$m_{\tilde{b_1}}$ and $m_{\tilde{b_2}}$ denote the masses of the
corresponding top and bottom squark mass eigenstates.

The CP--odd mass--squared matrix, in the basis $\{ H_1,H_2,N \}$ is
given by $\tilde{M}^2 + \delta \tilde{M}^2$, where $\tilde{M}^2$ and
$\delta \tilde{M}^2$ are given by
\begin{equation}
\tilde{M}^2 = \lambda x \left( \begin{array}{ccc}
      \tan \beta [A_\lambda+kx]
    & [A_\lambda+kx]
    & \frac{\nu_2}{x} [A_\lambda-2kx] \\
      \mbox{} [A_\lambda+kx]
    & \cot \beta [A_\lambda+kx]
    & \frac{\nu_1}{x} [A_\lambda-2kx] \\
      \frac{\nu_2}{x} [A_\lambda-2kx]
    & \frac{\nu_1}{x} [A_\lambda-2kx]
    & 3\frac{k A_k}{\lambda}+\frac{\nu_1 \nu_2}{x^2} [A_\lambda+4kx]
\end{array} \right)
\end{equation}
and
\begin{equation}
\delta \tilde{M}^2 =
\left( \begin{array}{ccc}
   \tan \beta   &       1         &   \frac{\nu_2}{x} \\
       1        &   \cot \beta    &   \frac{\nu_1}{x} \\
\frac{\nu_2}{x} & \frac{\nu_1}{x} & \frac{\nu_1 \nu_2}{x^2}
\end{array} \right) \Delta^2
\end{equation}

The charged Higgs mass--squared matrix, in the basis $\{ H_1,H_2 \}$,
is given by $M_c^2 + \delta M_c^2$, where $M_c^2$ and $\delta M_c^2$
are given by
\begin{equation}
M_c^2= \left( \begin{array}{cc}
                      \tan \beta &      1       \\
                           1     & \cot \beta
              \end{array} \right)
  [\lambda x (A_\lambda+kx)
         -(\lambda^2\nu^2-M_W^2)\frac{1}{2}\sin 2\beta],
\end{equation}
and
\begin{equation}
\delta M_c^2 = \left( \begin{array}{cc}
                       \tan \beta &      1     \\
                            1     & \cot \beta
                      \end{array} \right)
                         (\Delta_{c_1}^2 + \Delta_{c_2}^2),
\end{equation}
where $\Delta_{c_1}^2$ is the correction due to loops of top and
bottom quarks and is given by
\begin{equation}
\Delta_{c_1}^2 =
-\frac{3}{8 \pi^2} h_t h_b \frac{m_t m_b}{m_t^2 - m_b^2}
\left[  m_t^2 \left( \log \frac{m_t^2}{Q^2} -1 \right)
      - m_b^2 \left( \log \frac{m_b^2}{Q^2} -1 \right) \right],
\end{equation}
and $\Delta_{c_2}^2$ is the correction due to loops of top and bottom
squarks and is given by
\begin{equation}
\Delta_{c_2}^2 = \frac{3}{16 \pi^2}
   \sum_{i \in \{ \tilde{t_1}, \tilde{t_2}, \tilde{b_1}, \tilde{b_2} \} }
   m_i^2 \left( \log \frac{m_i^2}{Q^2} -1 \right)
   \frac{\partial^2 m_i^2}{\partial H_1^- \partial H_2^+ } |_{vevs},
\end{equation}
where
\begin{equation}
\frac{\partial^2 m_a^2}{\partial H_1^- \partial H_2^+ }|_{vevs} =
-\frac{B m_a^4 + C m_a^2 + D}{\Delta_a} |_{vevs},
 \end{equation}
and
\begin{equation}
\Delta_a = \prod_{ a' \ne a} (m_a^2 - m_{a'}^2).
\end{equation}
The coefficients $B$, $C$ and $D$ are messy functions of the various
Yukawa couplings and soft masses which we do not reproduce here. It is
possible to reduce these expressions to an elegant form in certain
limits, such as $h_b \rightarrow 0$ (see \cite{notation}).

This completes the Higgs mass matrices, together with radiative
corrections, in the NMSSM. To find the corresponding Higgs masses in
the MSSM it is sufficent to perform the following operations. First,
strike out elements of the matrices relating to the singlet field
$N$. Second, replace every occurrence of the term $(A_\lambda + kx)$ by
$-B$. Third, replace every occurrence of the term $\lambda x$ by
$\mu$. Finally, any remaining dependence on $\lambda$ is removed by
setting $\lambda$ to zero. These rules also apply to the matrices which
follow.

\subsection{Squark Mass Matrices}

Squark contributions to radiative corrections to the squark
mass--squared matrices do not appear to have been considered in the
literature before. Here we present the results. The calculations are
straightforward, so we do not go into the details. However, there is
one subtlety in the calculation which we do mention.

If we wish to calculate radiative corrections to the squark
mass--squared matrix due to loops of squarks, then we must calculate
the field--dependent mass--squared matrix, in which we retain
explicitly any dependence in the matrix on the squark fields. Since we
shall take derivatives of this matrix with respect to the squark
fields only, we may set all other fields to their VEVs. This means
that the $4 \times 4$ mass--squared matrix decomposes into two $2
\times 2$ matrices when the charged Higgs fields are set to their
VEVs, namely zero. In general, the field--dependent squark
mass--squared matrix, ${\cal M}^2$, has a non--trivial colour
structure. If $\alpha$ and $\beta$ are $SU(3)$ colour indices, then
${\cal M}_{\alpha}^{2\beta}$, where the colour indices are
now explicit, has the form
\begin{equation}
{\cal M}_{\alpha}^{2\beta} = {\cal A}^2 \delta_\alpha^\beta
+ {\cal B}_{\alpha}^{2\beta}.
\end{equation}
This means that the two $2 \times 2$ matrices become two $6 \times 6$
matrices when the colour structure is considered. On the face of it,
this implies that analytic calculation of the eigenvalues is
impossible, so that we may not analytically determine the radiative
corrections due to loops of squarks.

We may circumvent this problem in the following way. Since the matrix
${\cal B}^2$ transforms as $3 \otimes \bar{3}$ under $SU(3)_c$,
and $3 \otimes \bar{3} = 1 \oplus 8$, we may extract the singlet part
of ${\cal B}^2$ by pulling out the trace. Thus ${\cal M}^2$
may be written as
\begin{equation}
{\cal M}^2 = {\cal M}^2_{singlet} + {\cal M}^2_{octet}
\end{equation}
where
\begin{eqnarray}
{\cal M}^2_{singlet} & = &
({\cal A}^2 + \frac{1}{3}{\cal B}_{\gamma}^{2\gamma})
\delta_\alpha^\beta, \\
{\cal M}^2_{octet} & = &
{\cal B}_{\alpha}^{2\beta} - \frac{1}{3}{\cal B}_{\gamma}^{2\gamma}
\delta_\alpha^\beta.
\end{eqnarray}
The octet part is a pure bilinear function of the squark fields. We
know, since colour is to be left unbroken, that the radiative
corrections must turn out to be have trivial colour structure. Can the
octet contribute in any way? The only way to obtain a singlet is from
the singlet part of $8 \otimes 8$. However, such a term would be
quadrilinear in squark fields, and we need only take at most two
derivatives to obtain the radiative corrections. At the end of the
calculation, therefore, when the squark fields are set to their VEVs,
such a contribution will vanish. Hence we may discard the octet part
and retain only the singlet part in the calculations. We then regain
two $2 \times 2$ matrices with trivial colour structure, and now the
calculations may proceed in the usual fashion.

The top squark and bottom squark mass--squared matrices, in the basis
of gauge eigenstates $\{ \tilde{t}, \bar{\tilde{t^c}} \}$ and $\{
\tilde{b}, \bar{\tilde{b^c}} \}$, are given by $M_{stop}^2 + \delta
M_{stop}^2$ and $M_{sbot}^2 + \delta M_{sbot}^2$, respectively, where
\begin{eqnarray}
M_{stop}^2 & = &
     \left( \begin{array}{cc}
     m_Q^2 + h_t^2 \nu_2^2 & h_t (-A_t \nu_2 + \lambda x \nu_1) \\
     h_t (-A_t \nu_2 + \lambda x \nu_1) & m_T^2 + h_t^2 \nu_2^2
     \end{array} \right)   \\
M_{sbot}^2 & = &
     \left( \begin{array}{cc}
     m_Q^2 + h_b^2 \nu_1^2 & -h_b (-A_b \nu_1 + \lambda x \nu_2) \\
     -h_b (-A_b \nu_1 + \lambda x \nu_2) & m_B^2 + h_b^2 \nu_1^2
     \end{array} \right)   \\
\delta M_{stop/sbot}^2 & = &
     \frac{3}{16\pi^2} \sum_{a \in \{ \tilde{t_1}, \tilde{t_2},
                                      \tilde{b_1}, \tilde{b_2} \} }
     m_a^2 \left( \log \frac{m_a^2}{Q^2} -1 \right) M^{a}_{stop/sbot}
\end{eqnarray}
and the $M^{a}_{stop/sbot}$ are given by
\begin{eqnarray}
M^{\tilde{t_1},\tilde{t_2}}_{stop} & = &
 \frac{1}{6}
   \left( \begin{array}{cc} h_t^2 & 0 \\ 0 & h_t^2 \end{array} \right)
 \pm
 \frac{1}{6} \frac{m_Q^2-m_T^2}{m_{\tilde{t_1}}^2 - m_{\tilde{t_2}}^2}
   \left( \begin{array}{cc} h_t^2 & 0 \\ 0 & -h_t^2 \end{array} \right)
                                                     \nonumber \\
 & \mp & \frac{h_t (-A_t \nu_2 + \lambda x \nu_1)}
      {m_{\tilde{t_1}}^2 - m_{\tilde{t_2}}^2}
   \left( \begin{array}{cc} 0 & h_t^2 \\ h_t^2 & 0 \end{array} \right) \\
M^{\tilde{b_1},\tilde{b_2}}_{stop} & = &
 \frac{1}{6}
   \left( \begin{array}{cc} h_b^2 & 0 \\ 0 & h_t^2 \end{array} \right)
 \pm
 \frac{1}{6} \frac{m_Q^2-m_B^2}{m_{\tilde{b_1}}^2-m_{\tilde{b_2}}^2}
   \left( \begin{array}{cc} h_b^2 & 0 \\ 0 & -h_t^2 \end{array} \right)
          \\
M^{\tilde{t_1},\tilde{t_2}}_{sbot} & = &
 \frac{1}{6}
   \left( \begin{array}{cc} h_t^2 & 0 \\ 0 & h_b^2 \end{array} \right)
 \pm
 \frac{1}{6} \frac{m_Q^2-m_T^2}{m_{\tilde{t_1}}^2-m_{\tilde{t_2}}^2}
      \left( \begin{array}{cc} h_t^2 & 0 \\ 0 & -h_b^2 \end{array} \right)
          \\
M^{\tilde{b_1},\tilde{b_2}}_{sbot} & = &
 \frac{1}{6}
   \left( \begin{array}{cc} h_b^2 & 0 \\ 0 & h_b^2 \end{array} \right)
 \pm
 \frac{1}{6} \frac{m_Q^2-m_B^2}{m_{\tilde{b_1}}^2 - m_{\tilde{b_2}}^2}
   \left( \begin{array}{cc} h_b^2 & 0 \\ 0 & -h_b^2 \end{array} \right)
                                               \nonumber \\
 & \pm & \frac{h_b (-A_b \nu_1 + \lambda x \nu_2)}
      {m_{\tilde{b_1}}^2 - m_{\tilde{b_2}}^2}
  \left( \begin{array}{cc} 0 & h_b^2 \\ h_b^2 & 0 \end{array} \right)
\end{eqnarray}

\subsection{Fermionic Mass Matrices}

We now turn to the fermionic sector. The neutralino mass matrix, in
the basis of 2--spinors given by $\{ i\tilde{B^0}, i\tilde{W_3^0},
\tilde{H_1^0}, \tilde{H_2^0}, \tilde{N} \}$ is
\begin{equation}
\left( \begin{array}{ccccc}
-M_1 & 0 & -{\frac{1}{\sqrt{2}}} g_1 \nu_1 & {\frac{1}{\sqrt{2}}} g_1
\nu_2 & 0 \\ 0 & -M_2 & {\frac{1}{\sqrt{2}}} g_2 \nu_1 &
-\frac{1}{\sqrt{2}} g_2 \nu_2 & 0 \\ -\frac{1}{\sqrt{2}} g_1 \nu_1 &
\frac{1}{\sqrt{2}} g_2 \nu_1 & 0 & -\lambda x & -\lambda \nu_2 \\
\frac{1}{\sqrt{2}} g_1 \nu_2 & -\frac{1}{\sqrt{2}} g_2 \nu_2 & -\lambda
x & 0 & -\lambda \nu_1 \\ 0 & 0 & -\lambda \nu_2 & -\lambda \nu_1 & -2
k x
\end{array} \right)
\end{equation}
The mass of the gluino, which does not mix with the other neutralinos
because colour is unbroken, is simply given by $M_3$.

The chargino mass matrix is given by
\begin{equation}
\left( \begin{array}{cc}
M_2 &  g_2 \nu_1 \\
-g_2 \nu_2 & \lambda x
\end{array} \right)
\end{equation}

\newpage

\section{Figure Captions}

\noindent
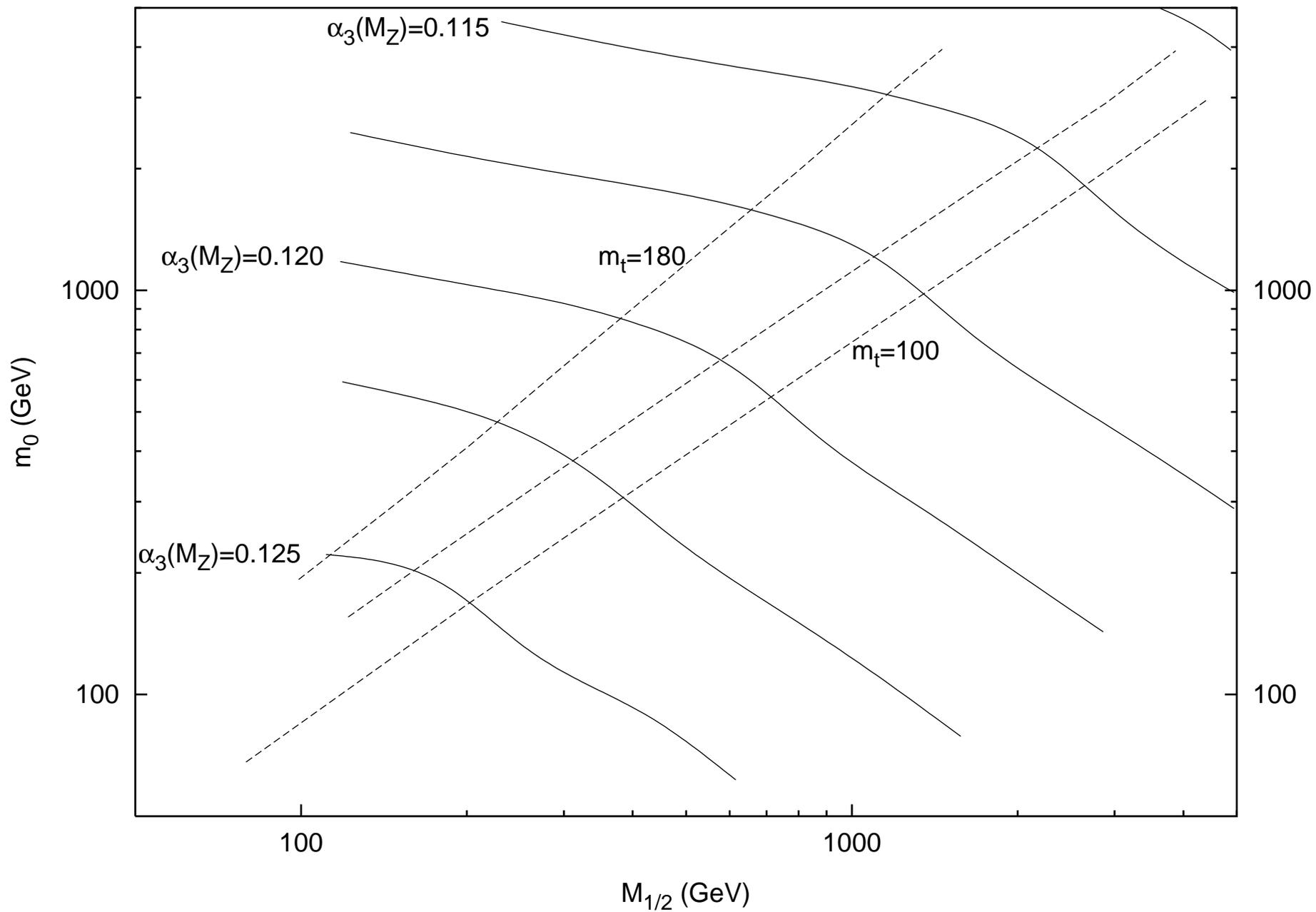
{\bf Figure 1 :}
Contours of $\alpha_3(M_Z)$ and $m_t$ in the $M_{1/2}-m_0$ plane in the
MSSM with $A_0=0$, $\tilde B_0=-1$, $\mu_0=m_0$, and several different
values of $m_0$. Ranging over $h_{t0}$ and $\tilde m_0$ allows us to
cover much of the plane, but note the regions where this is not
possible (because with our choice of $\tilde \mu_0$, $\tilde B_0$  and
$\tilde A_0$ correct electroweak symmetry breaking is not guaranteed to
occur everywhere). Contours are for $\alpha_3(M_Z)$=0.1125, 0.115,
0.1175, 0.120, 0.1225, 0.125 (solid lines) and for $m_t=$100, 140,
180GeV (dashed).

\noindent
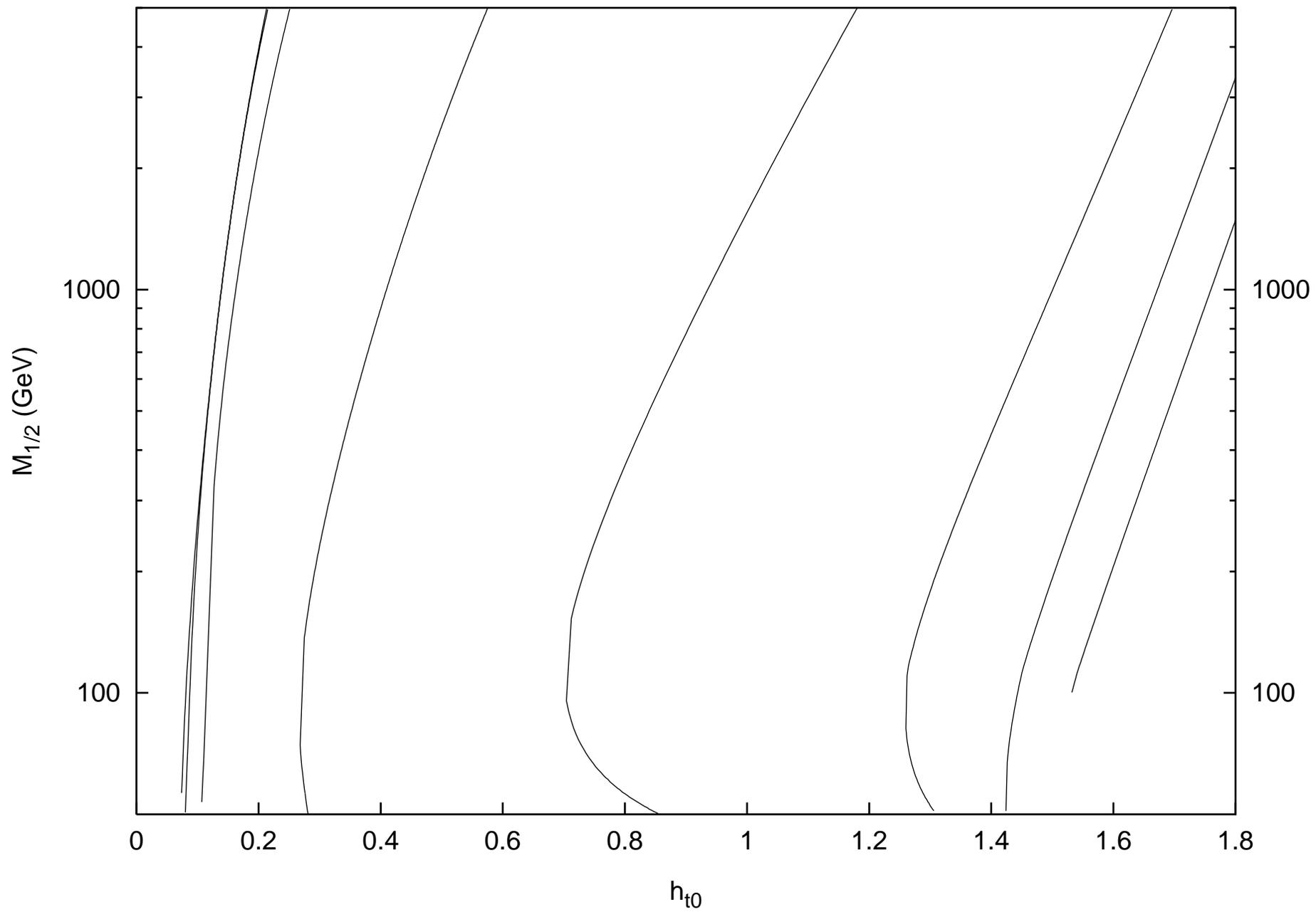
{\bf Figure 2 :}
$M_{1/2}$ against $h_{t0}$ in the MSSM, for values of the other
parameters $A=0$, $\tilde B=-1$, $\mu=m_0$, and
$\tilde m_0$=0.1,0.2,0.5,1,2,5,10,20 (from left to right). This
is typical behaviour in this model.

\noindent
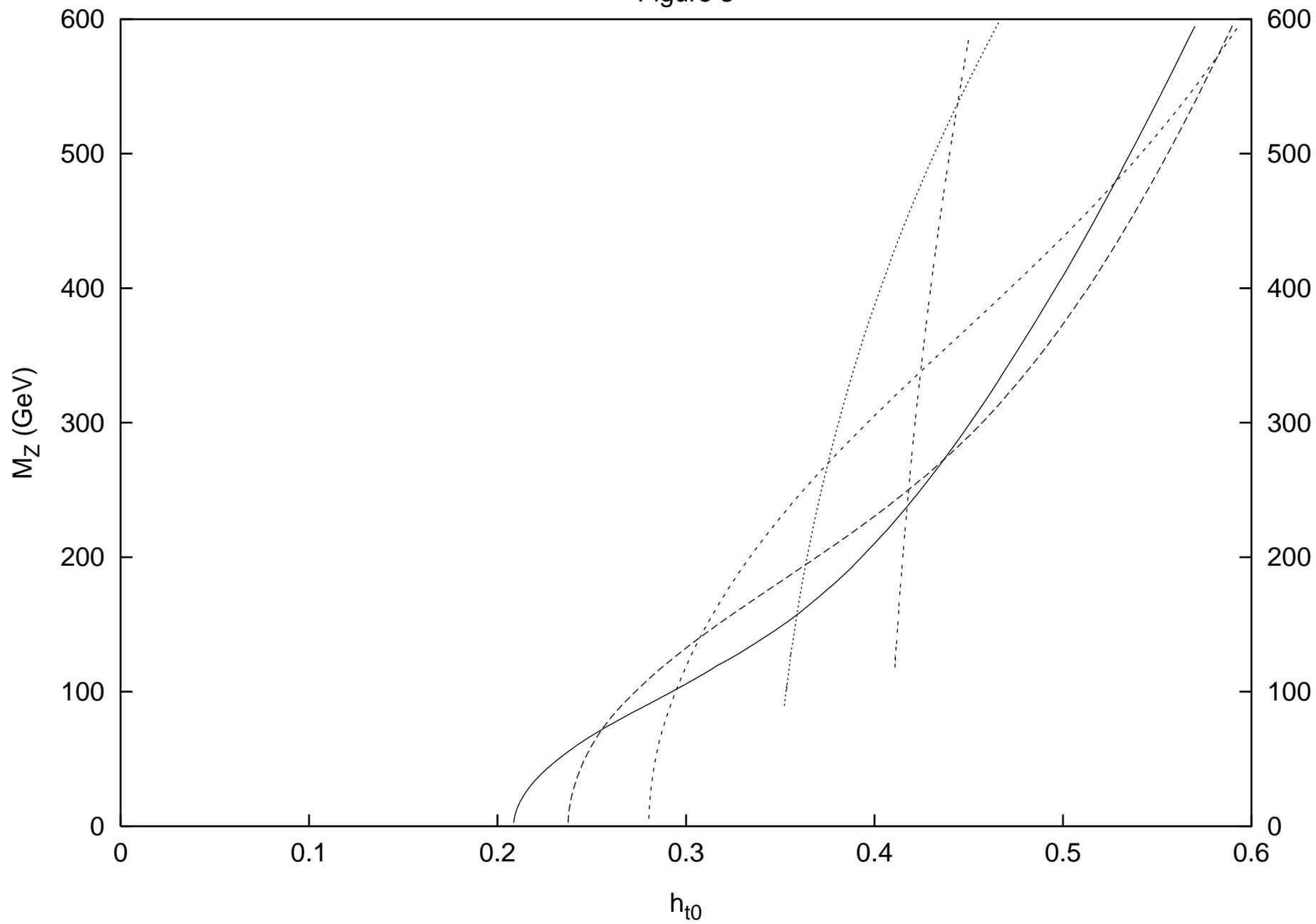
{\bf Figure 3 : }
$M_Z$ against $h_{t0}$  in the MSSM, for different choices of
$M_{1/2}=$ 50 (solid lines), 100 (dashed), 200 (short dashed), 500
(dotted), 1000 (dot-dashed) GeV for the fixed parameters $A_0=0$,
$\tilde\mu_0=1$, $\tilde B_0=-1$, $\tilde m_0=1$. As expected, for
$M_Z=91$ GeV, the curves corresponding to larger $M_{1/2}$ are
significantly steeper that those for smaller $M_{1/2}$, indicating a
higher degree of fine-tuning.

\noindent
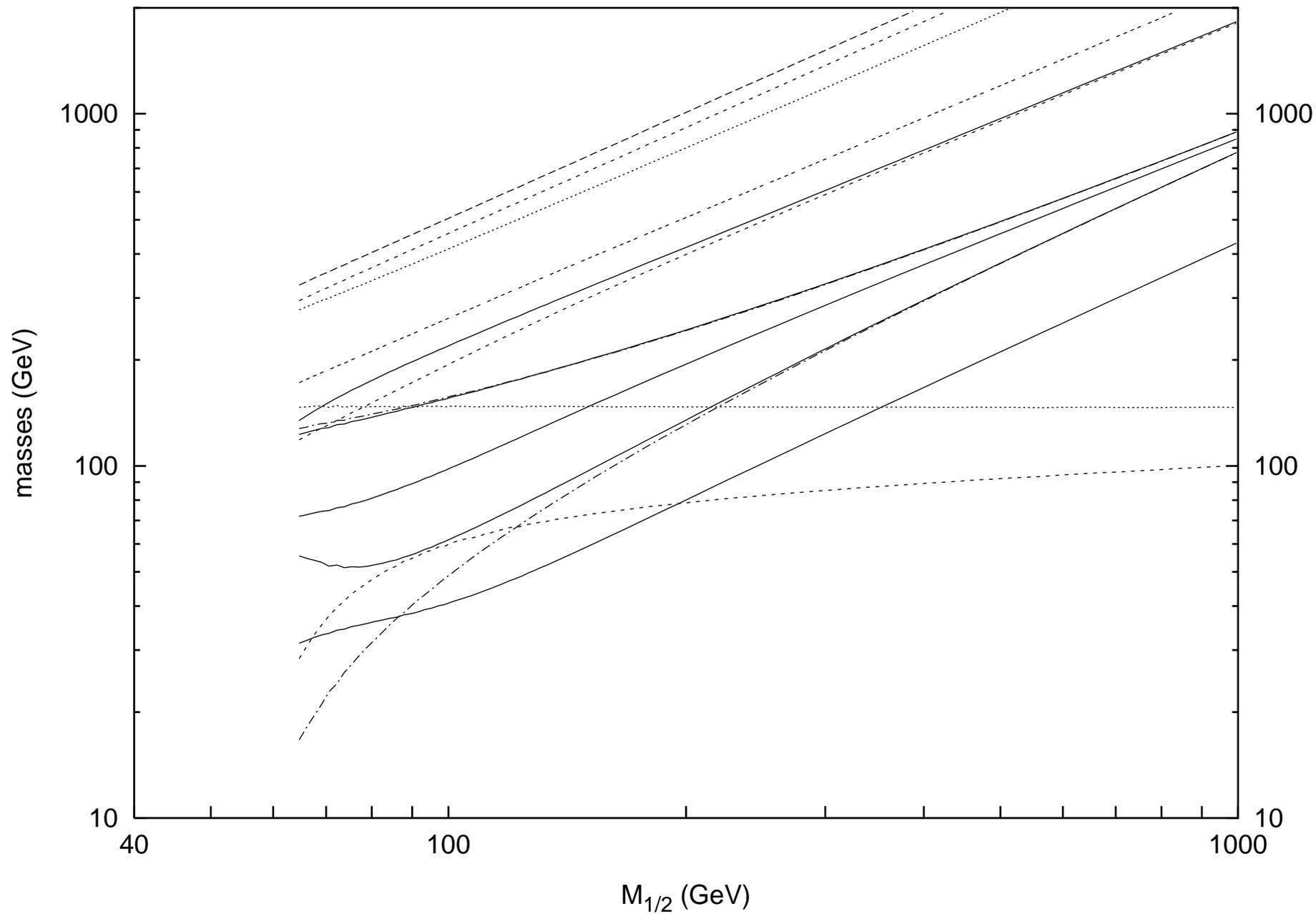
{\bf Figure 4: }
Masses of particles as a function of $M_{1/2}$ for the NMSSM in the
region with $k_0>\lambda_0>h_{t0}$. The input parameters are $A_0=0$,
$\tilde{m}_0=5$, $\lambda_0=0.8$, $k_0=2$, $h_{t0}=0.37-0.47$.
Neutralinos (solid lines), charginos (dot-dashed lines),
CP-even Higgs (short dashed lines), lighter stop and top quark (both
dotted lines), left-handed sleptons (long dashed lines), and gluino
(quadruple dashed lines) are displayed.

\noindent
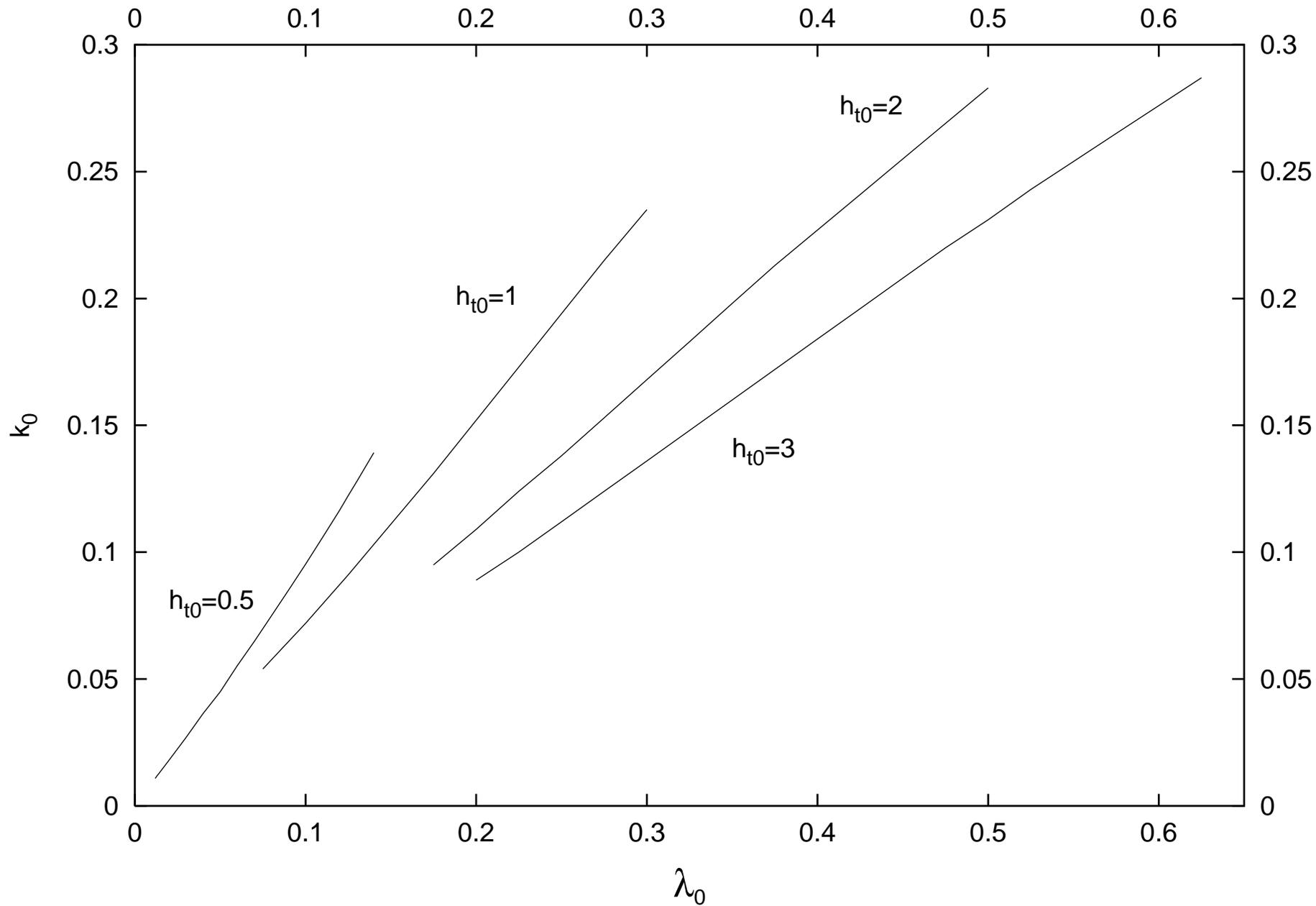
{\bf Figure 5 :}
A contour plot showing allowed values of $\lambda_0$ and $k_0$ for
given values of $h_{t0}=0.5,1,2,3$ in the $\lambda_0$-$k_0$ plane,
corresponding to $M_{1/2}=500$ GeV, $\tilde{m}_0=2$ with $A_0/m_0=-3$.

\noindent
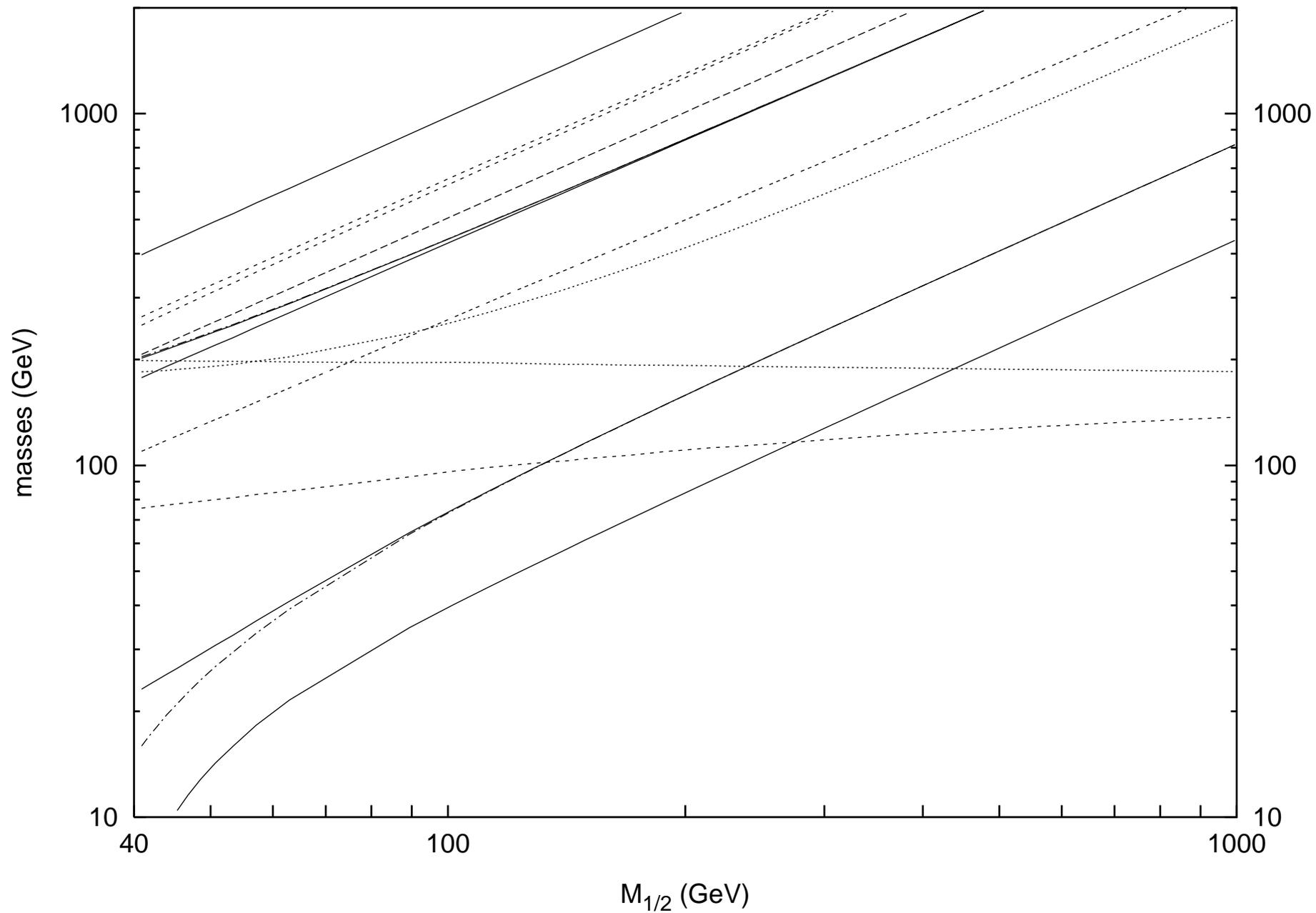
{\bf Figure 6:}
Masses of particles as a function of $M_{1/2}$ for the NMSSM in the
region with $h_{t0}>\lambda_0>k_0$ and for parameters lying in the
``safe'' region where arbitrarily small $M_{1/2}$ can be achieved. The
input parameters are $h_{t0}=2$, $A_0/m_0=-3$, $\tilde{m}_0=5$,
$\lambda_0=0.4$, $k_0=0.275-0.3$. Neutralinos (solid lines), charginos
(dot-dashed lines), CP-even Higgs (short dashed lines), lighter stop
and top quark (both dotted lines), left-handed sleptons (long dashed
lines), and gluino (quadruple dashed lines) are displayed. One of the
CP-odd Higgs and the charged Higgs bosons (not shown) are roughly
degenerate with heavier CP-even Higgs bosons. The remaining CP-odd
Higgs boson (not shown) is heavier than the heaviest neutralino. The
lightest CP-even Higgs (which has standard model-like couplings),
lighter chargino and lightest two neutralinos are all in the LEP2 range
for $M_{1/2}\stackrel{<}{\sim}100$ GeV.

\noindent
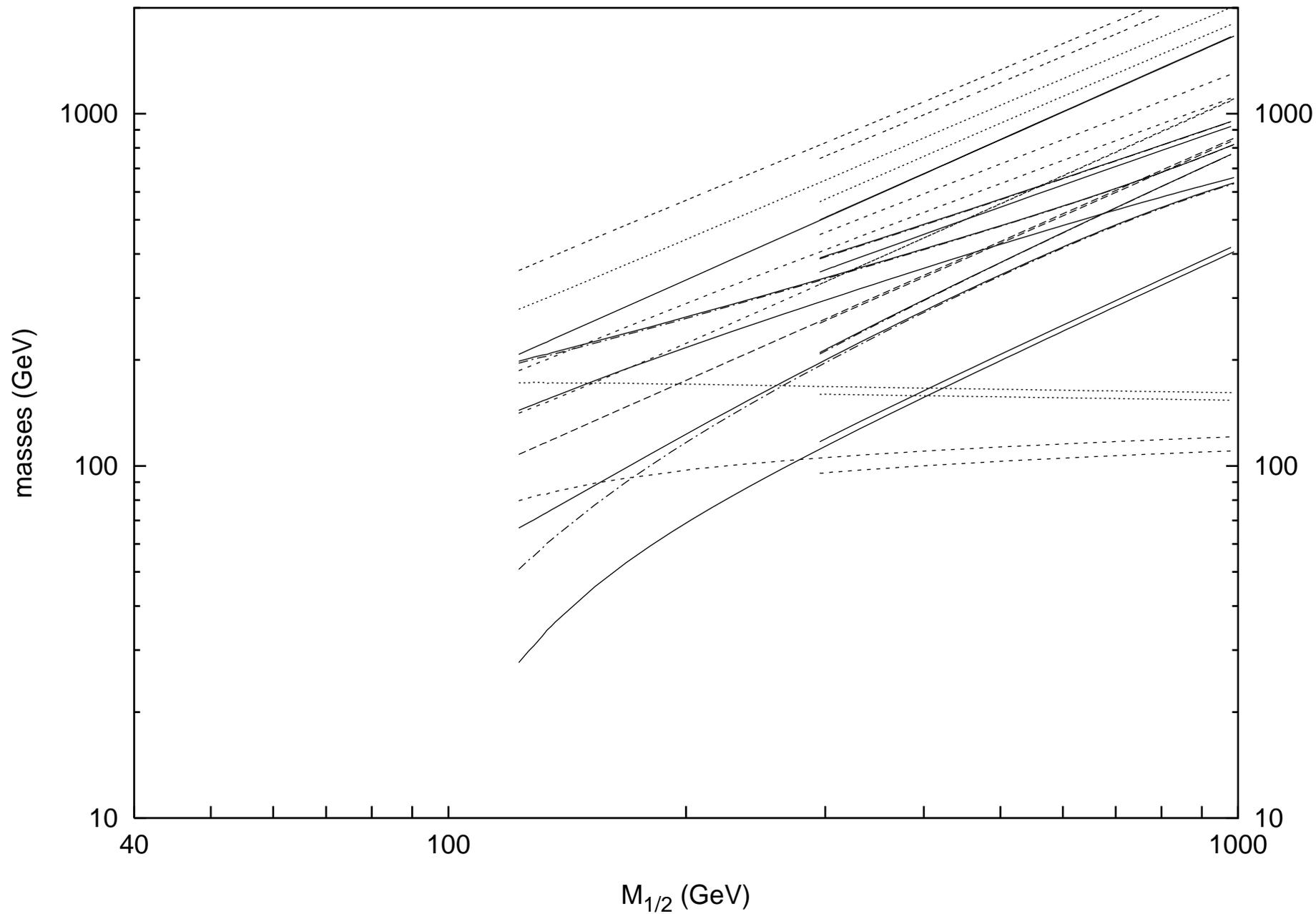
{\bf Figure 7:}
Masses of particles as a function of $M_{1/2}$ for the NMSSM in the
region with $h_{t0}>\lambda_0>k_0$ and for parameters lying outside the
``safe'' region where arbitrarily small $M_{1/2}$ can be achieved. The
input parameters are $h_{t0}=0.5$, $A_0/m_0=-4$, $\tilde{m}_0=0.5$,
$\lambda_0=0.1$, $k_0=0.07-0.13$. In addition to using $Q=150$ GeV as
usual, we also show results for $Q=25$ GeV, corresponding to minimum
values of $M_{1/2}=300$ GeV and $M_{1/2}=125$ GeV, respectively.
Neutralinos (solid lines), charginos (dot-dashed lines), CP-even Higgs
(short dashed lines), lighter stop and top quark (both dotted lines),
left-handed sleptons (long dashed lines), and gluino (quadruple dashed
lines) are displayed. One of the CP-odd Higgs (not shown) is roughly
degenerate with the second heaviest CP-even Higgs boson. The remaining
CP-odd Higgs boson (not shown) is roughly equal to the lighter stop
mass. The lightest CP-even Higgs (which has standard model-like
couplings as in Fig.2) has a mass $\stackrel{<}{\sim}100$ GeV.

\noindent
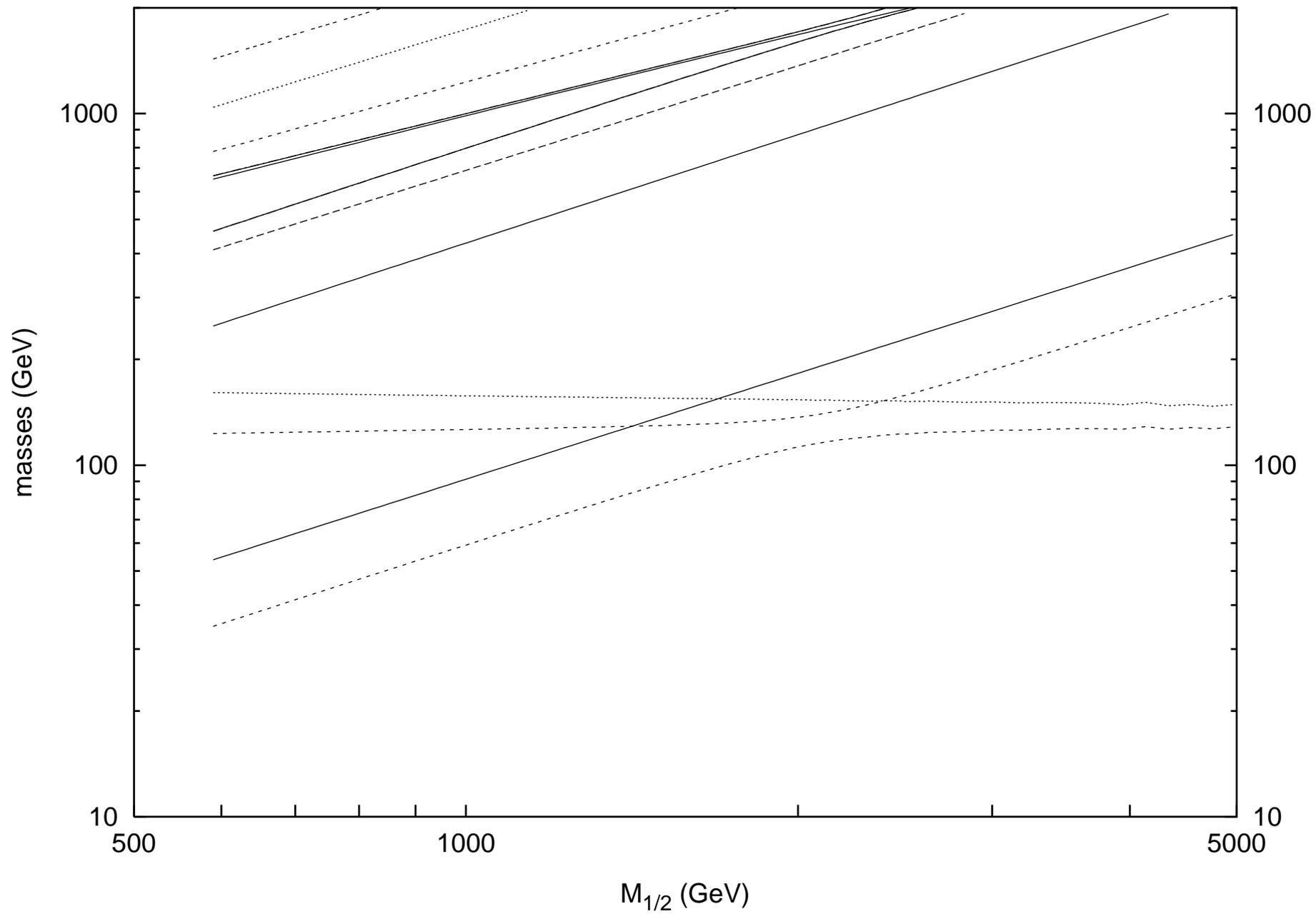
{\bf Figure 8:}
Masses of particles as a function of $M_{1/2}$ for parameters selected
to lie in the deep MSSM region. The input parameters are $h_{t0}=0.5$,
$A_0/m_0=-5$, $\tilde{m}_0=0.02$, $\lambda_0=0.005$,
$k_0=0.0002-0.0004$. Neutralinos (solid lines), charginos (dot-dashed
lines), CP-even Higgs (short dashed lines), lighter stop and top quark
(both dotted lines), left-handed sleptons (long dashed lines), and
gluino (quadruple dashed lines) are displayed.

\noindent
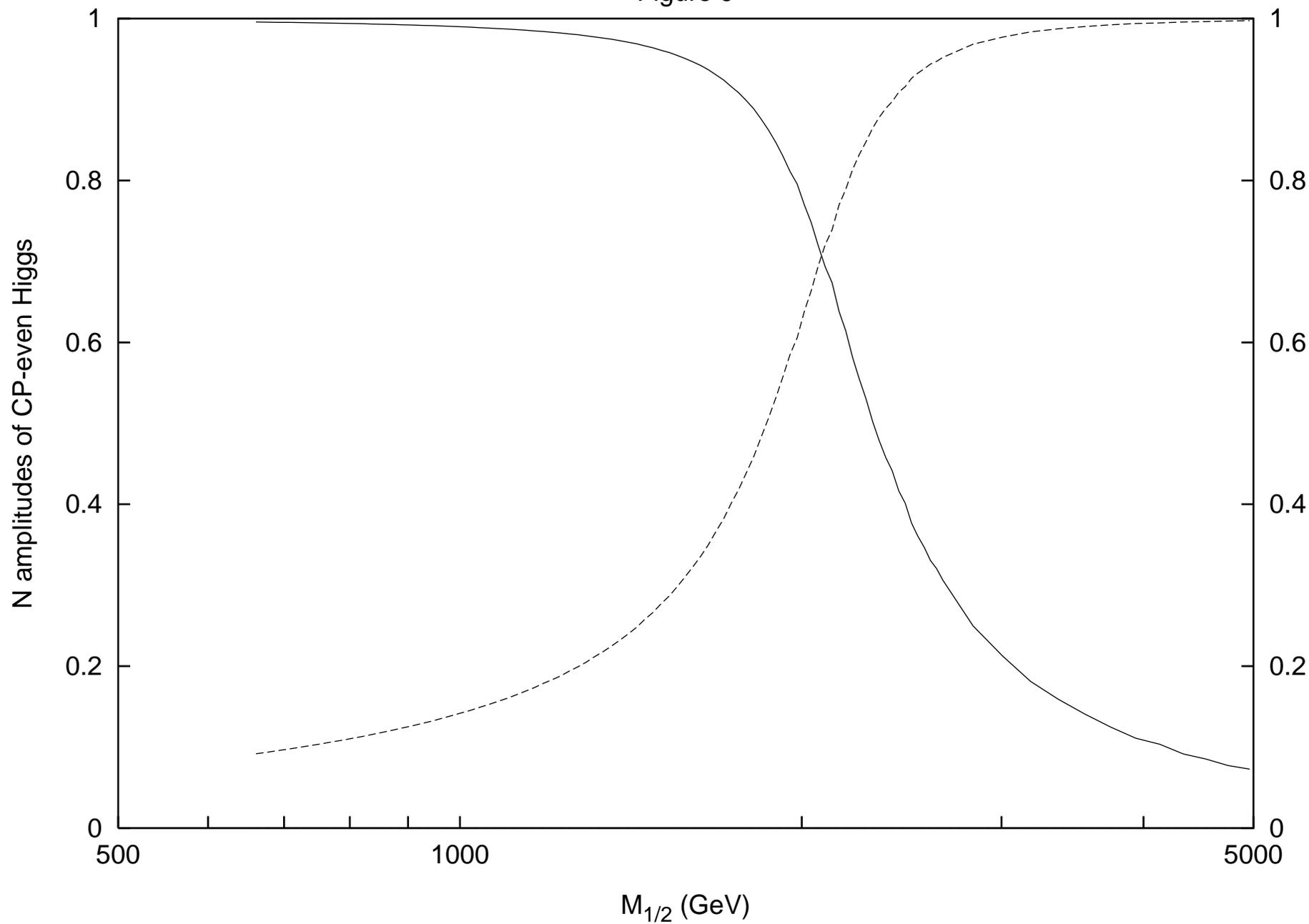
{\bf Figure 9:}
Amplitude of $N$ contained in the lightest (dashes) and second lightest
(dot-dash) CP-even Higgs bosons in Fig.8. The heaviest CP-even Higgs
boson has N-amplitude zero.

\noindent
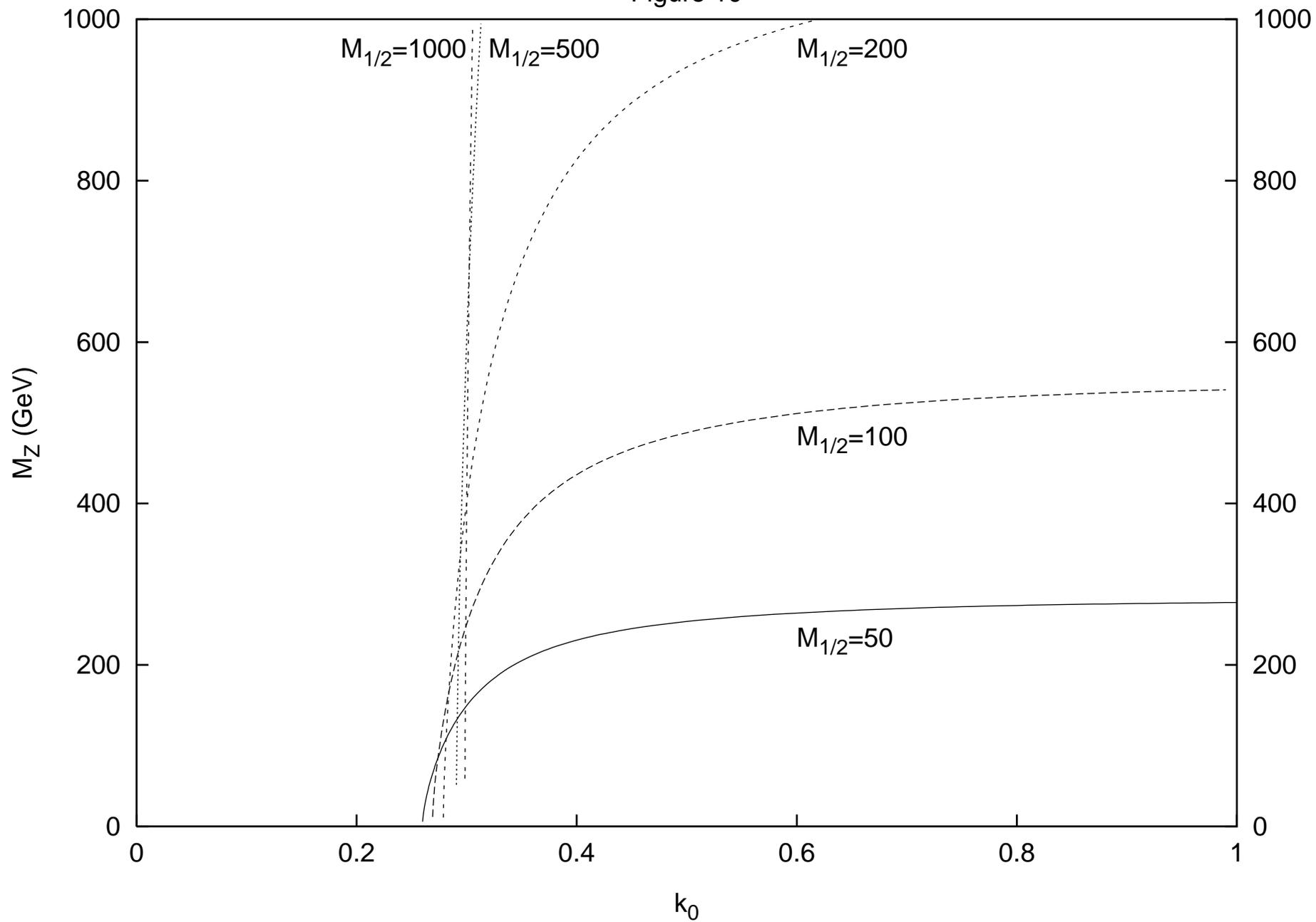
{\bf Figure 10:}
$M_Z$ against $k_0$ for the NMSSM with  $h_{t0}=2$, $A_0/m_0=-3$,
$\tilde m_0=5$, $\lambda_0=0.4$. $M_{1/2}=$ 50, 100, 200, 500, 1000 GeV

\noindent
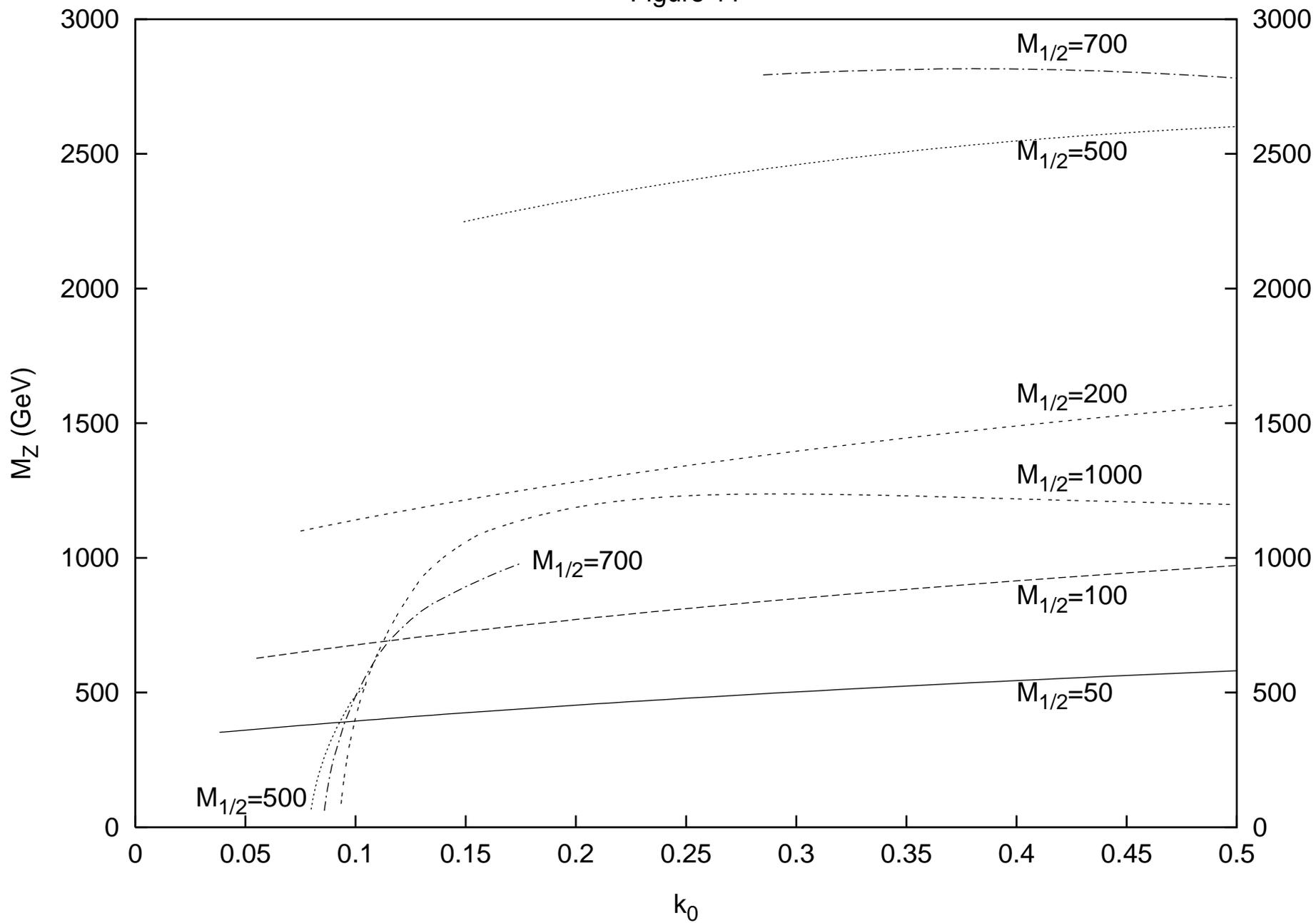
{\bf Figure 11:}
$M_Z$ against $k_0$ for the NMSSM with  $h_{t0}=0.5$, $A_0/m_0=-4$,
$\tilde m_0=0.5$, $\lambda_0=0.1$ $M_{1/2}=$ 50, 100, 200, 500, 700,
1000 GeV. Note the discontinuities in the lines, particularly for
$M_{1/2}=$500 and 700GeV.

\noindent
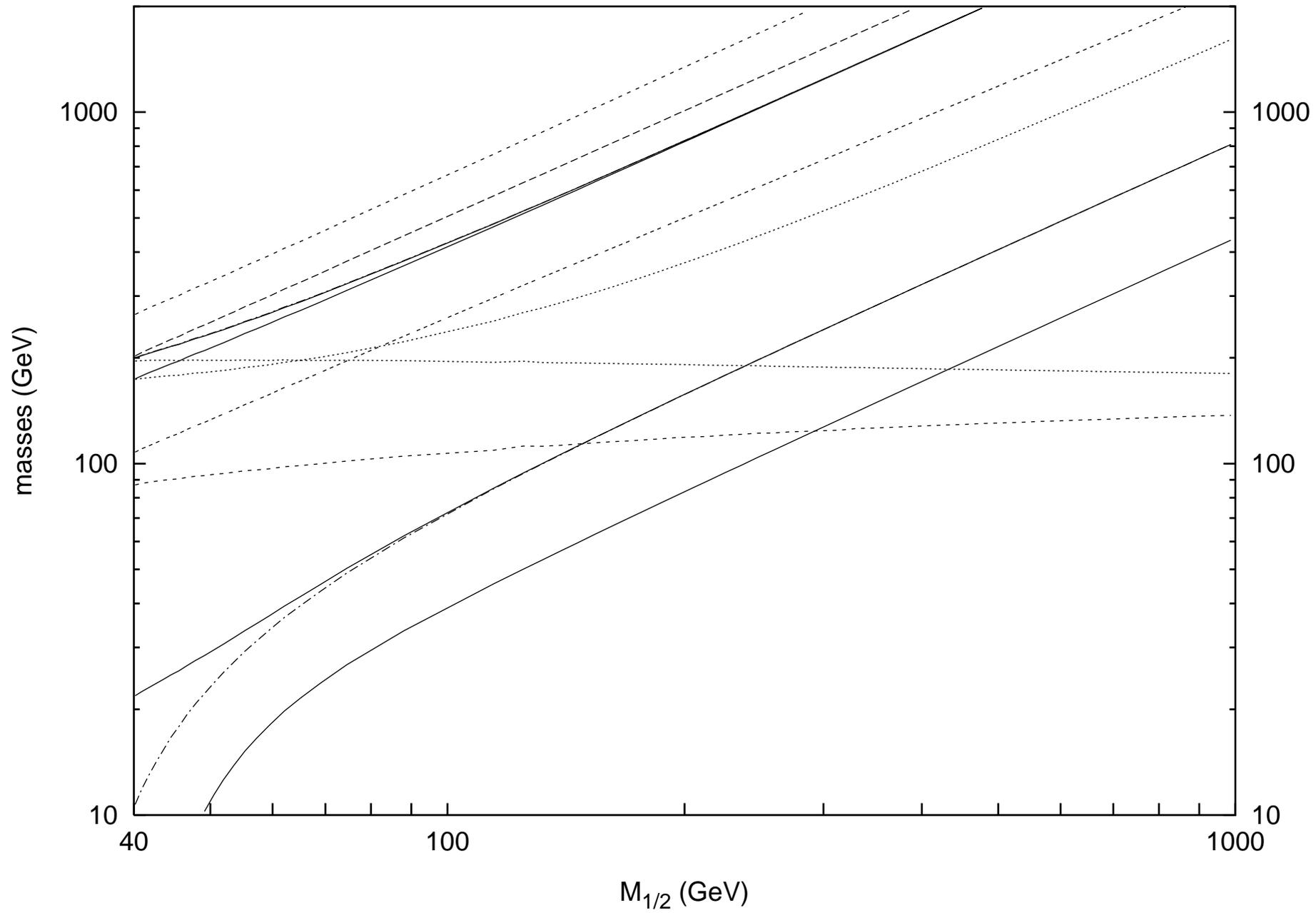
{\bf Figure 12:}
Masses of particles as a function of $M_{1/2}$, with parameters chosen
so as to mimic Figure 6. The input parameters are $h_{t0}=2$,
$A_0/m_0=-3$, $\tilde{m}_0=5$,
$\tilde\mu_0=-7.5$, $\tilde B_0=8.8-10.2$.
Neutralinos (solid lines), charginos (dot-dashed
lines), CP-even Higgs (short dashed lines), lighter stop and top quark
(both dotted lines), left-handed sleptons (long dashed lines), and
gluino (quadruple dashed lines) are displayed.

\noindent
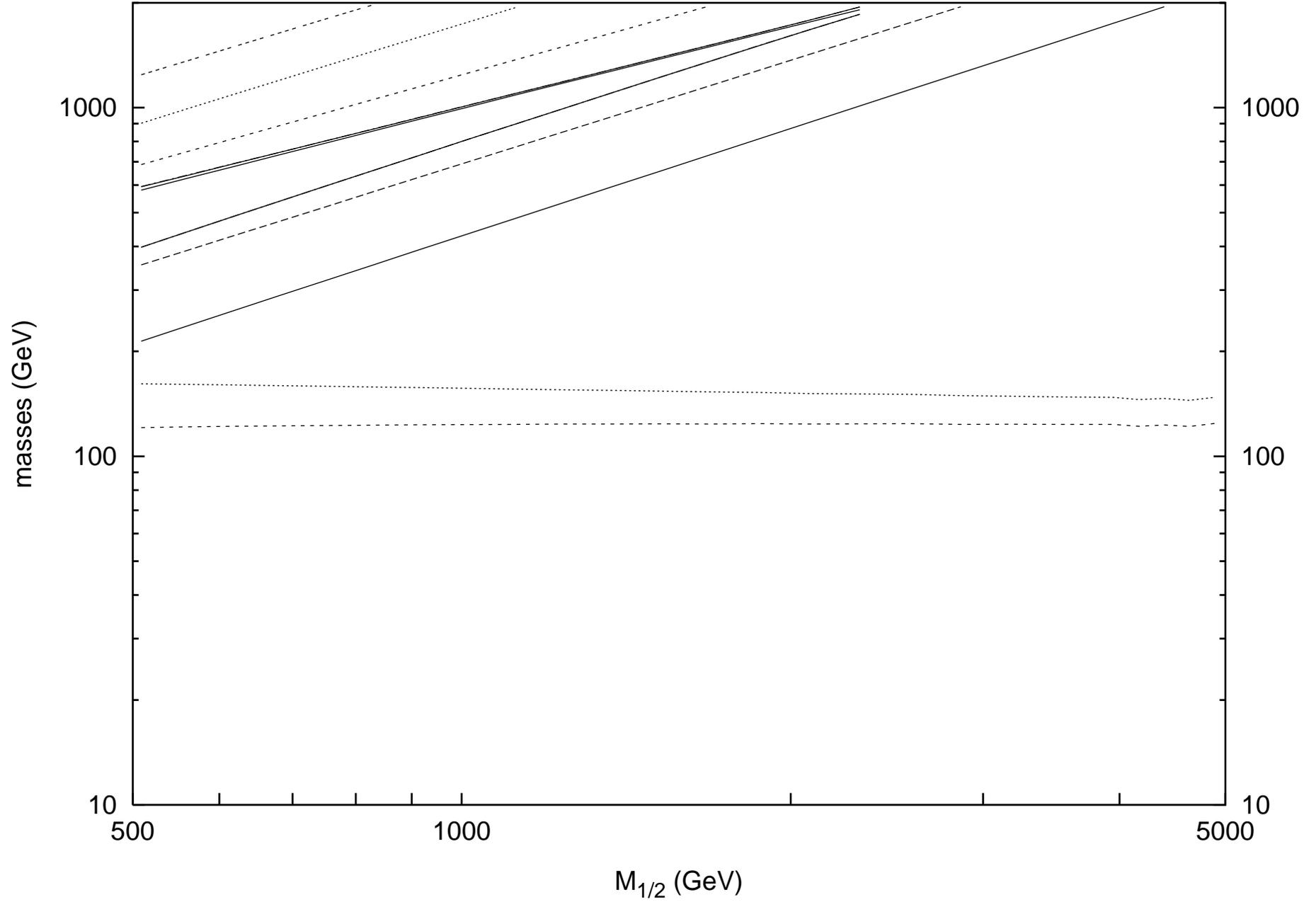
{\bf Figure 13:}
Masses of particles as a function of $M_{1/2}$, with parameters chosen
so as to mimic Figure 8. The input parameters are $h_{t0}=0.5$,
$A_0/m_0=-5$, $\tilde{m}_0=0.02$, $\tilde\mu_0=1.1$ to 0.5, $\tilde
B_0$=-0.015 to -0.135.  Neutralinos (solid lines), charginos
(dot-dashed lines), CP-even Higgs (short dashed lines), lighter stop
and top quark (both dotted lines), left-handed sleptons (long dashed
lines), and gluino (quadruple dashed lines) are displayed. Line styles
are as in Figure 6.

\noindent
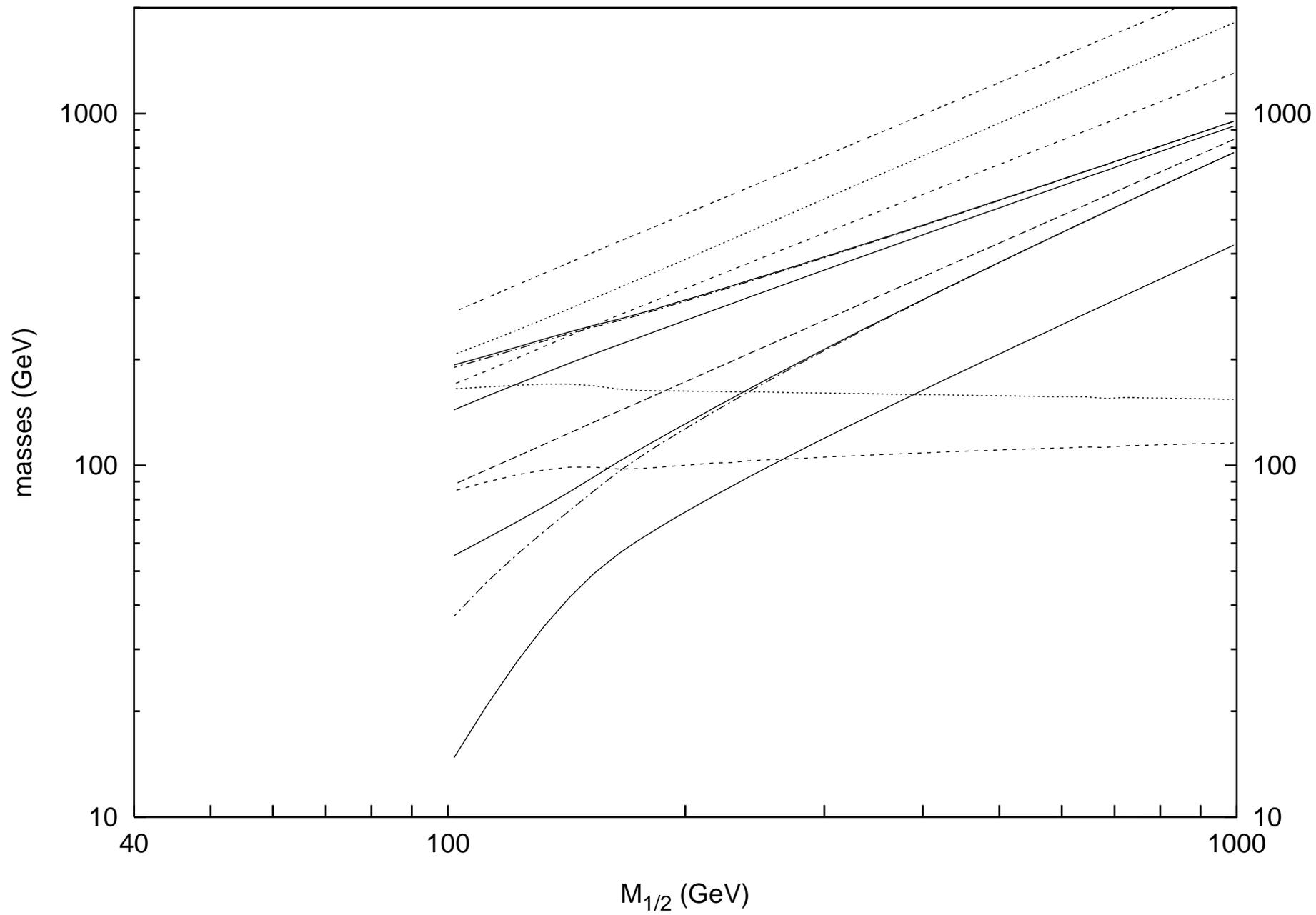
{\bf Figure 14:}
Masses of particles as a function of $M_{1/2}$, with parameters chosen
so as to mimic Figure 7. The input parameters are $h_{t0}=0.5$,
$A_0/m_0=-4$, $\tilde{m}_0=0.5$, $\tilde\mu_0=-1.3$ to -0.9, $\tilde
B_0=1.1$.  Neutralinos (solid lines), charginos (dot-dashed lines),
CP-even Higgs (short dashed lines), lighter stop and top quark (both
dotted lines), left-handed sleptons (long dashed lines), and gluino
(quadruple dashed lines) are displayed.

\noindent
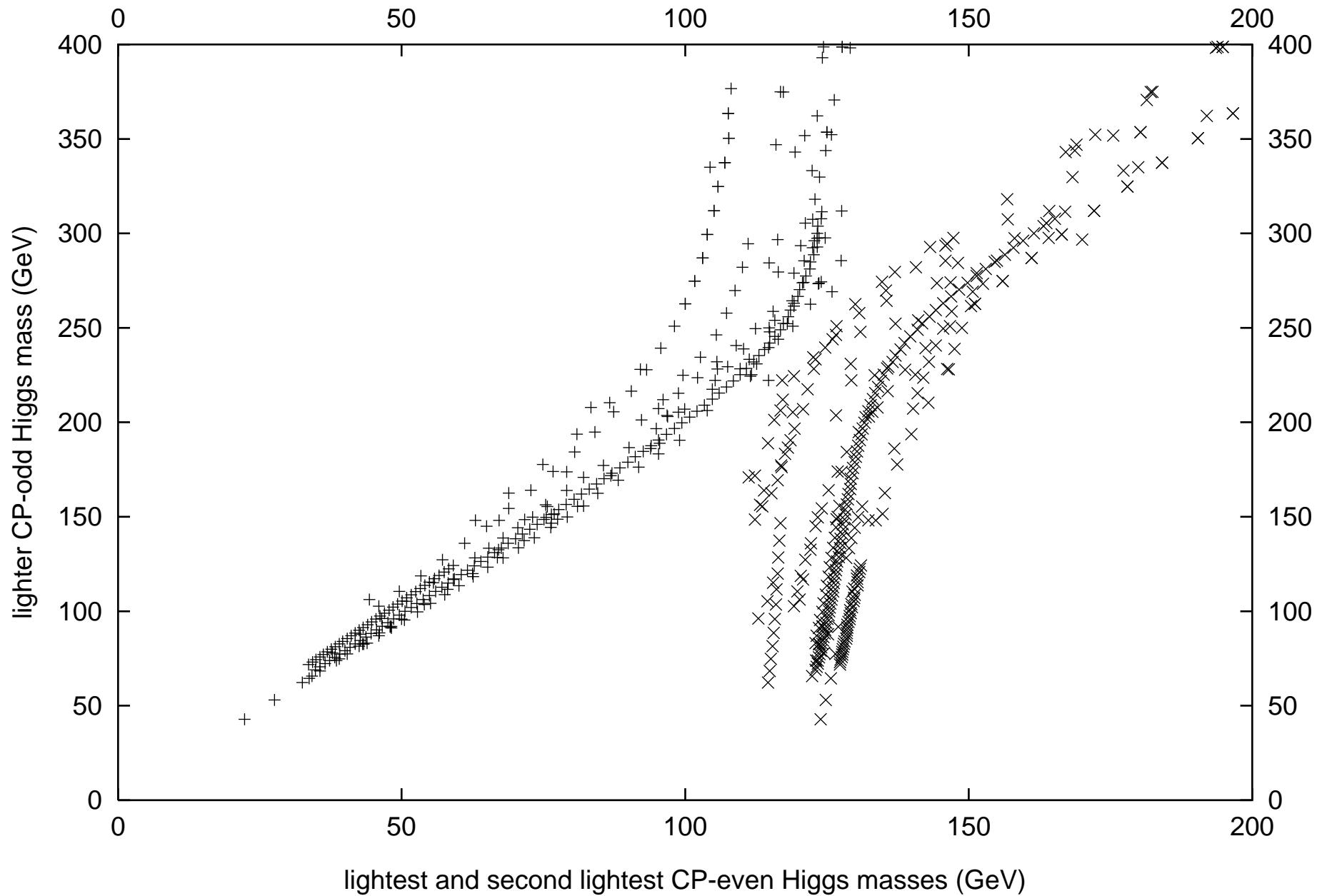
{\bf Figure 15:}
Scatter plot showing the mass of the lightest CP-odd Higgs boson
against those of the lightest (+) and second lightest CP-even Higgs
bosons ($\times$) in the NMSSM.

\noindent
{\bf Figure 16a:}
Masses of the lightest CP-even Higgs boson in the NMSSM plotted against
$R_{ZZh_1}$ (+),  and the second lightest CP-even Higgs
boson, plotted against $R_{ZZh_2}$ ($\times$). The input
parameters are $h_{t0}=0.5$, $A_0/m_0=-5$, $\tilde{m}_0=0.02$,
$\lambda_0=0.005$, $k_0=0.0002-0.0003$. The output parameters
include $M_{1/2}\approx 600-2600$ GeV, $m_t\approx 161-152$ GeV, $\tan
\beta >10$. Also shown is the region excluded by LEPI
(short-dashed line), the discovery reach of LEPII at an energy of 175
GeV and an integrated luminosity of 500 pb$^{-1}$ (long-dashed line)
and the discovery reach of LEPII at an energy of 205 GeV and an
integrated luminosity of 300 pb$^{-1}$ (full line).

\noindent
{\bf Figure 16b:}
$R_{\bar{d}dh_1}$ (+),
$R_{\bar{u}uh_1}$ (circles),
$R_{\bar{d}dh_2}$ ($\times$),
$R_{\bar{u}uh_2}$ (squares),
against the corresponding Higgs boson mass,
for the data shown in Fig.16a.

\noindent
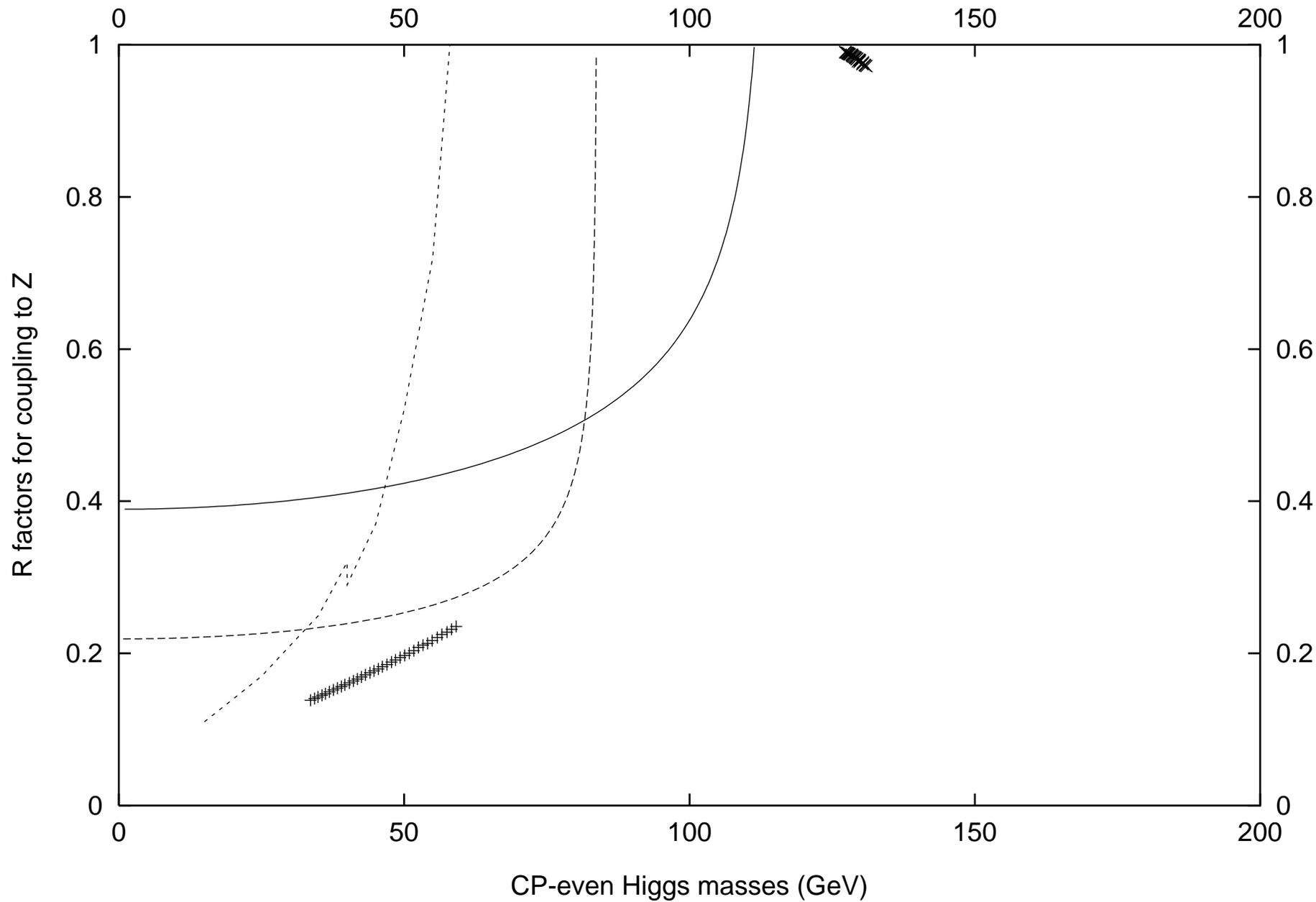
{\bf Figure 17:}
Masses of the lightest CP-even Higgs boson in the NMSSM plotted against
$R_{ZZh_1}$ (+), and the second lightest CP-even Higgs boson, plotted
against $R_{ZZh_2}$ ($\times$), for an example of scenario (i) defined
in the text. The input parameters are $h_{t0}=0.5$, $|A_0/m_0|=5$,
$\tilde{m}_0=0.01$, $\lambda_0=0.005$, $k_0=0.00012-0.00014$. The
output parameters include $M_{1/2}\approx 1-2$TeV, $m_t\approx
156-153$ GeV, $\tan \beta \approx 10-20$. Also shown is the region
excluded by LEPI (short-dashed line), the discovery reach of LEPII at
an energy of 175 GeV and an integrated luminosity of 500 pb$^{-1}$
(long-dashed line) and the discovery reach of LEPII at an energy of 205
GeV and an integrated luminosity of 300 pb$^{-1}$ (full line).

\noindent
{\bf Figure 18a:}
Masses of the lightest CP-even Higgs boson in the NMSSM plotted against
$R_{ZZh_1}$ (+), and the second lightest CP-even Higgs boson, plotted
against $R_{ZZh_2}$ ($\times$), for some examples of scenario (ii)
defined in the text. The parameters for examples of Higgs bosons
visible in the first phase of LEPII (within the dashed region) are:
$h_{t0}=0.4-0.5$, $A_0/m_0=-3$ to -5, $\tilde{m}_0=0.02$,
$\lambda_0=0.01$, $k_0=0.0004-0.0006$ with output parameters including
$M_{1/2}\approx 1-2$TeV, $m_t\approx 140-160$ GeV, $|\tan\beta|\approx
9-18$. The parameters for examples of Higgs bosons visible in the final
phase of LEPII (within the solid region) are $h_{t0}=0.4-0.5$,
$A_0/m_0=-3$ to -5, $\tilde{m}_0=0.02-0.2$, $\lambda_0=0.01-0.02$,
$k_0=0.0005-0.006$. The output parameters in this case include
$M_{1/2}\approx 200-2000$ GeV, $m_t\approx 140-165$ GeV,
$|\tan\beta|\approx 6-12$.

\noindent
{\bf Figure 18b:}
$R_{\bar{d}dh_1}$ (+),
$R_{\bar{u}uh_1}$ (circles),
$R_{\bar{d}dh_2}$ ($\times$),
$R_{\bar{u}uh_2}$ (squares),
against the corresponding Higgs boson mass,
for the data shown in Fig. 18a.

\noindent
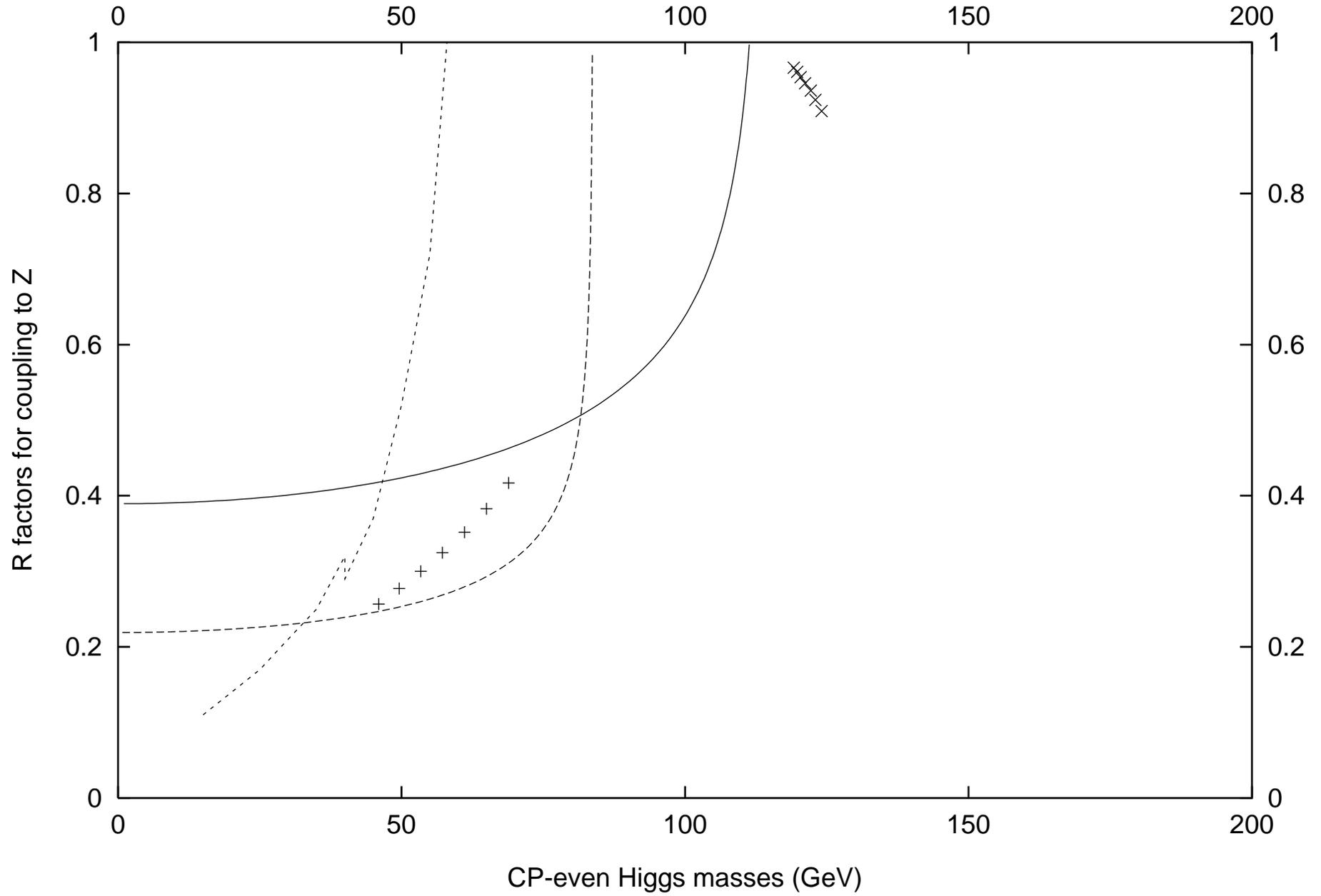
{\bf Figure 19:}
Masses of the lightest CP-even Higgs boson in the NMSSM plotted against
$R_{ZZh_1}$ (+),  and the second lightest CP-even Higgs
boson, plotted against $R_{ZZh_2}$ ($\times$), for a situation
close to scenario (iii) defined in the text. The parameters are
$h_{t0}=0.4$, $A_0/m_0=-5$, $\tilde{m}_0=0.02$, $\lambda_0=0.01$,
$k_0=0.00055-0.00060$. The output parameters in this case include
$M_{1/2}=900-1200$ GeV, $m_t\approx 143-145$ GeV, $\tan \beta \approx
10$. The second Higgs bosons in this case have masses of about 120 GeV,
outside the range of LEPII.


\clearpage

\begin{thebibliography}{99}


\bibitem{SUSY}
For reviews see {\it e.g.} \\
H.P. Nilles, Phys. Rep. {\bf 110} (1984) 1; \\
H.E. Haber, G.L. Kane, Phys. Rep. {\bf 117} (1985) 75.

\bibitem{GQW}
H, Georgi, S. Glashow, Phys. Rev. Lett. {\bf 32} (1974) 438; \\
H. Georgi, H.R. Quinn, S. Weinberg,
 Phys. Rev. Lett. {\bf 33} (1974) 451.

\bibitem{Amaldi}
U. Amaldi, W. de Boer, H. Furstenau,
 Phys. Lett. {\bf 260 B} (1991) 447; \\
J. Ellis, S. Kelley, D. V. Nanopoulos,
 Phys. Lett. {\bf 260 B} (1991) 131; \\
P. Langacker, M. Luo,
 Phys. Rev. {\bf D44} (1991) 817.

\bibitem{ir}
L. Ibanez, G. G. Ross, Phys. Lett. {\bf 110B} (1982) 215; \\
H. P. Nilles, Phys. Lett. {\bf B115} (1982) 193; \\
K.~Inoue, A.~Kakuto, H.~Komatsu, S.~Takashita,
 Prog. Theor. Phys. {\bf 68} (1982) 927; \\
L. Alvarez-Gaum{\'e}, J. Polchinski, M. Wise,
 Nucl. Phys. {\bf B221} (1983) 495;
J. Ellis, D. Nanopoulos, K. Tamvakis,
 Phys. Lett. {\bf B121} (1983) 123.

\bibitem{RR}
G.G. Ross, R.G. Roberts,
 Nucl. Phys. {\bf B 377} (1992) 571.

\bibitem{all}
R. Arnowitt, P. Nath,
 Phys. Lett. {\bf B 287} (1992) 89;
 Phys. Rev. {\bf D69} (1992) 725;
 Phys. Lett. {\bf B 289} (1992) 368;
 Phys. Lett. {\bf B 299} (1993) 58;
 Phys. Rev. Lett. {\bf 70} (1993) 3696; \\
V. Barger, M. S. Berger, P. Ohman,
 Phys. Rev. {\bf D47} (1993) 1093;
 Phys. Rev. {\bf D49} (1994) 4908; \\
W. de Boer, R. Ehret, D. I. Kazakov,
 IEKP-KA/94-05; \\
M. Carena, S. Pokorski, C.E.M. Wagner,
 Nucl. Phys. {\bf B406} (1993) 59; \\
D.J. Castano, E.J. Piard, P. Ramond,
 Phys.Rev. {\bf D49} (1994) 4882; \\
G. Kane, C. Kolda, L. Roszkowski and J. Wells,
 Phys. Rev. {\bf D49} (1994) 6173;\\
S. Kelley, J. L. Lopez, D. V. Nanopoulos, H. Pois and K. Yuan,
 Phys. Lett. {\bf B273} (1991) 423;
 Nucl. Phys. {\bf B 398} (1993) 3; \\
J. Lopez, D.V. Nanopoulos and H. Pois,
 Phys. Rev. {\bf D47} (1993) 2468; \\
J. Lopez, D.V. Nanopoulos, H. Pois, X. Wang and A. Zichichi,
 Phys. Lett. {\bf B306} (1993) 73; \\
S. P. Martin, P. Ramond,
 Phys. Rev. {\bf D48} (1993) 5365; \\
M. Olechowski, S. Pokorski,
  Nucl. Phys. {\bf B404} (1993) 590;\\
R. G. Roberts and L. Roszkowski,
 Phys. Lett. {\bf B309} (1993) 329; \\
I. Jack, D.R.T. Jones and K.L. Roberts,
hep-ph/9505242.

\bibitem{NMSSM1}
P. Fayet, Nucl. Phys. {\bf B90} (1975) 104.

\bibitem{NMSSM2}
H.P. Nilles, M. Srednicki, D. Wyler,
 Phys. Lett. {\bf B120} (1983) 346;\\
J.M. Frere, D.R.T. Jones, S. Raby,
 Nucl. Phys. {\bf B222} (1983) 11;\\
J.-P. Derendinger, C.A. Savoy,
 Nucl. Phys. {\bf B237} (1984) 307;\\
L. Durand, J. L. Lopez, Phys. Lett. {\bf B217} (1989) 463; \\
M. Drees, Intern. J. Mod. Phys. {\bf A4} (1989) 3645.

\bibitem{NMSSM3}
J. Ellis, J. Gunion, H. Haber, L. Roszkowski and F. Zwirner,
 Phys. Rev. {\bf D39} (1989) 844.

\bibitem{mu}
L. Hall, J. Lykken, S. Weinberg,
 Phys. Rev. {\bf D27} (1983) 2359; \\
J.E. Kim, H.P. Nilles,
 Phys. Lett. {\bf B138} (1984) 150; \\
K. Inoue, A. Kakuto, T. Takano,
 Prog. Theor. Phys. {\bf 75} (1986) 664; \\
A. A. Ansel'm, A. A. Johansen,
 Phys. Lett. {\bf B200} (1988) 331; \\
G. Giudice, A. Masiero,
 Phys. Lett. {\bf B206} (1988) 480.


\bibitem{stabone}
S. Ferrara, D.V. Nanopoulos, C. A. Savoy,
 Phys. Lett. {\bf B123} (1983) 214; \\
J. Polchinski, L. Susskind,
 Phys. Rev. {\bf 26} (1982) 3661; \\
H.P. Nilles, M. Srednicki and D. Wyler,
 Phys. Lett. {\bf B124} (1982) 337; \\
A. B. Lahanas, Phys. Lett. {\bf B124} (1982) 341; \\
L. Alvarez-Gaume, J. Polchinski, M. B. Wise,
 Nucl. Phys. {\bf B221} (1983) 495.

\bibitem{stabtwo}
U. Ellwanger, Phys. Lett. {\bf B133} (1983) 187; \\
J. Bagger, E. Poppitz, Phys. Rev. Lett. {\bf 71} (1993) 2380; \\
J. Bagger, E. Poppitz, L. Randall, preprint EFI-95-21, hep-ph/9505244.

\bibitem{ell}
U. Ellwanger, M. Rausch de Traubenberg, C. A. Savoy,
 Phys. Lett. {\bf B315} (1993) 331.

\bibitem{ell2}
Ph. Brax, U. Ellwanger, C. A. Savoy,
 Phys. Lett. {\bf B347} (1995) 269.

\bibitem{ell3}
U. Ellwanger, M. Rausch de Traubenberg, C. A. Savoy,
 LPTHE Orsay 95-04, LPT Strasbourg 95-01,
 SPhT Saclay T95/04, hep-ph/9502206.

\bibitem{ekw0}
T. Elliott, S. F. King, P. L. White,
 SHEP 93/94-19, hep-ph/9406303,
 Phys Lett {\bf B} (to appear).

\bibitem{walls}
A. Vilenkin, Phys. Rep. {\bf 121} (1985) 263.

\bibitem{NMSSMwalls}
G. B. Gelmini, M. Gleiser, E. W. Kolb,
 Phys. Rev. {\bf D39} (1989) 1558; \\
W. H. Press, B. S. Ryden, D. N. Spergel,
 Ap. J. {\bf 357} (1990) 293; \\
B. Rai, G. Senjanovic, Phys. Rev. {\bf D49} (1994) 2729; \\
S.A. Abel, P.L. White, preprint OUTP-9517P, CCL-TR-95-005,
 hep-ph/9505241; \\
S.A. Abel, S. Sarkar, P.L. White, in preparation.

\bibitem{nocpviol}
J. C. Rom\~{a}o, Phys Lett {\bf B173} (1986) 309.

\bibitem{effpot1}
S. Coleman and E. Weinberg, Phys Rev {\bf D7} (1973) 1888; \\
S. Weinberg, Phys Rev {\bf D7} (1973) 2887; \\
M. Sher, Phys Rep {\bf 179} (1989) 273.

\bibitem{effpot2}
G. Gamberini, G. Ridolfi and F. Zwirner, Nucl Phys {\bf B331} (1990) 331.

\bibitem{pierce}
D. Pierce, A. Papadopoulos,
 Phys. Rev. {\bf D50} (1994) 565;
 Nucl. Phys. {\bf B430} (1994) 278.

\bibitem{gencon}
M. Pohl, in ``Proceedings of the XXVII International Conference on
High Energy Physics'', eds. P. J. Bussey, I. G. Knowles,
IOP Publishing (Bristol) 1994.

\bibitem{hagelin}
J. Hagelin, S. Kelley, T. Tanaka, Nucl. Phys. {\bf B415} (1994) 293; \\
J. Hagelin, S. Kelley, T. Tanaka, Mod. Phys. Lett. {\bf A8} (1993) 2737; \\
G. C. Branco, G. C. Cho, Y. Kizukuri, N. Oshimo,
 Phys. Lett. {\bf B337} (1994) 316;\\
F. Gabbiani, A. Masiero, Nucl. Phys. {\bf B322} (1982) 235.

\bibitem{inobnd}
L.~Roszkowski, Phys. Lett. {\bf B252} (1990) 471.

\bibitem{NMSSMchicons}
F.~Franke, H.~Fraas, A.~Bartl, Phys. Lett. {\bf B336} (1994) 415.

\bibitem{ppbp}
L. Roszkowski, Phys. Lett. {\bf B262} (1991) 59;
J. L. Lopez, D. V. Nanopoulos, K. Yuan,
 Nucl. Phys. {\bf B370} (1992) 445; \\
S. Kelley, J. L. Lopez, D. V. Nanopoulos, H. Pois, K. Yuan,
 Phys. Rev. {\bf D47} (1993) 2461.

\bibitem{NMSSMcosmology}
K. Olive, D. Thomas, Nucl. Phys. {\bf B335} (1991) 192; \\
R.A. Flores, K. Olive, D. Thomas, Phys. Lett. {\bf B263} (1991) 425; \\
S.A. Abel, S. Sarkar, I.B. Whittingham, Nucl. Phys. {\bf B392} (1993) 83.

\bibitem{LEP1}
The ALEPH collaboration, Phys. Lett. {B313} (1993) 312; \\
The DELPHI collaboration, Nucl. Phys. {B373} (1992) 3; \\
The OPAL collaboration, Z. Phys. {C64} (1994) 1; \\
The L3 collaboration, Z. Phys. {C57} (1993) 355.

\bibitem{NMSSMhiggscon}
F.~Franke, H.~Fraas, preprint WUE-ITP-95-003, hep-ph/9504279.

\bibitem{bsg}
J. L. Hewett, preprint SLAC-PUB-6521, hep-ph/9406302.

\bibitem{CDF}
The CDF Collaboration, Phys. Rev. Lett. {\bf 73} (1994) 225;
 Phys. Rev. {\bf D50} (1994) 2966;
 FERMILAB-PUB-95/022-E; \\
The D0 Collaboration, FERMILAB-PUB-95/028-E.

\bibitem{ft}
R. Barbieri, G.F. Giudice, Nucl. Phys. {\bf B306} (1988) 63.

\bibitem{CC}
B. de Carlos, J.A. Casas, Phys. Lett. {\bf B309} (1993) 320.

\bibitem{langpol}
P. Langacker, N. Polonsky, Phys. Rev. {\bf D47} (1993) 4028;\\
M. Carena, S. Pokorski, C.E.M. Wagner, Nucl. Phys. {\bf B406} (1993) 59.

\bibitem{MSSMbound}
H. E. Haber and R. Hempfling, Phys Rev Lett {\bf 66} (1991) 1815; \\
Y. Okada, M. Yamaguchi and T. Yanagida,
 Prog. Theor. Phys. {\bf 85} (1991) 1.

\bibitem{tlNbound}
J. Espinosa, M. Quiros, Phys. Lett. {\bf B279} (1992) 92;
 Phys. Lett. {\bf B302} (1993) 51; \\
G. Kane, C. Kolda, J. Wells, Phys. Rev. Lett. {\bf 70} (1993) 2686.

\bibitem{NMSSMbound}
U. Ellwanger, M. Rausch de Traubenberg, Z Phys {\bf C53} (1992) 521; \\
U. Ellwanger, M. Lindner, Phys Lett {\bf B301} (1993) 365; \\
U. Ellwanger, Phys Lett {\bf B303} (1993) 271; \\
P. N. Pandita, Phys Lett {\bf B318} (1993) 338; Z Phys {\bf C59}
(1993) 575; \\
T. Elliott, S. F. King and P. L. White, Phys Lett {\bf B305} (1993) 71;
{\bf B314} (1993) 56.

\bibitem{copw}
B. Anathanarayan, G. Lazarides and Q. Shafi,
Phys. Lett. {\bf B300} (1993) 245;
L. Hall, R. Rattazzi and U. Sarid,
Phys. Rev. {\bf D50} (1994) 7048;
M.~Carena, M.~Olechowski, S.~Pokorski, C.~Wagner, ``Electroweak
Symmetry Breaking and Bottom-Top Yukawa Unification'',
preprint MPI-Ph/93-103; \\
E.G. Floratos, G. K. Leontaris, Phys.Lett. {\bf B336} (1994) 194;
 preprint IOA-320-95, hep-ph/9503455;
B. Anathanarayan and P.N. Pandita,
hep-ph/9503323.

\bibitem{ekw}
T. Elliott, S. F. King, P. L. White, Phys Rev {\bf D49} (1994) 2435.

\bibitem{Haber}
H. Haber, Santa Cruz preprint SCIPP 94/39, hep-ph/9501320.


\bibitem{hhg}
J. F. Gunion, H. E. Haber, G. Kane, S. Dawson,
 ``The Higg's Hunter's Guide'' Addison-Wesley (1990).


\bibitem{mikeandco}
D.R.T. Jones, Phys. Rev. {\bf D25} (1982) 581; \\
M. E. Machacek, M. T. Vaughn, Nucl. Phys. {\bf B236} (1984) 221;\\
M. E. Machacek, M. T. Vaughn, Nucl. Phys. {\bf B222} (1983) 83; \\
S. P. Martin, M. T. Vaughn, Phys.Rev. {\bf D50} (1994) 2282.

\bibitem{notation}
J. Ellis, G. Ridolfi, F. Zwirner, Phys Lett {\bf B257} (1991) 83;
{\bf B262} (1991) 477; \\
A. Brignole, J. Ellis, G. Ridolfi, F. Zwirner, Phys Lett {\bf B271}
(1991) 123.


\end{thebibliography}
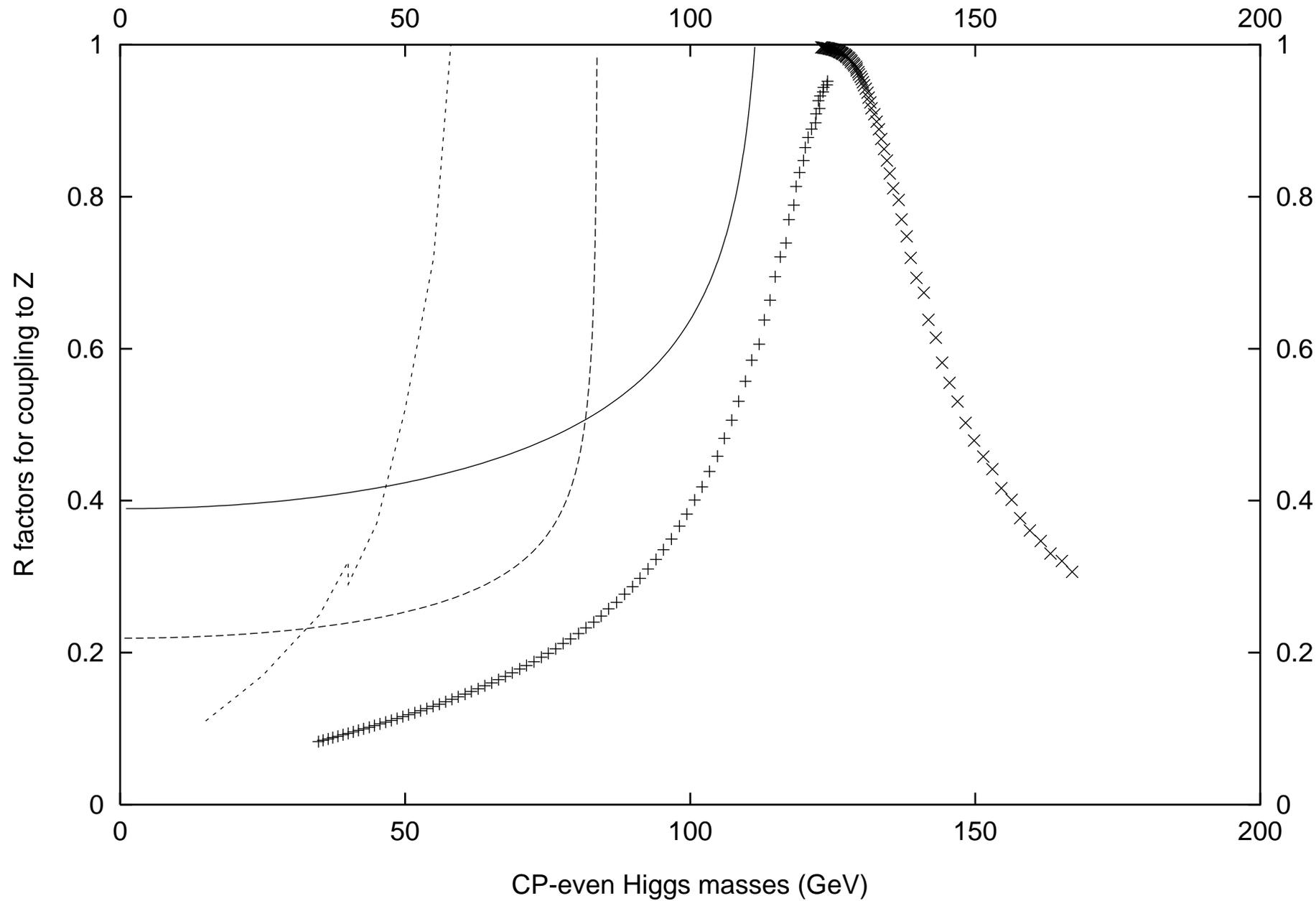
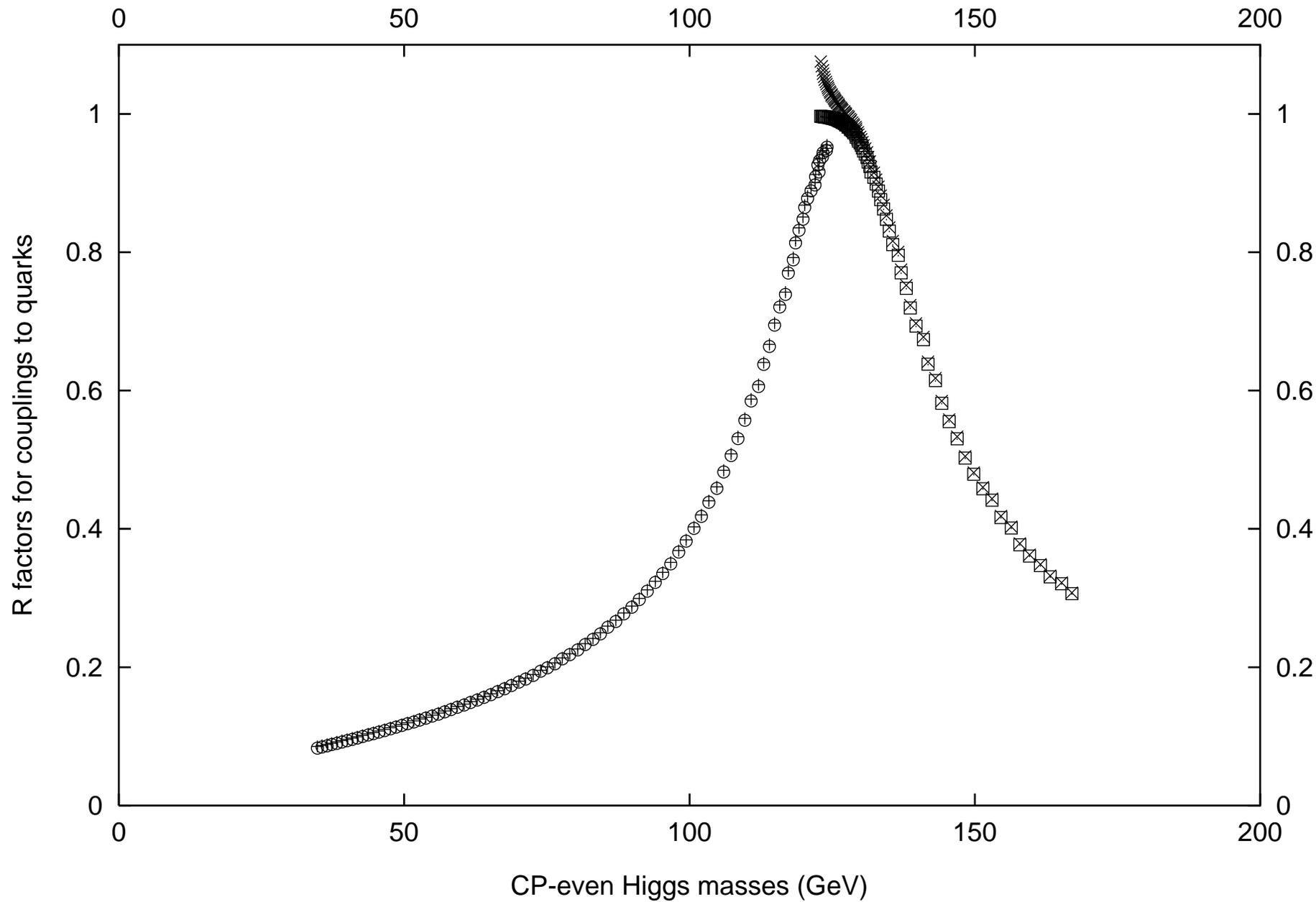
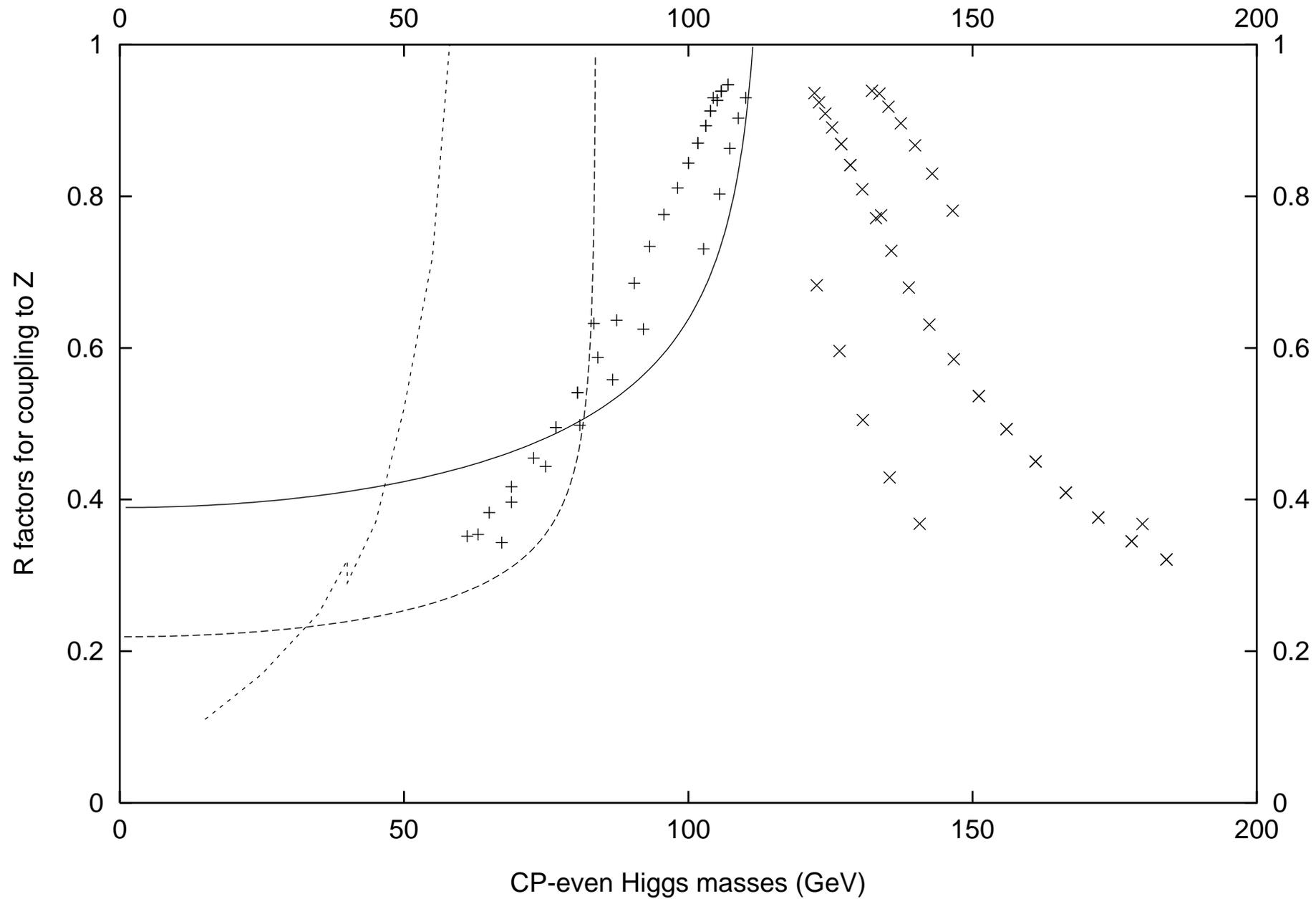
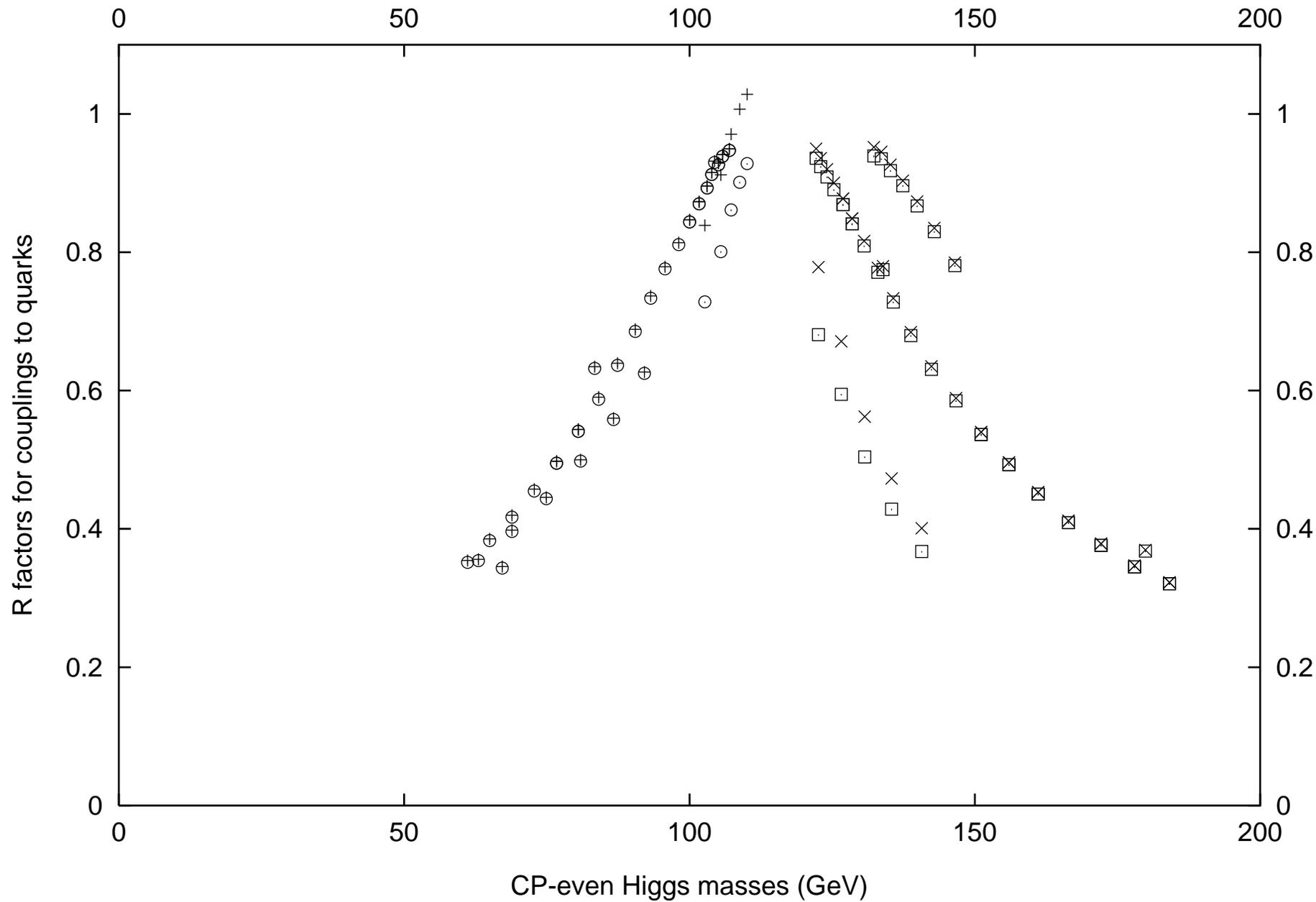
\end{document}